\documentclass[aps,pre,onecolumn,superscriptaddress]{revtex4-1}  
\usepackage{graphicx}  
\usepackage{dcolumn}   
\usepackage{bm}        
\usepackage{amssymb}   
\usepackage{amsmath}
\usepackage{epsfig}
\usepackage[utf8]{inputenc}
\usepackage{hyperref}

\begin{document}

\title
{
Quasistatic kinetic avalanches and self-organized criticality in deviatorically loaded granular media}

\begin{abstract}
  The behavior of granular media under quasi-static loading has recently been shown to attain a stable evolution state corresponding to a manifold in the space of micromechanical variables. This state is characterized by sudden transitions between metastable jammed states, involving the partial micromechanical rearrangement of the granular medium.  Using numerical simulations of two-dimensional granular media under quasistatic biaxial compression, we show  that the dynamics in the stable evolution state is characterized by scale-free avalanches well before the macromechanical stationary flow regime traditionally linked to a self-organized critical state. This, together with the non-uniqueness and the non-monotony of macroscopic deformation curves, suggests that the statistical avalanche properties and the susceptibilities of the system cannot be reduced to a function of the macromechanical state. The associated scaling exponents are non-universal and depend on the interactions between particles. For stiffer particles (or samples at low confining pressure) we find  distributions of avalanche properties compatible with the predictions of mean-field theory. The scaling exponents decrease below the mean-field values for softer interactions between particles. These lower exponents are consistent with observations for amorphous solids at their critical point. We specifically discuss the relationship between microscopic and macroscopic variables, including the relation between the external stress drop and the internal potential energy released during kinetic avalanches. 
\end{abstract}

\author{Jordi Bar\'o}
\email{jbaro@crm.cat}
\affiliation{Department of Physics and Astronomy,
University of Calgary.
2500 University Drive NW
Calgary, Alberta  T2N 1N4, Canada}

\affiliation{Centre for Mathematical Research,
Campus de Bellaterra, Edifici C	 
08193 Bellaterra, Barcelona, Spain}

\author{Mehdi Pouragha}
\affiliation{Civil Engineering Department, 
University of Calgary.
2500 University Drive NW
Calgary, Alberta  T2N 1N4, Canada}
\affiliation{Department of Civil and Environmental Engineering, 
Carleton University.
1125 Colonel By Drive, Ottawa, ON K1S 5B6 Canada}

\author{Richard Wan}
\affiliation{Civil Engineering Department, 
University of Calgary.
2500 University Drive NW
Calgary, Alberta  T2N 1N4, Canada}

\author{J\"orn Davidsen}%
\affiliation{Department of Physics and Astronomy,
University of Calgary.
2500 University Drive NW
Calgary, Alberta  T2N 1N4, Canada} 
\affiliation{Hotchkiss Brain Institute,
University of Calgary.
3330 Hospital Drive NW,
Calgary, Alberta  T2N 4N1, Canada}
  
\maketitle

\section{Introduction}

Mechanical systems with evolving internal structure often exhibit intermittency or avalanche dynamics under quasistatic mechanical driving (see~\cite{Sethna2001,Salje2014,Nicolas2018} for reviews). 
Specifically, intermittent athermal dynamics dominated by slip dislocations are observed in the plastic deformation of single crystals~\cite{Miguel2001,Zaiser2006,Friedman2012} as well as amorphous materials such as  bulk metallic~\cite{Wang2009, Antonaglia2014} and colloidal~\cite{Schall2007} glasses, high entropy alloys~\cite{Hu2018,Rizzardi2018}, 
and granular assemblies~\cite{Dahmen2011,Denisov2016,Berthier2019a}, with their prevalent dynamics being successfully captured through discrete element (DEM) simulations in idealized athermal quasistatic (AQS) conditions considering Lennard-Jones (LJ) particles --- simulating  molecular dynamics in glassy materials~\cite{Maloney2004, Salerno2012, Salerno2013, Parisi2017, Zhang2017, Shang2019} --- as well as soft spheres \cite{Lerner2009,Karmakar2010} --- simulating assemblies of larger spherical aggregates --- and elastic interactions --- simulating granular matter \cite{Gimbert2013, Pouragha2016, Karimi2018}. 
In granular materials, where potentials are exclusively repulsive, intermittency occurs upon the activation of \textit{fragile} jammed states~\cite{Cates1998,Liu1998}, where the system is consolidated in marginally stable solid-like configurations that become unstable under infinitesimal stress variations. Jamming states can only exist within certain conditions \cite{Liu1998}. Under external mechanical driving, and in absence of thermal fluctuations, jamming is observed to be limited below a macroscopic shear stress threshold, $\sigma_c$, often referred to as a critical point, which becomes stationary after an initial transient regime~\cite{Bi2011}. When the system is pushed above $\sigma_c$, jammed states are unlikely and the excess of free energy (corresponding to the excess of shear stress) is dissipated through particle rearrangements.
Under \textit{strain} driving, such dissipation reestablishes the mechanical stability at stress values below $\sigma_c$.
Such a sequence of quasistatic metastable states leads to an elastic/plastic stress-deformation response and the macromechanical state of the system eventually hovers around $\sigma_c$, which can be considered a case of \textit{self-organized criticality} (SOC)~\cite{Bak1991,Chen1991,Zapperi1995,Watkins2016} characterized by critical avalanche dynamics exhibiting scale-free behavior. \\
 
The hypothesis of SOC in the deformation of amorphous materials inspired the development of the theory of \textit{elasto-plasticity}, bridging the gap between macromechanical rheology models and SOC ~(see \cite{Nicolas2018} for an extensive review). 
Elastoplastic models (EPM), as defined in soft-matter physics \cite{Nicolas2018}, represent amorphous materials as lattices of bistable -- either elastic or plastic -- representative volume elements or sites. Under strain or stress driving, each site experiences an affine elastic deformation. When the local stress at a site overcomes a local threshold, representing a local mechanical instability, the element undergoes a local non-affine plastic-deformation event involving a predefined stress propagator interacting with other elements and triggering an avalanche as a cascading event.  
Conceptual mean field models interpret the residual stresses to local instabilities as scalars undergoing a stochastic process \cite{Hebraud1998}. Within this later category, the simplified and mathematically solvable slip mean field theory (SMFT) \cite{Ben-Zion1993,Dahmen2009, Dahmen2011,Dahmen2017}, considering all-to-all homogeneous site interactions, provides a remarkably apt conceptual model for a wide spectrum of processes including the macromechanics and the statistical features of externally measured avalanche dynamics in granular media~\cite{Dahmen1998,Mehta2006,Dahmen2011,Geller2015,Denisov2016,Zhang2017}. Such results reinforced the idea of a predominant mean-field universality class in the avalanche statistics \cite{Corwin2005,Dahmen2011, Geller2015,Denisov2016,Zhang2017, Parisi2017, Nicolas2018}.
In SMFT, SOC appears in the thermodynamic limit as an asymptotic approach to a regime at stress value $\sigma_c$ characterized by a divergence of scales in avalanche statistics. Following scaling relations \cite{Dahmen2017}, the macromechanical evolutionary path of strain-driven ($\epsilon$) amorphous solids leading to the steady-state flow regime is uniquely represented in the SMFT by the stress-strain relation:
\begin{equation}
\sigma(\epsilon) \approx \sigma_c \left({1-\exp\left({-\epsilon}\right)}\right)
\label{eq:critSMFT}
\end{equation}

Such a generalized macromechanical perspective is by construction insufficient to describe the dynamics and statistics of slip avalanches as a function of state \cite{Nicolas2018}. Both experiments and DEM simulations display a non-monotonous evolution towards $\sigma_c$, implying that the relation $\sigma (\epsilon)$ does not represent a function of state \cite{Karmakar2010}. Furthermore, initial configurations determine the $\sigma (\epsilon)$ response during such a transient regime and do not converge to $\sigma_c$ through the same evolutionary path (e.g. Fig.~\ref{fig:macromech}).
A widespread scientific endevour aims to understand the limitations and discrepancies of atomistic DEM, coarse-grained full tensorial \cite{Lin2014, Budrikis2017, Karimi2018, Korchinski2021}, scalar \cite{Chen1991, Baret2002, Picard2005}, and mean-field \cite{Hebraud1998, Picard2005, Agoritsas2015, Lin2016, Ferrero2019} EPM under the lens of the micromechanics of avalanche dynamics~\cite{Rodney2011, Budrikis2017,Nicolas2018}. 
Other studies focusing on the internal topology of the granular material suggest that the state of compact granular media is better parameterized by a micromechanical description of the particle configuration and the network of contact forces, which can be only partially represented in terms of its macromechanics. Specifically, the role of internal structures to the mechanics of granular media has been discussed in early conceptual models~\cite{Cates1998,Liu1998}, experiments~\cite{Miller1996,Mueth1998, Liu1995,Cates1998,Howell1999, Ando2012experimental,Gendelman2016, Berthier2019a,Berthier2019b}, DEM~\cite{Tordesillas2007, Guo2013signature,Kuhn1999structured,Pouragha2017non} and new mechanical-statistical theories of granular matter~\cite{Edwards1989,Blumenfeld2009, Blumenfeld2012,Bi2015,Baule2018,Bililign2019,Ball2019,Bi2011}.
Indeed, the definition of fragile matter is intrinsically linked to the existence of force-chain structures \cite{Cates1998} and experiments reveal how the nucleation of failure events is directly linked to \emph{local} topological measurements such as the betweenness centrality~\cite{Berthier2019b}. 
Parallel DEM studies identified a stable evolution state (SES) in a space of independent structural variables. The SES is reached after an adaptation transient, reached considerably before the stationary flow regime observed in the macromechanical variables~\cite{Pouragha2016}. As an extended micromechanical representation of the jamming transition, the SES describes a subset of micromechanical states upon which the internal structure of granular media starts to evolve leading to stick-slip like events or avalanches. In particular, the same spontaneous shear localization leading to the non-monotonous macromechanical state evolution can degenerate into different evolutionary paths towards the flow regime even under similar intial conditions \cite{masson2001,Desrues2004strain,Desrues1989shear,Pouragha2017strain}, reaching the same topological SES at different values of stress. The parametric study in~\cite{Pouragha2016} shows that once SES is reached, regardless of the initial condition and material parameters, the granular assembly starts to exhibit irreversible deformations that are predominantly governed by micro-avalanches. 

Here, we analyze the statistics of kinetic avalanches identified during AQS biaxial compression in DEM simulations of granular assemblies. We interpret the results in the context of different hypothesis regarding the coarse-grained description of amorphous materials.
We show how the distribution of avalanche sizes is independent of the macroscopic state of the system once SES is reached, even when the macromechanical paths from different initial conditions (e.g. loose vs. dense assemblies) differ significantly. More importantly, the SES itself exhibits the properties of a self-organized critical state. Thus, while this critical state includes the final steady-state regime at the customary \textit{critical} stress value $\sigma_c$, it is reached significantly earlier. Because of that, rather than using the unique term \textit{critical point} to designate both, we will distinguish between the final \textit{steady state regime} at $\sigma_c$ and \textit{criticality} as the statistical mechanics concept refering to a divergence of susceptabilities.
We also show that while the statistical avalanche behavior is stationary within SES, each individual avalanche modifies noticeably the state of the system, introducing a characteristic delay, a.k.a. a \textit{pseudogap}, in the interevent times between consecutives avalanches. This leads to stronger correlations between the states at which avalanches are triggered than between the states at which avalanches stop and suggests that the SES defines an evolving stability limit leading to the steady-state regime at the macroscopic scale rather than a stability basin towards which the system is relaxed after each avalanche. Moreover, we find non-universality in the power-law exponents defining the statistics of the energy transfers involved in avalanches. In this case, we expose a dependence of the exponents on the softness of the interactions between particles.
The external stress drop is found to be approximately proportional to the drop of internal elastic energy caused by configurational changes, but, due to the presence of soft lateral walls, this is not the case for the energy dissipated by frictional forces and damping. The relation between internal energy drops and the peaks of kinetic energy are consistent with a specific set of assumptions within the SMFT formulation. This consistency allows us to relate the non-universality of our results with a potential transition between the expectations of SMFT and elasto-plasticity based on the estimated \textit{critical exponents}.

The manuscript is organized as follows: 
In section~\ref{sec:DEM} we introduce our numerical DEM simulations of granular assemblies, review the definition of SES and explain how the definition of kinetic avalanches is used to implement the quastistic biaxial compression driving. In section~\ref{sec:results} we present the statistical analysis of the avalanche properties, focusing on the average relations between avalanche-size measurements, the existing correlations between events and their full distributions with the corresponding fitted exponents. In section \ref{sec:discuss} we discuss the conditions and properties of avalanche criticality and compare them with the mean-field solution of elastoplastic models. Final conclusions are presented in section \ref{sec:conclude}.\\

\section{Background and Methods}
\label{sec:DEM}

\subsection{Computational Model}

The software package PFC2D~\cite{itasca2008version} has been used here for discrete element (DEM) simulation  of 2d granular assemblies under AQS compression. A linear contact model with constant stiffness is considered along normal and tangential directions with their respective stiffnesses, $k_n$ and $k_t$ being the same and equal to $10^{9}N/m$. The tangential force is  restricted by a Coulomb friction law $f_t \le \mu f_n$ with $\mu=0.5$ for all simulations. A numerical damping of $5\%$ is also considered which acts as a drag-like force on particles by artificially ramping changes in velocities (see \cite{itasca2008version} for exact definition). The radii of circles are uniformly distributed over $r_{\min}$ and $r_{\max}$ with $r_{\max}/r_{\min} = 1.5$, around an average radius of $1~mm$. To avoid overcrowding the figures and for a better readability, the results in this paper are presented
in their dimensional forms with SI units. The material and loading parameters adopted for numerical simulations are summarized in Table~\ref{tb:samples}.

\begin{table}
	\begin{small}
		\begin{tabular}{c|cccc}
			\hline
			name			
		&	\parbox[c]{1.55cm} {num. of\\particles\\$N$}
		&   \parbox[c]{1.55cm} {confining\\pressure \\$\sigma_x (N/m)$}
		&   \parbox[c]{2.00cm} {driving\\rate\\$\dot{\epsilon}_y (\times 10^{-9}s^{-1})$}
		&	\parbox[c]{1.55cm} {initial\\porosity\\$\phi_0$}\\
			\hline
			D20kSc5 &	19520	&	$10^5$	&	2.3	&	0.156		\\
			L20kSc5 &	19353	&	$10^5$	&	2.3	&	0.190		\\
			D5kSc5	&	6374    &	$10^5$	&	2.4	&	0.159		\\
			L5kSc5	&	5504    &	$10^5$	&	2.4	&	0.192		\\
			D2kSc5	&	1593    &	$10^5$	&	1.3	&	0.165		\\
			D5kSc6  &	6374    &  	$10^6$	&	2.4	&	0.154		\\
			D5kFc5  &	6374    &	$10^5$	&	7.0	&	0.159		\\
			D5kFc6	&	6374    &	$10^6$	&	7.0	&	0.154		\\
			D5kFc7	&	6374    &	$10^7$	&	7.0	&	0.120		\\
			\hline
		\end{tabular}
	\end{small}
	\caption{\label{tb:samples} Simulation details for the nine numerical samples. The confining pressure is kept constant, the driving rate is halted during the propagation of avalanches. The initial porosity is measured at the beginning of the simulation.}
\end{table}

\begin{figure}
\centering
\includegraphics[width=0.5\columnwidth]{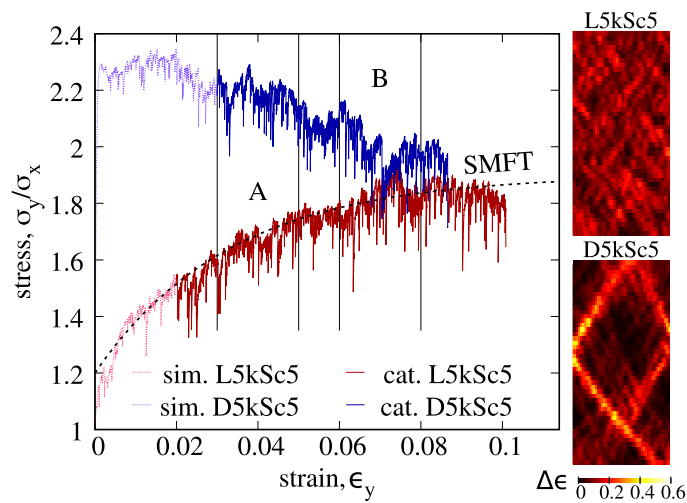} 
    \caption{\label{fig:macromech} (color online) Stress-strain response from simulation results (\textit{sim.}) of samples D5kSc5 and L5kSc5. Avalanche catalogs (\textit{cat.}) only consider the intervals represented in wider and darker lines. The black dotted line represents a fitted SMFT prediction (\ref{eq:critSMFT}) to L5kSc5. 
The statistical analysis represented in Fig.~\ref{fig:PUKsplit} correspond to the subcatalogs constrained within strain intervals $A$ and $B$ of the evolution. Right figures represent the changes in deviatoric strain ($\Delta \epsilon$) of both samples during the whole simulation, showing strong localization in D5kFc5.}
\end{figure}

The rectangular sample has an initial aspect ratio of 2-to-1 which is confined in between frictionless walls. During the initial so-called compaction stage, the particles are slowly inflated, as a common numerical technique, to obtain the target isotropic pressure on the walls. During this stage the interparticle friction is artificially modified  to achieve the desired initial density. The friction is set to $0.5$ at the end of compaction. Next, the deviatoric load is applied by moving the top and bottom walls towards each other with a straining rate of $\dot{\epsilon_{y}}$ while the stress on the lateral walls, $\sigma_x$, is kept constant through a servo-control mechanism. The wall velocity is sufficiently small to keep the inertia number well below the quasi-static limit~\cite{Da2005rheophysics}. Moreover, to ensure a quasi-static loading, the driving is halted during surges in kinetic energy, as is explained in detail later. The typical stress-strain responses and internal displacement maps are exemplified by a pair of initially loose and dense assemblies in Fig.~\ref{fig:macromech}. A relatively monotonic and homogeneous average trend is seen for loose assembly (L5kSc5) while the dense assembly (D5kSc5) experiences a peak of stress before dropping to the steady-state regime at larger axial strains. This previous transient regime is linked to the formation of shear localization bands (see right-bottom map in Fig.~\ref{fig:macromech}) 

\subsection{Internal dynamics}

At any given instance, the energy of the granular system comprises the following parts: The kinetic energy of particles, $E_K$, the internal elastic energy stored at contact points, $U$, the dissipation due to frictional and drag forces, $D=D_f+D_d$, and the external work by vertical and lateral stresses, $W_y$ and $W_x$, respectively. The balance of energies is given by $W_y+W_x=U+D+E_K$.
For a system of $N$ particles, the total instantaneous kinetic energy can be expressed as $ E_K (t) = \frac{1}{2}\sum_{i=1}^N ( m_i v_i^2 + I_i \omega_i^2) $, where  $m_i$ and $I_i$ are the mass and moment of inertia of $i^{th}$ particle, respectively, and $v_i$ and $\omega_i$ are translational and angular velocities.
Considering that the velocity of the lateral walls is, on average, proportional to the vertical driving rate, a characteristic value for total kinetic energy of a homogeneous sample can be defined as: $ K_{D}:= \frac{1}{6} N \langle m \rangle (h\dot{\epsilon_y})^2$, where $h$ is the sample's height. We use $K_{D}$ as a reference parameter  to define the energy thresholds for avalanche detection. 

The microscopic configuration of contact points creates a heterogeneous landscape of elastic potentials. Even under monotonic driving the system exhibits a non-linear mechanical response involving sharp local accelerations, complex internal velocity fields and global bursts of internal kinetic energy that cannot be simplified in terms of the external measures such as $\dot{\epsilon_y}^2$ \cite{Radjai2002}. Considering a quastistically slow driving, the system is episodically trapped in a series of metastable jammed configurations well represented by the potential-energy \cite{Bonfanti2019}. The external AQS driving pushes the system through local mechanical instabilities that trigger local dynamics in the form of frictional sliding between particles at contact points, or the collective buckling of chained contact forces \cite{Cates1998,Tordesillas2007,Nicot2017force}. These initial instabilities can develop into collective internal avalanches whose propagation is almost independent of the driving speed and is controlled, instead, by particle-scale dissipative forces due to the drag-like forces and inter-particle friction.

\subsection{Stable evolution state}

\begin{figure}
    \includegraphics[width=0.5\columnwidth]{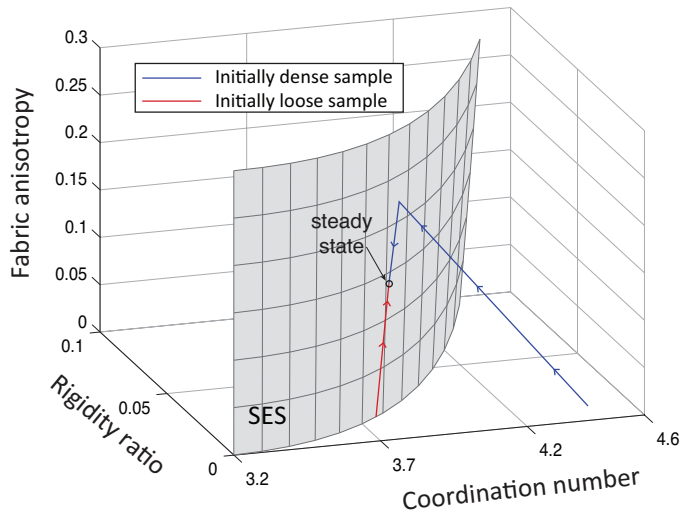}
    \caption{\label{fig:SES}(color online) Schematic visualization of SES surface and the material state path for initially loose and dense samples. The micromechanical state of the loose sample (red) is situated on the SES surface all along the deviatoric loading history. The response of the dense sample (blue) starts outside of SES but rapidly approaches the surface and remains on it afterwards. The meta-stable, stick-slip like response in stress is observed only when the micromechanical state of granular assembly reaches the SES. }
 \end{figure}

The mechanical evolution of the assembly is determined by a sequence of episodes of quiescent deformation and sudden avalanches involving configurational rearrangements. Recent studies \cite{Pouragha2016,Pouragha2017non,Pouragha2017strain, Pouragha2018mu} suggested that mechanically stable configurations are constrained by a stable evolution state surface (SES) defined as a manifold observed in the space of three micromechanical variables: coordination number (average number of contacts per particle), rigidity ratio (average contact deformation relative to the particle size, which scales with $(\sigma_x+\sigma_y)/k_n$), and fabric anisotropy (a measure for directional distribution of contacts), as represented in Fig.~\ref{fig:SES}. The concept of SES  has been proven to be effective in developing  new micromechanical constitutive models for granular materials~\cite{Pouragha2018mu}.

Considering a dense sample, the initial state of the assembly is located on the stable side (right side in Fig.~\ref{fig:SES}) of the SES. As the sample is loaded deviatorically, the microstructure quickly evolves through loss and gain of contacts approaching the SES limit~\cite{Pouragha2017non}. Upon reaching the SES, the assembly appears to have just enough constraints to retain metastable static equilibrium \cite{Pouragha2016,Bonfanti2019}. As such, the additional deviatoric straining introduced by the monotonic driving acts as an initial perturbation pushing the assembly beyond the stability limit and triggering the avalanche as a sudden increase in the internal kinetic energy. Kinematic momentum is known to modulate the jamming transition in shear processes \cite{Lamaitre2009,Lu2007,Nicot2012inertia, Gimbert2013} and alters the stability conditions during the transition. The kinematic momentum caused by avalanches is rapidly dissipated, hence acting as a transient overshooting mechanism \cite{Maimon2004} pushing the system away from the fragile SES towards more stable (non-fragile) configurations under quasistatic conditions, as discussed in more detail later in Section~\ref{sec:correl}.

The resulting sequence of avalanches involves rearrangements adjusting the number and orientation of contacts bringing gradually the micromechanical state of the granular assembly towards the steady state at $\sigma_c$, where the deviatoric straining is sustained with no appreciable change in the micromechanical state of the assembly.
The evolution within the SES towards the final steady state is not necessarily unique and depends on the original configuration and loading path. This is most clearly observed for the initially loose cases whose initial state falls on SES, see Fig.~\ref{fig:SES}.

\subsection{Quasistatic driving and kinetic avalanches}
\label{sec:avalDef}

\begin{figure}
\centering
    \includegraphics[width=0.45\columnwidth]{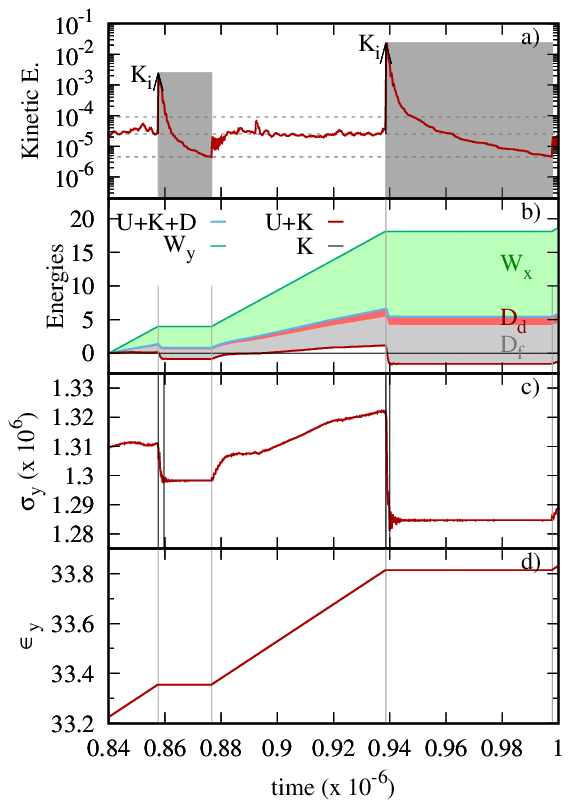}
    \caption{\label{fig:avalDef}(color online) Close-up example (D5kFc6) of the macromechanical evolution of the system during the span of two consecutive avalanches as a function of time. a) Evolution of the total kinetic energy ($E_K(t)$) used to define the avalanches and the driving protocol, which halts the driving above $K_h$ (upper dashed line) and restarts it when  $E_K(t)>K_r$ (lower dashed line).
      The middle dashed line indicates the expected average kinetic energy during driving. The intervals of the avalanche relaxation time without driving is indicated by the gray area. b) Energy transfer during driving and avalanche propagation. The supplied work ($W_{y}$) is represented as a green line. The green area represents the part of the work invested in lateral expansion of the system against the confining pressure ($W_{x}$). The blue line represents the rest of the work transformed into heat (dissipation $D$ and transient kinetic energy $E_K$) and the final variation of the internal energy ($U$). Energy is dissipated by a medium drag-like force ($D_{d}$) and friction between particles ($D_{f}$) represented as red and grey areas respectively. c) Avalanches are always accompanied by a drop in the internal energy that matches a drop in the external vertical stress ($\sigma_{y}$). We use the derivative of this stress drop to measure the duration of the avalanche. d) Shows the driving stopping during the relaxation time of the avalanche. The relation between $\sigma_{y}$ (c) and $\epsilon_{y}$ (d) is represented in Fig.~\ref{fig:macromech} for two other simulations.} 
\end{figure}

To guarantee a quasitatic condition and in particular, a clean separation between independent avalanches, in our simulations, the driving is halted as soon as an avalanche is detected. Such a detection occurs when the total kinetic energy ($E_K(t)$) exceeds a selected threshold value  $K_{h} = 3 K_{D}$. A similar strategy was used by \cite{Salerno2012,Salerno2013} for the DEM simulations of amorphous materials. As an example, Fig.~\ref{fig:avalDef} shows the assembly magnitudes in a typical simulation (D5kFc6) during the time interval containing 2 avalanches. The jumps in $E_K(t)$, up to several decades (Fig.~\ref{fig:avalDef}.a), signals the onset of avalanches triggered by the sudden release of elastic potential energy (Fig.~\ref{fig:avalDef}.b). Such a release of elastic energy is signified also by the drop in the boundary stress, $\sigma_{y}$ (Fig.~\ref{fig:avalDef}.c), while the lateral strain is adjusted to keep $\sigma_x$ constant. 
The surplus of kinetic energy is dissipated over time by frictional sliding and drag-like forces (Fig.~\ref{fig:avalDef}.a,b).  
Since the simulations are athermal, the heat transfer due to dissipation compensates the loss in internal energy.
The numerical results exhibit dissipation intervals (grey areas in Fig.~\ref{fig:avalDef}.a) considerably longer than the duration $T$ of the first rebound ($\dot{U}\geq 0$) of the potential energy release (Fig.~\ref{fig:avalDef}.c). Once $E_K(t)$ is damped below $K_{r}= 0.15 K_{D}$, the external driving is resumed. The factors defining $K_{h}$ and $K_{r}$ from the base $K_{D}$  have been selected to optimize the resolution and the computational speed. The values of $K_{h}$ are just above noise fluctuations in all of our simulations, and $K_{r}$ is low enough to represent a precise measure of the dissipated energy guaranteeing the quasistatic limit.

We obtain an avalanche catalog by labeling individual avalanches with a sorted index $i$ and represent them as a marked point process. Instead of the instant in simulation time $t_0$ when the $i$-th avalanche initiates at $E_K(t)\geq K_{h}$, we select $\epsilon_y(t_0)$ as our temporal scale, which is stationary during the propagation of the avalanche because of the quasistatic driving. We also record the ``size'' of each avalanche as measured by different means. In particular, we focus in this study on the duration $T$ of the potential energy release $\Delta U:=U(t_0)-U(t_0+T)$, the stress drop $\Delta \sigma_y := \sigma_y(t_0)-\sigma_y(t_0+T)$, the peak in kinetic energy $K=E_K^{\max}-K_D$, the total dissipated energy until the quasi-static driving resumes, $D=D_f+D_d$, which accounts for the friction at the contact points ($D_f$) and the drag-like forces ($D_d$), the work against the lateral walls $W_{x} = \int_{t_0}^{t_0+T} V(t)  \sigma_x \dot{\epsilon}_x dt$, where $V(t):=V_0 \epsilon_x(t)\epsilon_y(t)=V_0(1+\epsilon_v)$ is the  2d volume of the assembly whose variation during a given avalanche can be safely neglected.
The term \textit{kinetic avalanches} is used here to specify the avalanche detection method and driving protocol, based on the internal kinetic energy. However, following the observations presented below, we will argue that most avalanche detection protocols based on other physical measures such as external stress, or internal elastic energy drop would render similar results.

In the following sections, we study avalanches occurring within the SES, including avalanches before reaching the steady state (see intervals A in Fig. \ref{fig:macromech}). The behavioural characteristics of all assemblies at SES are argued to be the same  \cite{Pouragha2016}. We will verify that avalanche statistics are stationary within the SES regime from the analysis of subcatalogs, such as the intervals A and B represented in Fig.~\ref{fig:macromech}.

\section{Results}
\label{sec:results}

\subsection{Energy partition and avalanche scaling relations}
\label{sec:epart} 

As illustrated for the interval represented in Fig.~\ref{fig:avalDef}.b, the energy supplied to the system as external work along the y-axis 
($W_y = \int V \sigma_y d\epsilon_y = \int V_0 (1+\epsilon_v) \sigma_y d\epsilon_y$)
is mostly dissipated during the avalanche processes ($D$) or transferred to the lateral walls ($W_x$). The difference corresponds to variations in the elastic potential energy $\Delta U$. Instead of following a random sequence of drops and loads, $U$ self-organizes towards a stationary value --- determined by the flow regime in terms of elastoplasticity, or the SES in terms of granular microstructure --- and fluctuates around that value through a sequence of loading and avalanche cycles.\\

\begin{figure}
\centering
     \includegraphics[width=0.45\columnwidth]{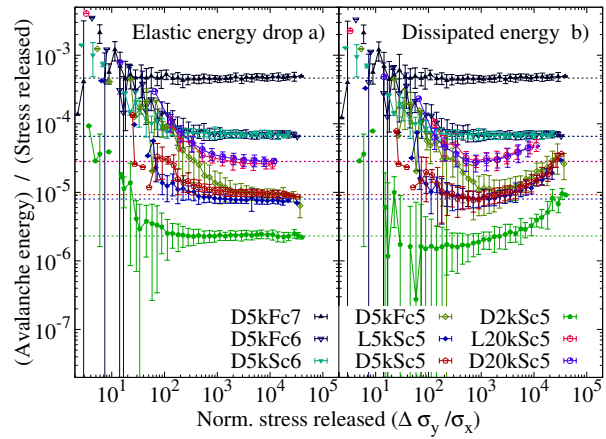}
    \caption{\label{fig:stressEnergy}(color online) Relation between normalized stress released ($\Delta \sigma_y /\sigma_x$) and (a) average elastic potential released ($\langle \Delta U \rangle (\Delta \sigma_{y})$) or (b) dissipated energy ($\langle D \rangle (\Delta \sigma_{y})$) for stress released. Error-bars correspond to one standard deviation. Dashed lines represent an asymptotic relation between the energy magnitudes and $N\sigma_x^{0.85}$ which fits the results as discussed in section \ref{sec:epart}.}
\end{figure}

This metastable cascade appears also in the external macromechanics of the assembly, i.e. in the stress drop ($\Delta \sigma_{y}$). For kinetic avalanches, Fig.~\ref{fig:stressEnergy} tests the linearity between the external $\Delta \sigma_y$ and the internal elastic potential drop averaged at $\Delta \sigma_y$ windows, $\langle \Delta U \rangle (\Delta \sigma_y)$ (left), as well as the average dissipated energy, $\langle D \rangle (\Delta \sigma_y)$ (right). Recalling that the lateral stress is kept constant, $\sigma_x$ is considered as the characteristic stress measure and, as such, the normalized stress release ($\Delta \sigma_y/\sigma_x$) is used for describing the avalanche size.
On average, and for sufficiently large avalanches, $\Delta U$ is approximately proportional to $\Delta \sigma_{y}$ in all simulations. 
  This is different from the behavior of $\langle D \rangle$: For most simulations, $\langle D \rangle$ is not proportional to $\Delta \sigma_y$, though such a proportionality has been observed in other works~\cite{Salerno2012}. The difference between  $\langle D \rangle (\Delta \sigma_y)$ and  $\langle \Delta U \rangle (\Delta \sigma_y)$ reflects the contribution of the lateral walls to the energy balance ($W_x$), which appears to be more prominent for large avalanches but disappears at high confining pressures.
  In addition, we observe a scaling factor, matching an apparent relation $\Delta U \propto N \sigma_x^{1.85(5)}\Delta\sigma_y/\sigma_x$. This relation is represented as horizontal dashed lines in Fig.~\ref{fig:stressEnergy}.
This scaling relation is reasonably close to the expectations from a dimensional analysis: Following a linear elastic relation, the change in elastic energy can be written in terms of the wall stresses and the normalized stress released:
\begin{equation}\label{e:dU}
\Delta U = V \left({\frac{\sigma_y}{\sigma_x}\frac{1}{E_y}-\frac{\nu_{xy}}{E_x}}\right)\sigma_x^2 \frac{\Delta \sigma_y}{\sigma_x},
\end{equation} 
where $V:= l_x l_y\propto N \pi \langle r^2 \rangle/(1-\phi)$ is the system's volume with $\phi$ being the porosity, $E_x$ and $E_y$ the macroscopic Young's moduli, and $\nu_{xy}$ the associated Poisson ratio. The macroscopic Young's moduli also depends on coordination number ($Z$) and the total volume, $E \propto Z\/V^{-1} \propto Z(1-\phi)$~\cite{Chang1995estimates,Pouragha2018elastic}. Taking into account that the stress ratio $(\frac{\sigma_y}{\sigma_x})$ fluctuates in no more than 1/3 of its average value within the catalog (Fig.~\ref{fig:macromech}), Eq.~\ref{e:dU} can be simplified to give the following scaling:
\begin{equation}\label{e:dU2}
\Delta U \propto Z^{-1}(1-\phi)^{-2} \sigma_x^2 \frac{\Delta \sigma_y}{\sigma_x}.
\end{equation} 	

The remaining mismatch between the exponent ${1.85(5)}$ used in the scaling of Fig.~\ref{fig:stressEnergy} and the exponent 2 in Eq.~(\ref{e:dU2}) is found in an effective relation between $Z (1-\phi)^{-2}$ and $\sigma_x$ which renders $\Delta U \propto  \sigma_x^{1.91(4)} {\Delta \sigma_y}/{\sigma_x}$ (see Fig. S2 in supplementary material).\\

\subsection{Distribution of avalanche sizes}
\label{sec:dist}

\begin{figure}
\centering
    \includegraphics[width=0.5\columnwidth]{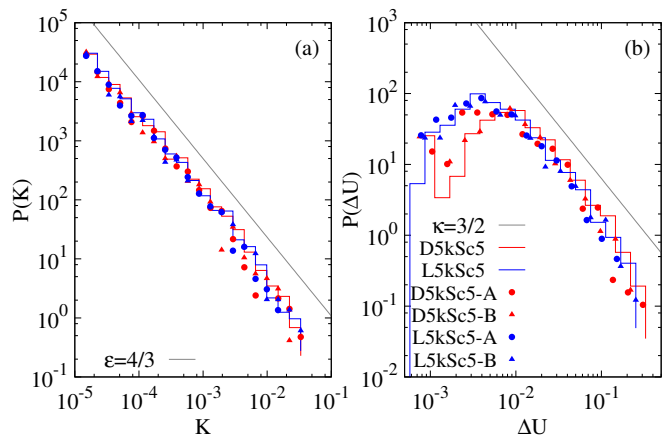}
    \caption{\label{fig:PUKsplit} (color online) Distributions of (a) kinetic energy peaks ($K$) (b) internal energy drops ($\Delta U$), for L5kSc5 and D5kSc5 separated into stages A and B (see Fig.~\ref{fig:macromech}) compared with the total sequences in the SES regime. Exponent values ($P(K) \sim K^{-4/3}$ and $P(\Delta U) \sim \Delta U^{-3/2}$) expected from SMFT are shown for comparison (black lines). }
\end{figure}

In addition to the scaling relations between different measures of avalanche sizes, the distribution of these sizes close to criticality also satisfy scaling relations of interest for the comparison with other avalanche phenomena and modelling approaches (see section \ref{sec:discuss}).
 Fig.~\ref{fig:PUKsplit} shows the binned avalanche distributions for $K$ and $\Delta U$ on a log-log scale for a loose and a dense assembly of $N\sim 5000$ particles under similar external and internal conditions during different time intervals. In particular, the external stress in the assemblies is considerably different during interval $A$, and evolves towards almost indistinguishable stress values ($B$) as both systems approach the steady-state regime (see Fig.~\ref{fig:macromech}). 
Note that in a macromechanical formulation, the state of the material is different during strain hardening (interval $A$ for the loose case), strain softening (interval $A$ for the dense case), and close to the steady state (interval $B$). However, despite such differences in stress state and the softening/hardening rate,  no pronounced differences between the distributions are present in  Fig.~\ref{fig:PUKsplit}. Both distributions of $\Delta U$ and $K$ appear to be compatible with the scale-free hypothesis, i.e., the distributions follow a power law decay, $P(\Delta U) \sim \Delta U ^{-\kappa}$ and $P(K) \sim K ^{-\varepsilon}$, to a good approximation over a wide range of scales, although finite size effects and limited statistics might play a role. The collapse of data with different initial densities and stress values further confirms that the state of the material associated with meta-stable avalanches is not sufficiently represented by macromechanical state variables, such as stress ratio and porosity, as commonly used in coarse-grained models. Considering that avalanche statistics can be represented in terms of state variables one needs a higher-dimensional space capable of representing all states with coinciding statistical properties, such as intervals $A$ and $B$ of dense and loose assemblies, in a unique manifold. In this regard, here we propose the SES as a viable candidate for such representation, since all evaluated intervals fall inside the SES manifold.\\

\begin{figure}
\centering
    \includegraphics[width=0.6\columnwidth]{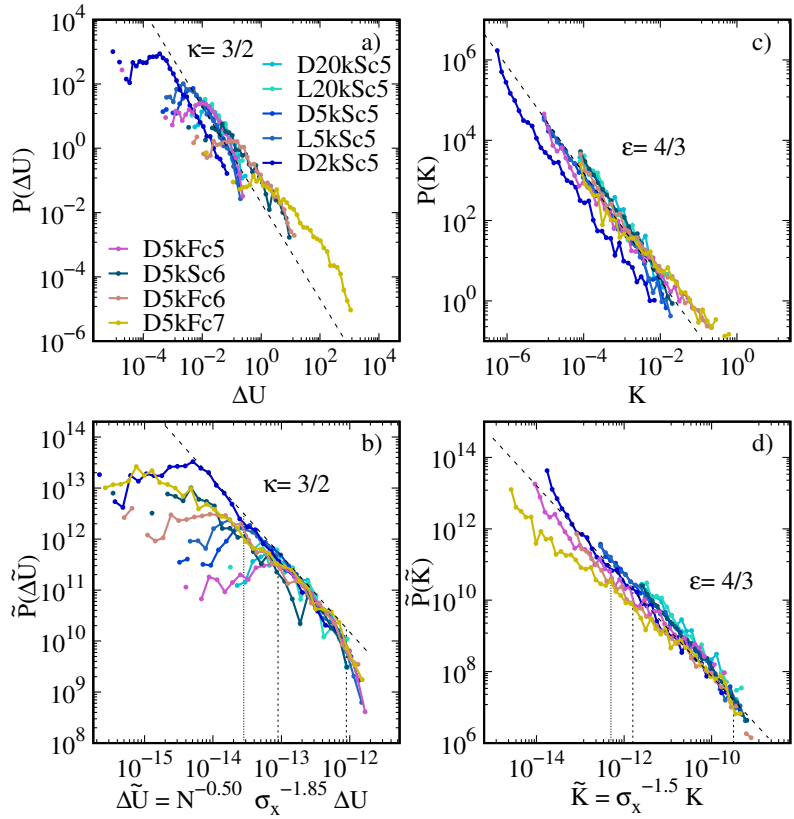}
\caption{\label{fig:distroUK} (color online) Distributions of (a,b) internal energy drops ($\Delta U$); (c,d) kinetic energy peaks ($K$) for all simulations. All distributions are shown on a regular scale (a,c) and rescaled to fit the apparent scaling relations in the x-axes (b,d). The y-axes are scaled to collapse the distributions to a single curve in the region of large values. Vertical dashed lines denote the intervals used to estimate the power-law exponents (dotted lines correspond to the lower limit for $\sigma_x\geq 10^6$).}
\end{figure}

To further investigate the collapse of the avalanche data, we focus on the standard scaling ansatz for self-similar behavior:
\begin{equation}
\label{eq:PLdistro}
P(x)dx = x^{-\tau_x} \Phi_x(x/x^*) dx,
\end{equation}
where $x$ is a measure of the avalanche size, $\tau_x$ is the associated critical exponent, $\Phi_x$ is a universal scaling function and $x^*$ is a characteristic avalanche size. This characteristic $x^*$ should be controlled by the finite system size as well as by other parameters. We attempt the following general ansatz motivated by the usual relations found in finite size scaling of scale-free phenomena~\cite{privman1990}
\begin{equation}
\label{eq:scaleX}
x^*(N, \sigma_x, \dot{\epsilon_y}) = \tilde{x}^* N^{\gamma_N^x}\sigma_x^{\gamma_\sigma^x}\dot{\epsilon}_y^{\gamma_\epsilon^x}.
\end{equation}
\begin{figure}
\centering
     \includegraphics[width=0.5\columnwidth]{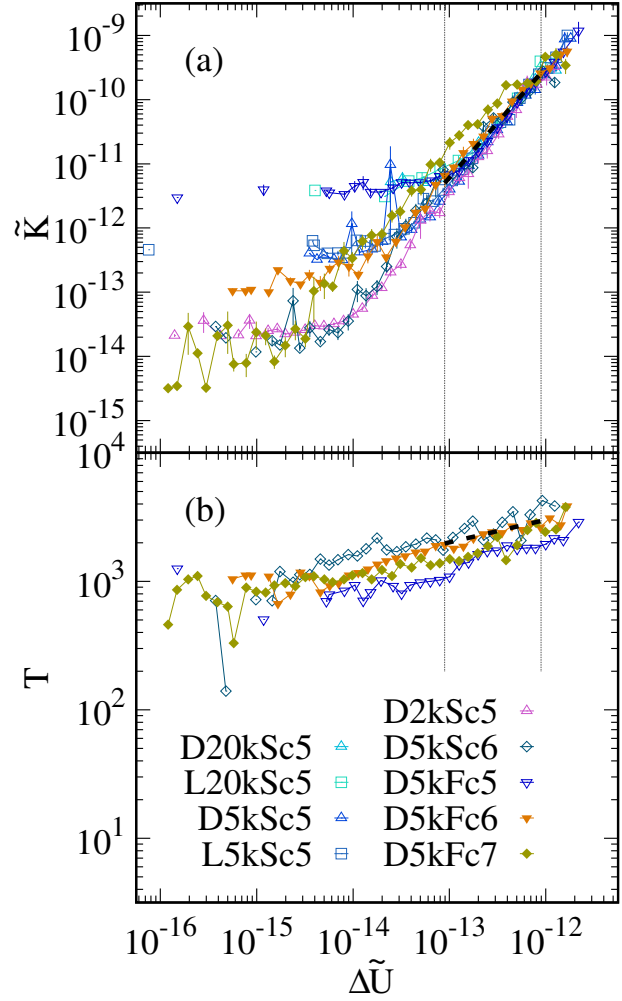}
    \caption{\label{fig:UKrelations}(color online) Average relations between the potential energy drop ($\widetilde {\Delta U}$), kinetic peak ($\widetilde {K}$) and duration of the energy drop ($T$). Units have been rescaled according to the exponents represented in Fig.~\ref{fig:distroUK}.b,d, see main text for details. Black dashed lines: fitted power-law in the intervals represented in Fig.~\ref{fig:distroUK}.b,d. 
(a) Mean kinetic peak ($\langle \widetilde {K} | \widetilde {\Delta U} \rangle $) for all simulations. Averages scale to a single relation with exponents close to an average $\gamma = 1.71(9)$. (b) Mean duration ($\langle T|\widetilde {\Delta U} \rangle $) for simulations D5kFc5, D5kSc6, D5kFc6, D5kFc7, averages scale as $\langle T | \widetilde {\Delta U} \rangle \sim \widetilde {\Delta U}^{0.177(9)}$.}
\end{figure}
If the scale-free hypothesis is valid, $\Phi_x$ should be universal and all the distributions of $x$ obtained with different parameters should collapse when rescaled by the characteristic event size, i.e. the distribution of $\tilde {x} := x/x^*$ should be independent of system size and parameters.
Fig.~\ref{fig:distroUK} shows the  distribution of $\Delta U$ (a,b) and $K$ (c,d) without scaling (a,c) and scaled (b,d) according to Eq.~\ref{eq:scaleX} using the best exponents found to collapse the exponential cutoffs into a single curve on visual inspection. Note that the y-axes are normalized independently, since the normalization of the distributions are non-trivially affected by the lower cutoffs. 
We find a reasonable collapse for ${\gamma_\sigma^{\Delta U}}= 1.85$ for potential energy drops and ${\gamma_N^{\Delta U}}= 0.50$ corresponding to a fractal dimension $d_f^{\Delta U} := d \times \gamma_N^{\Delta U} =1$, where $d$ corresponds to the dimension of the specific system, which is two for our model. The dependence on $\sigma_x$ is reasonable from the analysis in section~\ref{sec:epart}, if we assume that the intrinsic measure $\Delta \sigma_y /\sigma_x$ is also characteristic of the avalanche propagation at a given simulation size~\cite{Lerner2009}.
For the peaks of kinetic energy ($K$) we find a good agreement with ${\gamma_\sigma^{K}}= 1.5$, and ${\gamma_N^{K}}= 0.0$. Therefore, $K$ appears to be an intensive quantity. We argue that the instantaneous $E_k$ is probably dominated by a few fast-moving particles at a size-independent characteristic speed. In agreement with the quastatistic hypothesis, we do not observe any apparent relation with the speed of the external driving ($\dot\epsilon_y$), such that ${\gamma_\epsilon^{x}}= 0$, except for a shortening of the scale-free regime. The latter is caused by the lower bound of the power-law behavior controlled by the applied threshold $K_h$ in the definition of the avalanches, which results in a sharp cutoff in $K$ which in turn translates into a smoother transition in $\Delta U$. The distribution of avalanche durations are presented in the Fig.~S1 of the Supplementary Material. Such distributions appear to be compatible with a power-law distribution $P(T)\sim T^{-\alpha}$ with exponent $\alpha\approx 2$. Yet, the power-law range is too short to provide a reliable estimate of the exponent.%
\begin{center}
\begin{table}
\begin{tabular}{rccccc}
\hline
&  stiffness & $\#$ events & $\kappa$ & $\varepsilon$ & $\gamma$ \\
\hline
D2kSc5   & (stiff limit) & 1684 & 1.62(10) & 1.32(10) & 1.95(5)\\
L5kSc5   &  " "          & 979  & 1.60(7)  & 1.34(5) & 1.85(10) \\
D5kSc5   &  " "          & 788  & 1.71(8)  & 1.33(6) & 1.83(4)\\
L20kSc5  &   " "         & 130  & 1.77(17) & 1.45(6) & 1.85(15)\\
D20kSc5  &  " "          & 236  & 1.49(11) & 1.36(4) & 1.69(6)\\
D5kFc5    &  " "         & 1215 & 1.46(6)  & 1.36(4) & 1.71(5)\\
\hline
D5kSc6   &  & 396 & 1.41(8)  & 1.14(11) & 1.65(8)\\
D5kFc6  &   & 851 & 1.32(5)  & 1.14(6) & 1.71(4)\\
\hline
D5kFc7 & (soft limit) & 633 & 1.08(3)  & 1.02(8) & 1.48(7)\\
\hline
\hline
SMFT$^{(1)}$ & &  &  1.5 & 
~
$1+\frac{\kappa - 1}{2-\varsigma \nu z} =$ 
1.33 ~~~ & $2-\varsigma \nu z=$1.5 \\
\hline
SMFT$^{(2)}$ & &  &  1.5 & $(\mu +1)/2=$1.5 & $2\varsigma \rho =$1 \\
\hline
2D EPM & & & ~ 1.25--1.28 ~ & $\sim$1.2 \cite{Budrikis2017} & $\sim$1.45\cite{Budrikis2017} \\
\hline
\end{tabular}
\caption{\label{tb:exponents} Summary of the number of kinetic avalanches and critical exponents estimated from the simulations. Errors in brackets represent the std. deviation estimate from the likelihood function given a Gaussian approximation.  For comparison, the predictions of SMFT theory considering $K=E$ (SMFT$^{(1)}$) and $K=E_m$ (SMFT$^{(2)}$) \cite{Dahmen2017} (see Section~\ref{sec:discuss}) are also shown together with published findings of 2D full tensorial FEM simulations~\cite{Budrikis2017} and EPM simulations~\cite{Nicolas2018}. Note that similar exponents are also found in some DEM simulations of shear processes~\cite{Liu2016,Zapperi1997} while other DEM simulations return values as low as $\kappa = 0.98(1)$ \cite{Shang2019}.}
\end{table}
\end{center} 

The relation between ${\Delta U}$ and $K$ of individual avalanches agrees with the scaling relation found in Fig.~\ref{fig:distroUK}. In particular, Fig.~\ref{fig:UKrelations} shows the average duration $\langle T | \widetilde{\Delta U} \rangle $  (a) and the rescaled kinetic peaks $\langle \widetilde{K} | \widetilde{\Delta U} \rangle $ (b) expressed in bins of $\widetilde {\Delta U}$.  Over a range of a factor 25 in $\Delta U$ (x-axis in Fig.~\ref{fig:UKrelations}) and 150 in $K$ (y-axis in Fig.~\ref{fig:UKrelations}.b), the conditional averages for all simulations collapse and agree with a power-law relation between avalanche energies:  $\langle \widetilde{K} | \widetilde{\Delta U} \rangle \propto \widetilde{\Delta U}^{\gamma}$ with $\gamma = 1.71(9)$. As a particular case, a clear lower exponent is found at high confining pressures (D5kFc7). In a similar way, we can assess the average relation between $\widetilde{\Delta U}$ and the duration of the avalanche $T$, rendering a power-law relation  $\langle T | \widetilde{\Delta U} \rangle \propto \widetilde{\Delta U}^{0.20(3)}$. Notice that the values of $T$ have not been scaled since no clear parameter nor size dependencies have been observed in its distribution.

Visually, the exponent values seem to be in reasonable agreement with the expected values of the slip mean field theory (SMFT), $\varepsilon = 4/3$, $\kappa = 3/2$,  represented as dashed lines in Fig.~\ref{fig:distroUK} and the $\gamma = 3/2$,  see Section \ref{sec:discuss} for an extended discussion. However, a closer inspection reveals a clear dependency on confining pressure. In order to estimate their actual values we evaluate the avalanche sizes in their rescaled form considering the exponents used in Fig.~\ref{fig:distroUK} and the ranges of scale invariance shown as black segments in Fig.\ref{fig:UKrelations}, to mitigate the effect of $K_h$ on $\Phi_x$. 

\begin{figure}
\centering
     \includegraphics[width=0.5\columnwidth]{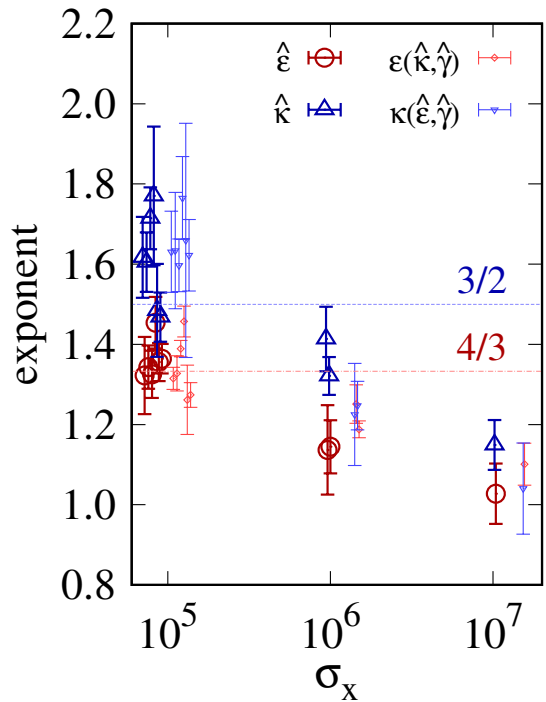}
    \caption{\label{fig:exponents}(color online) Estimated exponents $\kappa$ (blue triangles) and $\varepsilon$ (red circles) as function of the confining pressure within the intervals represented in Figs.~\ref{fig:distroUK} and \ref{fig:UKrelations}. Errorbars represent one standard deviation of the normal approximation to the likelihood function. Smaller points represent the estimation of each of the exponents with the help of Eq.~\eqref{eq:gamma} by using the other exponent and the estimated $\hat{\gamma}$ (see Table~\ref{tb:exponents}) to check the  consistency of the results. Dashed lines represent the expected values of $\kappa$ (blue) and $\varepsilon$ (red) from SMFT.} 
\end{figure}

The exponents $\kappa$ and $\varepsilon$ characterizing the distributions of $\Delta U$ and $K$, respectively, have been estimated using the maximum likelihood method following the recipe in \citep{Baro2012} within the intervals 
$-13.05 <\log_{10}({\widetilde {\Delta U}})<-12.05 $
 and 
$-11.3 < \log_{10}({\tilde {K}})<-9.8 $
represented in Fig.~\ref{fig:UKrelations}. For simulations at high confining pressures ($\sigma_x\geq 10^6$) we increase the fitting ranges to 
$-13.55 <\log_{10}({\widetilde {\Delta U}})<-12.05 $ and 
$-11.8 <\log_{10}({\widetilde {K}})<-9.8 $ to take advantage of the extended power-law regime to lower values (see individual distributions and selected estimation ranges in Fig.~S4,~S5 of the supplementary material). All estimated exponents are summarized in Table~\ref{tb:exponents} and represented in Fig.~\ref{fig:exponents} as a function of the confining pressure $\sigma_x$. 
Table~\ref{tb:exponents} includes the least square fit of $\gamma$ for each experiment and Fig.~\ref{fig:exponents} represents as well the expected exponents considering the approximated scaling relation $\widetilde{K}(\widetilde{\Delta U}) \approx \langle \widetilde{K} | \widetilde{\Delta U} \rangle \propto \widetilde{\Delta U}^{\gamma}$, which renders the scaling relation between exponents: 

\begin{equation}
\label{eq:gamma}
\gamma = \frac{1-\kappa}{1-\varepsilon}.
\end{equation}
This relation can be deduced from Eq.~(\ref{eq:PLdistro}).
The maximum likelihood estimation method reveals that both exponents $\hat{\kappa}$ and  $\hat{\varepsilon}$ decrease with the confining pressure, from $\hat{\kappa} = 1.58(11)$ and $\hat{\varepsilon} = 1.36(4)$  for $\sigma_x = 10^{5}$ to $\hat{\kappa} = 1.14(6)$ and $\hat{\varepsilon} = 1.02(8)$  for $\sigma_x = 10^{7}$. The exponent $\hat{\gamma}$ also decreases from $1.82(10)$, close to the value $1.71(9)$ found by averaging all the conditional average dependencies of Fig.~\ref{fig:UKrelations}, to the value $1.48(7)$ for $\sigma_x=10^{7}$ (golden lines in Fig.~\ref{fig:UKrelations}). 
In general, the maximum likelihood method provides a more robust estimation of the exponent $\varepsilon$ than $\kappa$. This is probably a consequence of  the extended scale-free regime in $K$, given the lower exponent values and the sharper cutoff at $K_h$.
The scaling relation captured by Eq.~\eqref{eq:gamma} is consistent with the estimated exponents when used to determine $\kappa(\hat{\varepsilon},\hat{\gamma})$ and $\varepsilon(\hat{\kappa},\hat{\gamma})$, especially at low confining pressures. 
The consistency between the two estimation methods asserts the robustness of the results and the existence of an actual dependence beyond statistical artifacts that could alter the exponent estimation within the fitting intervals.
We notice as well that this agreement survives at high confining pressures, close to the singular undetermined point at $\varepsilon \approx \kappa \approx 1$. Such values are close to the values reported in DEM simulations with Lennard-Jones (LJ) potentials \cite{Shang2019}. 
Note that lower scaling exponents ($\tau_x<1$) are not expected in the framework of scale invariance since they would imply a divergence of energy released for large system sizes. 
However, it is possible that \emph{estimated} exponents ($\hat{\tau}_x$) in \emph{finite} systems are lower even though the true scaling exponent is $\tau_x=1$. This difference corresponds, in that case, to the form of the scaling function $\Phi_x(x/x^*)$ \cite{Christensen2008}.
Variations of exponents as function of input parameters or conditions are fairly common in DEM \cite{Salerno2013,Karimi2016} and FEM \cite{Budrikis2017} simulations of amorphous materials exhibiting SOC, but they are seldom observed in conceptual models and they are usually linked to the superposition of different subcritical processes \cite{Amitrano2003,Lord-May2020}, the sweeping through criticality \cite{Sornette1994}, or rare  regions, a.k.a \textit{Griffiths phases} \cite{Vazquez2011}.\\

\subsection{Temporal correlations}

\begin{figure}
\centering
    \includegraphics[width=0.6\columnwidth]{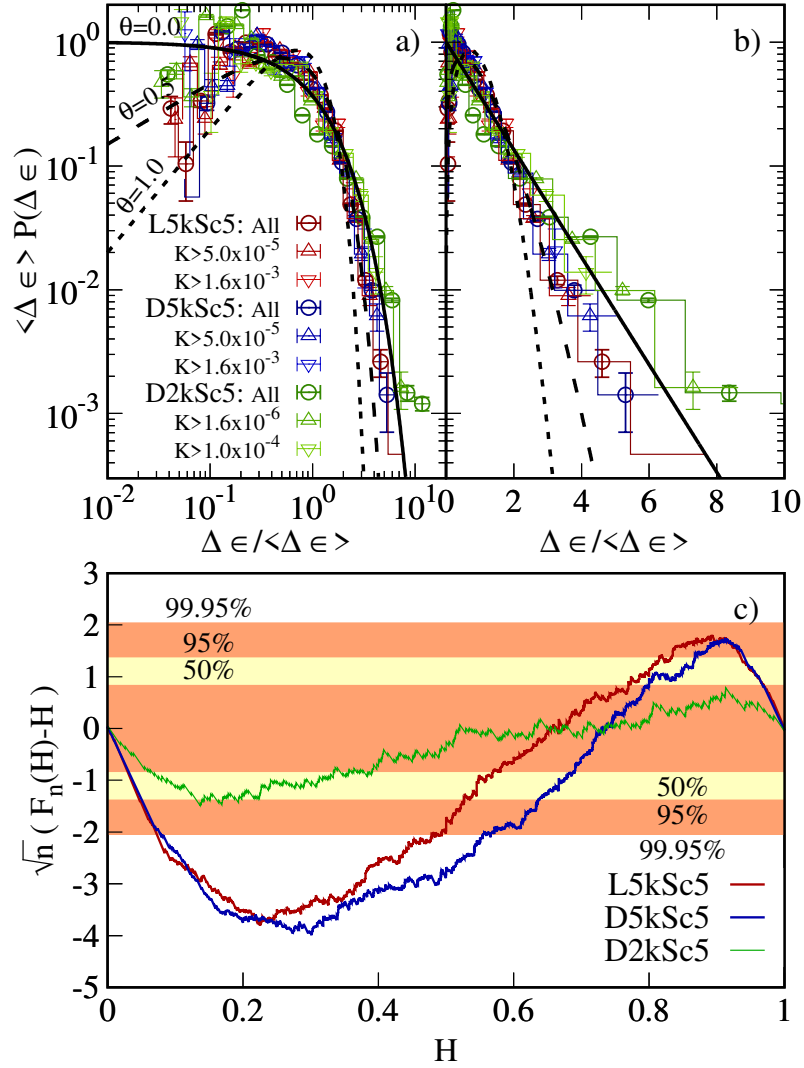}\\
    \caption{\label{fig:WT}(color online) (a,b) Distribution of interevent times as measured by the strain difference $\Delta \epsilon_{y}$ for different $K$-thresholds (corresponding to $100\%$, $\sim 50\%$, and $\sim 10\%$ of the events) and scaled by average $\langle \Delta \epsilon_{y}\rangle(K)$ in (a) log-log  and (b) linear-log scales, exemplified by three different simulations. For comparison, we show the Weibull renewal model (black lines) predicted by EPM (see Eq.~\eqref{e:pseudogapWT}) with $\theta=1.0$, $\theta=0.5$ and $\theta=0.0$, the latter corresponding to a homogeneous Poisson process. (c) Kolmogorov–Smirnov test over the $H$-values obtained by the Bi-test for Poisson rejection (see text) where $n$ is the number of interevent times. Color bands represent confidence intervals for rejection as indicated in the labels.}
\end{figure}

Considering the selected AQS driving protocol, the time between avalanches is best defined in terms of the strain component $ \epsilon_y$, since $\dot{\epsilon_y}$  is constant during external driving and $0$ during the avalanche propagation intervals, i.e., avalanches are instantaneous in terms of $\epsilon_y$. Thus, interevent times are represented as $\Delta \epsilon_y$. Correlations between avalanches are often observed in natural mechanical processes.
In particular, we are interested in the presence of temporal clustering, linked to triggering phenomena, which is often observed in natural systems \cite{Utsu1995,Weiss2004,Crane2008,Sornette2009,Baro2013,Hainzl2014,Baro2014,Davidsen2017, Kumar2020} but seldom reproduced in conceptual constitutive models unless explicitly introduced \cite{Dieterich1972, Hainzl1999,Mehta2006, Zoller2006, Ben-Zion2006,Jagla2014,Zhang2016, Baro2018a}.
Conversely, conceptual and numerical elastoplastic models identify a power-law relation in the distribution of the local residual-stresses or stability limits close to the steady state, a.k.a. \textit{pseudogap} \citep{Hebraud1998, Picard2005, Karmakar2010,  Lin2014, Agoritsas2015} leading to quasiperiodicity in the strain increments between consecutive avalanches \cite{Lin2016}. Recent studies revealed that the pseudogap is more prominent for large system sizes~\cite{Ruscher2019,Ferrero2019,Korchinski2021}, contrasting with the naive expectation of quasiperiodicity being caused solely by finite-size effects. We address this question in section \ref{sec:pseudo} through the analysis of $\Delta \epsilon_y$ presented here. 

Fig.~\ref{fig:WT} shows the statistical analysis of the time between avalanches as measured by the axial strain increment $\Delta \epsilon_y$ for selected samples L5kSc5, D5kSc5, D2kSc5 used here as examples. Each of the lines represent the strain increments between consecutive events with energy $K$ above different thresholds values. The distribution of interevent times in linear-log scale (Fig.~\ref{fig:WT}.b) follows an exponential distribution for long interevent times, suggesting that avalanches are independent of one another and follow a Poisson process with a stationary characteristic avalanche rate~\cite{Touati2009,Baro2013,Baro2017}. Similar results have been recently reported by Kuhn et al. (2019) \citep{Kuhn2019stress} for non-spherical particles.
In contrast, self-excitations usually involve strong correlations between large events ~\cite{Utsu1995, Baro2013} which would be easily identified as an excess of short interevent times.
No such an excess is observed in Fig.~\ref{fig:WT}(a). Therefore we can rule out the presence of self-excitations in our simulations.
On the contrary, most thresholds values reveal a deficit of short interevent times compared to a homogeneous Poisson process in all simulations. This implies regularity, i.e. the existence of characteristic (quasiperiodic) interevent times, which might be linked to finite size effects or the aforementioned pseudogap.

This deviation from Poisson behavior in the avalanche process can also be verified by the rejection method first proposed in \citep{Bi1989}, so-called Bi test, consisting of the evaluation of the statistical variable $h_{i}= \delta^{(1)}_i/(\delta^{(1)}_i + 0.5 \delta^{(2)}_i) $, where $\delta^{(1)}_i$ is the shortest of the intervals preceding or following event $i$ (i.e., $\delta^{(1)}_i := \min (\epsilon_{y, i+1}-\epsilon_{y, i}, \epsilon_{y, i}-\epsilon_{y, i-1})$) in terms of strain and  $\delta^{(2)}_i$ is the next interevent time in the same temporal direction of $\delta^{(1)}_i$ (i.e. $\delta^{(2)}_i :=\epsilon_{y, i-1}-\epsilon_{y, i-2}$ if $\delta^{(1)}_i=\epsilon_{y, i}-\epsilon_{y, i-1}$ and $\delta^{(2)}_i:=\epsilon_{y, i+2}-\epsilon_{y, i+1}$ if $\delta^{(1)}_i=\epsilon_{y, i+1}-\epsilon_{y, i}$  ). The Kolmogorov-Smirnov test over the  $h_{i}$ values shown in Fig.~\ref{fig:WT}.c displays a rotated 'S' profile characteristic of regular processes, in opposition to the  a rotated 'Z' profile, which is associated with clustering or self-excitation mechanisms \cite{Baro2014}. The 'S' profile is clear in all simulations (see Fig.~S3 in the supplementary material). The Poisson hypothesis is rejected at 95\% confidence interval in all simulations and at 99.95\% in most of them, including D5kSc5 and L5kSc5. 
In the case of D2kSc5 the $F_n(H)$ distribution falls short to reject the Poisson hypothesis at the 99.95\% confidence interval.
This discrepancy is in line with the better agreement  with a Poisson process observed for the distribution of waiting times of D2kSc5 (Fig.~\ref{fig:WT}). We verified numerically that the deficit of short interevent times is more pronounced in larger simulations. In particular, the fraction of rescaled interevent times below the expected first quartile of a normalized exponential distribution is $\sim 0.2$ for $\sim 2000$ particles, $\sim 0.13$ for $\sim 5000$ particles and $\sim 0.08$ for$\sim 20000$ particles with respect to the expected $0.25$ if the process were Poisson (see Fig.~S6 in the supplementary material). Unfortunately, since the threshold selection to define avalanches is size dependent, we cannot straightforwardly infer the relation $\langle \delta (N)\rangle$ which is expected to go as $N^{d/d_f-1}$ \cite{Lerner2009}.

\begin{figure}
\centering
    \includegraphics[width=0.6\columnwidth]{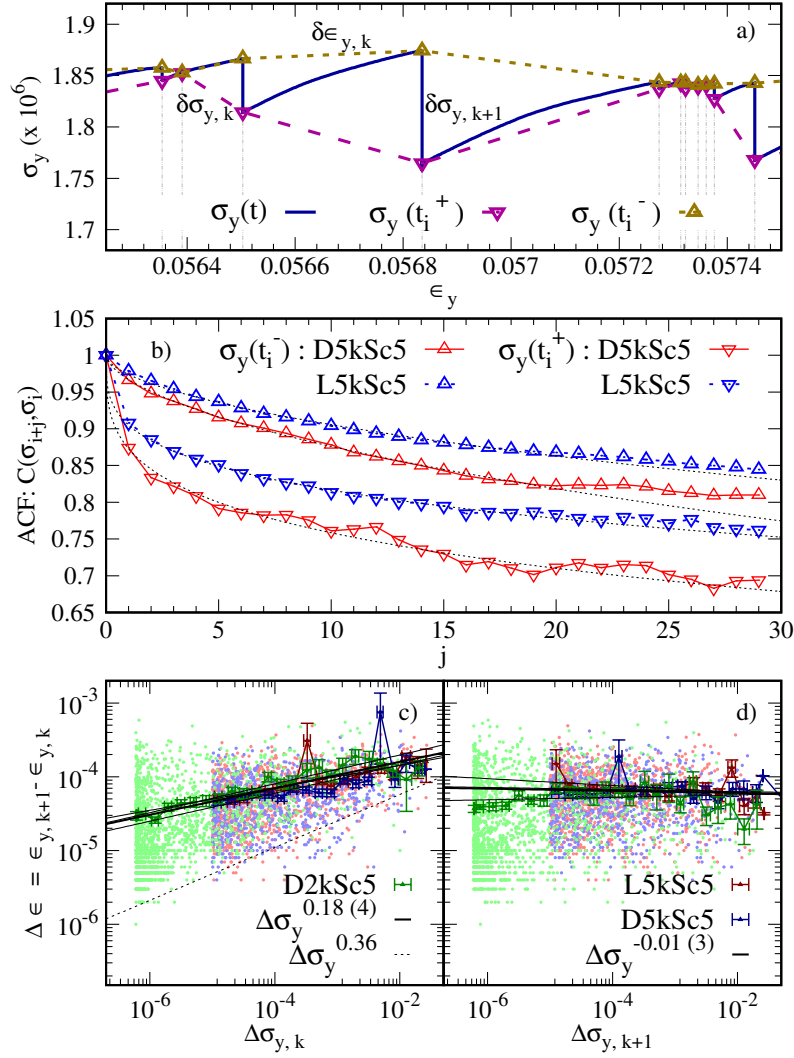}\\
    \caption{\label{fig:correl}(color online) (a) The macromechanical evolution of stress ($\sigma_{y}$) vs strain ($\epsilon_{y}$) (blue line). The dashed and dotted lines connect the values of stress on the onset ($t^{-}$) and at the end ($t^{+}$) of each avalanche $i$, respectively, for visual purposes. (b) Autocorrelation functions for the sequence of $\sigma_{y}(t^{-})$ and  $\sigma_{y}(t^{+})$, respectively. Lines represent best fits to a stretched exponential decay over the interval $j \in [1,30]$   (c,d)  Scatter plots of the relations between the size of the (c) previous and (d) next avalanche ($\Delta \sigma_{y}$) and the interevent strain $\Delta\epsilon_y$ for three different simulations. Conditional averages represented in colored lines and linear fitting in black. Dashed line in (c) represents an empirical minimal loading period $\Delta \epsilon \sim \Delta \sigma_{y}^{0.36(8)}$.}
\end{figure}

\subsection{Stress correlations}
\label{sec:correl}

 Fig.~\ref{fig:macromech} shows that the macromechanical state of the system is constrained within an evolving narrow stripe following a well-defined path at the large scale. 
However, the commonly used smooth average stress-strain relation is clearly a broad approximation. At the scale displayed in Fig.~\ref{fig:avalDef} the intrinsic metastable dynamics arising from the competition between the slow driving and the fast relaxation in the form of avalanches exhibits its non-smooth behaviour. As an example, Fig.~\ref{fig:correl}.a shows a close-up to the macromechanics of one of the simulations. We mark the stress ($\sigma_{y}$)  at the onset of the avalanche ($\sigma_y(t^{-})$) and at the end of the avalanche ($\sigma_y(t^{+})$). Fig.~\ref{fig:correl}.a also shows an example of interevent loading time $\Delta\epsilon_y^k = \epsilon_{y, k+1}-\epsilon_{y, k}$, preceded by an avalanche $k$ with stress drop $\Delta \sigma_{y, k}$ and followed by an avalanche $k+1$ with stress drop $\Delta \sigma_{y, k+1}$. 
The regularity in the distribution of interevent times exposed in Fig.~\ref{fig:WT} could be a consequence of a characteristic release or reloading of stress correlated to the interevent times, similar to predictions of the elastic rebound theory \cite{Reid1910} and related earthquake models \cite{Zoller2006}.
Fig.~\ref{fig:correl}.b shows the autocorrelation function (ACF) of the sequences of $\sigma_{y}(t^{+})$ and $\sigma_{y}(t^{-})$, respectively, for a pair of loose and dense assemblies with similar simulation  parameters. Since the macromechanics of the system are relatively smooth at the long range, as shown in Fig.~\ref{fig:macromech}, both autocorrelations persist for long times and remain at significant positive values over the shown range. In particular, correlations decay to a good approximation as a stretched exponential $ C \approx A\exp{(-j^{\xi})}$ for $j < 30$ with $\xi= 0.598(16)$ for $\sigma(t^-)$ and a faster $\xi= 0.317(18)$ for $\sigma(t^+)$. Therefore, $\sigma(t^{-})$ values are significantly more persistent than $\sigma(t^{+})$, suggesting that the state triggering the avalanches is more regular than the state arresting it. This hypothesis is confirmed by a comparison between the scatter plots of $\Delta \sigma_{y, k}$ and $\Delta \sigma_{y, k+1}$ as a function of $\Delta \epsilon_y$ (Fig.~\ref{fig:correl}.c and d). The duration of an interevent loading time does not affect the size of the next avalanche (Fig.~\ref{fig:correl}.d). Instead, the size of an avalanche correlates with the next interevent loading time (Fig.~\ref{fig:correl}.c).  Specifically, after an event with stress drop $\Delta \sigma_y$, the system needs to load a characteristic minimum amount of strain $\Delta \epsilon^{\min} \propto \Delta \sigma_y^{0.36}$ to trigger the next avalanche. This so called \textit{time-predictability} \cite{Shimazaki1980} is consistent with the lack of short interevent times (Fig.~\ref{fig:WT}.a). In terms of the SES representation, and at least for the testing procedure adopted in this study, the upper values of stress, i.e. $\sigma(t^{-})$, represent a better candidate for an objective reference state, and can be associated with SES. The fact that the interevent times are correlated with their preceding events implies that typically the avalanches can be considered as excursions bringing the system away from the critical SES limit towards more stable configurations. The following loading has then a tendency to return the system to the SES.

We have shown that both distributions of avalanche sizes and interevent times are stationary once the SES is reached. On the other hand, the effective macroscopic Young's moduli $E_y:=\dot \sigma_y(\epsilon_y)$ within the SES is non-stationary and evolves towards the steady state in non-unique evolutionary paths (Fig.~\ref{fig:macromech}). This implies that a time dependence exists in the average stress reloading amount between consecutive avalanches ($\langle \Delta \sigma_y^{r} \rangle := \langle \sigma_{y, k}(t^-)  - \sigma_{y, k-1}(t^+)  \rangle$) or the avalanche rates ($r:= N_{\mathrm{aval}}/\Delta \epsilon_y$), even within the SES. To verify this hypothesis we evaluate such average reloading sizes in addition to the the effective elastic modulus ($ \hat{E}_y := \Delta \sigma_y /\Delta \epsilon_y$) and average avalanche sizes ($\langle \Delta \sigma_y \rangle := \langle \sigma_{y, k}(t^+) - \sigma_{y, k}(t^-) \rangle$) in relatively broad intervals of strain, i.e. driving, of $\Delta \epsilon = 0.005$ containing around 70-100 avalanches each. We build a catalog considering all strain intervals in L5kSc5 and D5kSc5 within SES together, since the input parameters are similar and include a large variability in $E_y$, mostly positive in L5kSc5 and mostly negative  in D5kSc5 (see Fig.~\ref{fig:macromech}).  We report the following Pearson correlation coefficients with respect to $\hat{E}_y$:
\begin{equation}
\rho_{\hat{E}_y, r} = 0.37\;, \;\;
\rho_{\hat{E}_y,\langle \Delta \sigma_y^{r} \rangle } = 0.30 \;,\;\; 
\rho_{\hat{E}_y, \langle \Delta \sigma_y \rangle} = 0.058
\end{equation}
the latter being effectively null. Therefore, we argue that the effective  macroscopic  Young's modulus is determined by the rate of avalanches and the reloading between avalanches in a way that low rates and high stress reloading correlates with positive effective moduli, while the characteristic size of avalanches is effectively constant, as already argued in section \ref{sec:dist}. Some of the issues pertaining to time averaging such meta-stable quasistatic behaviours are discussed in \cite{combe2000strain}.

\section{Discussion}
\label{sec:discuss}

Non-linear mechanics and avalanches in amorphous solids are often studied under the lens of AQS \textit{shear} driving. Both atomistic and coarse-grained simulations usually consider overdamped systems of soft particles, usually modeled with attractive LJ potentials in DEM models, returning short-range or far-field homogeneous stress transfer functions in a coarse-grained representation, either isotropic when considering interactions along the maximum shear plane \cite{Ben-Zion1993,Dahmen2011}, or  polar Eshelby inclusions, when considering off-plane interactions  \cite{Hebraud1998,Picard2004,Albaret2016}.
The DEM simulations of compact granular assemblies presented here differ from some of the previous considerations. Most notably, the system is driven by biaxial compression, instead of shear, particles have inertia and the stress distribution is channeled in complex force-chain networks resulting in highly disordered long-range interactions.

\subsection{Biaxial compression}
\label{sec:amorph}
Biaxial and triaxial compression experiments~\cite{Lockner1993,Baro2013,Lennartz2014,Davidsen2017,Baro2018b}, and numerical simulations~\cite{Amitrano2003,Kun2014} such as the ones presented in this work, serve as analogues for geomechanical processes. Mechanics at the mesoscale are dominated by internal shear processes introduced by residual deviatoric stresses at contact points. The correspondence with macroscopic shear is especially clear when activity is localized in a shear band \cite{Lennartz2014,Kwiatek2014,Davidsen2017} (see deformation of D5kSc5 in Fig.~\ref{fig:macromech}) but not so much when deformation is homogeneously distributed within the sample (L5kSc5 in Fig.~\ref{fig:macromech}). Avalanche statistics shown in previous sections appear to be independent of internal details once the system reaches SES, whether system exhibits softening behaviour associated with localized failure (e.g. in the case of D5kSc5), or monotonic hardening leading to diffuse failure  (e.g. in the case of L5kSc5). This suggests that the similarities in avalanche statistics is linked to particle scale mechanisms, such as frictionally arrested stick-slip events and force-chain buckling, rather than the coarse-grained elasto-plastic macromechanics~\cite{tordesillas2009buckling}. On the other hand, the introduction of soft, instead of fixed, lateral walls affects the energy partition of the system dynamics in general (Fig.~\ref{fig:avalDef}) and the avalanche energy budget in particular. For purely strain-controlled loading programmes, such as in constant-volume tests, an average proportionality is reported between $D$ and $\Delta \sigma$~\cite{Salerno2012,Salerno2013}. In our study presented here, the equivalence between $D$ and $\Delta U$, and thus, the proportionality between the stress drop and the energies, does not stand, because of the free production of lateral work $W_x$. Specifically, Fig.~\ref{fig:stressEnergy} shows that such a proportionality is maintained in $\Delta U$ and, therefore, lost for $D$ in general. Recalling the different scaling exponents in  $\Delta U^* \propto \sigma_x^{1.85}$ and $W_x \propto \sigma_x$,  the equivalence $\Delta U \approx D$ is asymptotically recovered at high confining pressures, given $D=\Delta U + W_x\approx \Delta U$.\\

\subsection{Effect of inertia and the pseudogap}
\label{sec:pseudo}

Recent experiments  in granular aggregates reported interevent time distributions consistent with triggering~\cite{Kumar2020}. Inertia is known to affect statistics of both avalanche sizes and temporal correlations and could potentially constitute a mechanism of triggering explaining such observation.
The relatively slow dissipation of the kinetic energy with respect to the stress drops we observe (see Fig.~\ref{fig:avalDef}) might trigger secondary avalanches in the form of new spikes in $\dot U(t)$ and $E_K(t)$.   Similar phenomena have been reported in the signal profiles of acoustic emission events during the uniaxial compression of concrete samples~\cite{Vu2020}, 
We do observe, in rare occasions, a double peak in the kinetic energy within the same avalanche profile (not shown). Nevertheless, given the selected AQS driving protocol and low values of $K_r$, we can be certain that such double peaks are absorbed in a single avalanche event, recovering a practical independence between avalanches under the definition suggested in Ref.~\cite{Janicevic2016}. No evidence of self-excitation is found in either the Poisson-rejection test (Figs.~\ref{fig:WT}.c) or the distribution of interevent times shown in Fig.~\ref{fig:WT}.b, which exhibits an exponential decay.
On the contrary, we report a deficit of short interevent times linked to regularity in the macromechanical state of the system at the onset of an avalanche (Fig.~\ref{fig:correl}). This indicates that the process has a certain time-predictability~\cite{Shimazaki1980}, well identified as a characteristic minimum recharging strain as represented in Fig.~\ref{fig:correl}.c.\\


This regularity of interevent times is predicted by finite size effects as considered in some earthquake hazard assessment frameworks~\cite{Salditch2020}. In small systems, all avalanches dissipate macroscopic fractions of elastic energy, causing longer reloading times after large avalanches. This effect disapears in the thermodynamic limit, where all subcritical and critical avalanches are massless. Alternatively, a similar quasiperiodic behavior is predicted in elastoplasticity by the presence of a so-called '\textit{pseudogap}' in the density of residual-stresses, i.e. the local stress differences to failure through different mechanisms, usually represented on a coarse-grained scale~\cite{Hebraud1998,Picard2004,Karmakar2010,Lin2014, Lin2016, Ferrero2019}. 
Rather than a \textit{gap} i.e., absence of small residual stresses ($x$), the pseudogap is characterized by a power-law decay in the density of low values attributed to the non-monotonous evolution of residual-stresses towards the stability limit which corresponds to an absorbing boundary in a stochastic process representations \cite{Lin2016}: 
\begin{equation}
N(x) \sim {x}^{\theta} \;\; \mathrm{ with }\;\; \theta>0.
\label{e:pseudogap}
\end{equation}
The pseudogap introduces a characteristic delay in the interevent times between consecutive avalanches, defined here as the strain $\epsilon_y$ needed to activate the lowest $x$ value in Eq.~(\ref{e:pseudogap}), and rendering the Weibull renewal  behavior~\cite{Karmakar2010}:
\begin{equation}
P(\Delta\epsilon)
= \frac{1+\theta}{\langle \Delta\epsilon\rangle}
\left({\frac{\Delta\epsilon}{\langle \Delta\epsilon\rangle}}\right)^{\theta}
\exp\left( {- \left({\dfrac{\Delta\epsilon}{\langle \Delta\epsilon\rangle}}\right)^{\theta+1}}\right)  .
\label{e:pseudogapWT}
\end{equation}  

Such a behavior is shown in Fig.~\ref{fig:WT}(a,b) as the black lines.
Recent studies in DEM \cite{Ruscher2019}, FEM \cite{Korchinski2021} and conceptual \cite{Ferrero2019} models identified an additional system-size dependent plateau overlapping with the pseudogap, which vanishes in the thermodynamic limit. According to these results, quasi-periodicity would be more prominent in large systems while interevent times in small systems would be better described by a Poisson process. This system-size dependence, opposite to the expectations from finite size effects, qualitatively agrees with our results presented in Fig.~\ref{fig:WT} and is numerically verified in the supplementary material (Fig.~S3,~S6). However, the analysis presented in Fig.~\ref{fig:WT} is insufficient to determine the system-size dependency of the plateau or the exponent $\theta$ defining the pseudogap following the analysis presented in \cite{Lin2014}. Instead one can use hyperscaling relations to address the latter as we discuss in section \ref{sec:critExp}. Yet, Fig.~\ref{fig:WT} shows that the distribution of short interevent times are close to the expectations of the Weibull renewal model with an exponent between 0.5 and 1, with a generally higher exponent, suggesting a more prominent pseudogap in the large samples (Fig.~\ref{fig:WT}.a). At the region of long interevent times, the distributions are better reproduced by the Poisson hypothesis instead (Fig.~\ref{fig:WT}.b). This might indicate a change in the distribution of $x$, marking the end of the power-law regime described by Eq.~(\ref{e:pseudogap}).\\

Finally, inertial effects were found to affect the critical exponents determinining bulk avalanche statistics in DEM simulations of LJ particles~\cite{Salerno2012,Salerno2013,Karimi2016,Karimi2017}. 
This feature could as well be related to the observation that the states triggering avalanches, herein considered to be constrained to a SES manifold, are more regular than the arrested states (see section~\ref{sec:correl}).
The kinetic energy introduced by `fast' external driving  is known to affect mechanical equilibrium~\cite{Anthony2005,Lamaitre2009,Lu2007,Nicot2012inertia}.
We propose that the SES represents a collection of metastable configurational limits to the mechanical equilibrium that are marginally reachable under quasistatic conditions \cite{Bonfanti2019}. 
When an avalanche is initiated at SES, the excess kinetic energy prevents the relaxation of the system at a fragile state, e.g. the same configurational state at the onset of the avalanche. The configurational state arresting the avalanche, which is reached at $t_0+T$, needs to be stable while the system still retains a relatively high amount of kinetic energy (see Fig.~\ref{fig:avalDef}). Since the excess of kinetic energy is dampened below $K_r$ before resuming the driving, we believe that the particle inertia during the avalanche has to affect the distribution of residual stresses, and the properties of the pseudogap. 
This idea of stability \textit{overshooting} was conceptually captured in a deppining model with transient effects predicting a system-size dependence on the time-predictability of transitions~\cite{Maimon2004} qualitatively similar to the one observed in the pseudogap, verified at the stiff particles limit in our simulations. A similar effect is reproduced by the addition of \textit{dynamic weakening} in some EPM, and the SMFT in particular \cite{Dahmen2009}. However, the effect on avalanche sizes is far from being trivial. Numerical studies with LJ particles found non-monotonous relations between critical exponents and inertia, and regimes with characteristic and time-predictable avalanches \cite{Salerno2013}, being the latter predicted in SMFT as well~\cite{Dahmen2009}.\\

\subsection{Stiff granular particles}
Apart from particle inertia, most coarse-grained EPM also overlook the non-linearity of the internal stress fields \cite{Liu1995,Mueth1998,Howell1999}. 
The hypothesis of the uniformity in the stress propagation enabling the coarse-graining in elastoplastic models is known to fail in the presence of stiff particles \cite{Nicolas2018}. 
As mentioned, stress distribution in granular media is governed by the structural topology of contact points, generating complex force-chain networks, highly susceptible to small variations of stress. A cross-over between fragile and amorphous materials has long been suggested to exist as a function of softness or \textit{deformability} \cite{Cates1998}, even in the absence of attractive particle potential \cite{Liu1998,Lerner2009}. The variation in the power-law exponents in Fig.~\ref{fig:exponents} is in agreement with such a transition (see section \ref{sec:critExp} for a more detailed discussion).
Recent studies have suggested the introduction of internal structural variables such as compativity \cite{Edwards1989,Frenkel2008,Blumenfeld2012}, angoricity \cite{Henkes2007,Blumenfeld2009,Bi2015,Baule2018}, or the more recent concept of keramicity \cite{Bililign2019,Ball2019,Baule2018} complementing the stress-strain formulation traditionally considered in EPM. Based on our results, we propose the SES concept as an alternative micromechanical representation of the state of granular materials determined by the structure of internal contact network rather than the macromechanics~\cite{Pouragha2018mu}. \\

\subsection{Interpretation of avalanche sizes in terms of elastoplasticity}

The EPM formulation separates plastic avalanche events from the elastic deformation caused by the AQS driving. Although the separation of temporal scales is not as straightforward in DEM simulations, one can still find a correspondence between the variables defining avalanches in the different modeling approaches. In this sense, we are interested in the conceptual picture provided by the SMFT, which is arguably the simplest EPM framework to characterize avalanche statistics, and has shown good agreement with the statistics of avalanches in granular assemblies sharing in particular the same universal critical exponents~\cite{Dahmen2011,Denisov2016,Geller2015}.
In the SMFT, the avalanches are measured in terms of a global signal ($v(t)$ in short intervals) which accounts for the collective velocity of displacement of the constituting elements (a.k.a. cells) of the model, and is proportional to the rate of external stress drop under strain driving. Henceforth, the size $S := \int_{t_0}^{t_0+T} v(t)dt$ of an avalanche of duration $T$ starting at time $t_0$ is a measure proportional to stress drop. Similarly, $E := \int_{t_0}^{t_0+T} v^2(t)dt$ is proportional to the total kinetic energy dissipated, and $E_m :=  v_{\max}^2(t)dt$ is proportional to the maximum peak of the instantaneous kinetic energy within the avalanche. In order to compare DEM results with the expectations from elastoplasticity, we first need to establish reliable links between the avalanche sizes defined within the conceptual framework proposed in the SMFT and the measures defining avalanche sizes in DEM.\\

Considering EPM formulations, it is natural to assume an approximate proportionality between elastic energy and small deformations. Given its mean-field nature, the SMFT assumes an exact proportionlity ($\dot \sigma \propto  v(t)$) during avalanches, and predicts an approximate proportionality with the elastic energy release. This is consistent with the observed proportionality between internal elastic energy $\Delta U$ and $\Delta \sigma_y$ illustrated in Fig.~\ref{fig:stressEnergy}. We adopt the strain drop as the avalanche size $S := \Delta \sigma_{y}$ following the convention in Refs.~\cite{Salerno2012,Salerno2013}. Moreover, to remain consistent with SMFT, the avalanche duration $T$ has been defined in this work to be the time interval of the potential energy release, which is virtually indistinguishable from the time interval of the stress drop. \\

Although $K:=E_K^{\max}-K_D$ represents the peak in our simulations and, hence, it is reminiscent of $E_m$,
the  dissipation process is significantly slower compared to the rise of kinetic energy. In contrast, SMFT considers a strict overdamped limit where all inertial effects are instantly dissipated. Therefore, we argue that the pick of kinetic energy $K$ reached during the initial avalanche growth corresponds to the collective cumulative energy $E$ in terms of SMFT. In Fig.~\ref{fig:avalDef}.a we observe a separation of temporal scales between the kinetic energy rise time, roughly matching the temporal interval ($t_0$, $t_0+T$), and the characteristic dissipation time (see the grey areas in Fig.~\ref{fig:avalDef}.a). Almost no energy has been dissipated at $T$. We therefore adopt $K \propto E$ in terms of the variables in SMFT.
This assumption can be verified by comparing the values of $K$ with an auxiliary avalanche variable $\Delta (U^2):= \int_{t_0}^{t_0+T} dt (dU/dt)^2$ (see Fig.~S7 in supplementary material). If we assume a temporal proportionality between $\dot{U}(t)$ and $v(t):=dS/dt$, for consistency $(dU/dt)^2(t) \propto v(t)^2$. Therefore the proportionality $K \propto E$ implies $K \propto \Delta (U^2)$, as verified in Fig.~S7.
 However, for the sake of completeness, in the next section we compare our results considering both $K \propto E$ (herein referred as model SMFT$^{(1)}$) and $K \propto E_m$ (SMFT$^{(2)}$) to validate our hypothesis in light of the exponent values.\\

\subsection{Non-universal critical exponents}
\label{sec:critExp}

In the absence of self-excitation the spatial extension of avalanches serves as a proxy of local spatial correlations and, therefore, one can assess system susceptibilities from avalanche statistics. 
Since no divergence of activity rate is observed in the catalogs, criticality has to be linked to a divergence of characteristic scales in avalanche distributions. Hence the power-law distributions  $P(\Delta U)$ and $P(K)$ reported in section \ref{sec:dist} are a trademark of criticality. Close to criticality such distributions obey a scaling form: 
	\begin{equation}
	\begin{array}{c}
	P(\Delta U)\:d\Delta U = \Delta U^{-\kappa} \:\Phi_{\Delta U}(\Delta U/\Delta U^*)\: d\Delta U,\\
	P(K)\:dK = K^{-\varepsilon}\: \Phi_K(K/K^*)\: dK,
	\end{array} 
	\end{equation}
where $\kappa$ and $\varepsilon$ are the so-called critical exponents (Table \ref{tb:exponents}), $\Phi_{\Delta U},\Phi_{K}$ are universal scaling functions (often in the form of an exponential cut-off) and $\Delta U^*,K^*$ are characteristic avalanche sizes, related to the correlation length. When approaching criticality, the correlation length diverges such that for any finite system it eventually takes on the system size value. Therefore, $\Delta U^* \sim L^{d_f}$, where $L$ is the lateral size of the system. Considering that the system stays close to criticality, at the stiff particle limit ($\sigma_x=10^5$) our simulation results are consistent with $d_f\approx 1$ (see section \ref{sec:dist}). Values $d_f \sim 1$ have been commonly found in previous studies \cite{Maloney2004} and DEM simulations \cite{Salerno2013,Shang2019}. Other FEM simulations \cite{Karimi2017} reported a dependence of exponents $d_f$ and $\theta$ on a damping coefficient $\Gamma$ that could, potentially, include the contribution of confining pressure to the avalanche propagation. Here for the sake of the discussion we will consider $d_f$ to be invariant to $\sigma_x$.

In SMFT, the relations between the average sizes, energies, and durations take the following power-law form (see \cite{Dahmen2017}): 
\begin{equation}
\begin{array}{lcl}
\langle T | S \rangle \sim S^{\varsigma \nu z} & \;\mathrm{where}\; & \varsigma \nu z= 1/2\\
\langle E | S \rangle \sim S^{2-\varsigma \nu z} & \;\mathrm{where}\; & 2-\varsigma \nu z= 3/2\\
\langle E_m | S \rangle \sim S^{2\varsigma \rho} & \;\mathrm{where}\; & 2\varsigma \rho = 1\;.\\
 \end{array}
\end{equation}
Considering the proportionality between differential deformation and avalanche sizes, one can prove that the monotonic driving brings the system towards the aforementioned stationary SOC flow regime around $\sigma_c$ with scale-free statistics, leading to Eq.~\ref{eq:critSMFT}. Considering the exponents above, and the proportionalty between $\Delta U$ and $S$, criticality imposes power-law distributions with the following exponent relations (see \cite{Dahmen2017}): 
\begin{equation}
\begin{array}{lcl}
 P(S) \sim S^{-\kappa}& \;\mathrm{where}\; & \kappa = 3/2\\
P(E) \sim E^{-1-\frac{\kappa -1}{2-\varsigma \nu z}} & \;\mathrm{where}\; & 1+\frac{\kappa -1}{2-\varsigma \nu z} = 4/3\\
P(E_m) \sim E_m^{-\frac{1+\mu}{2}}& \;\mathrm{where}\; &\frac{1+\mu}{2} = 3/2 \; . \\
 \end{array}
\end{equation}
 
In Table \ref{tb:exponents} we compare our results with the predictions of SMFT in both interpretations of $K$ (models SMFT$^{(1)}$ and SMFT$^{(2)}$) and the typical range of exponents reported in numerical studies considering similar relation between observables \cite{Budrikis2017,Liu2016, Nicolas2018}. We notice that studies investigating the effect of dimensionality reported no significant differences between 2D and 3D systems \cite{Budrikis2017, Salerno2013}.  
The most remarkable observation is that the exponents in our simulations are non-universal. When the confining pressure is low, or contacts are relatively stiff ($\sigma_x=10^5$), the granular compact assemblies in all three exponents $\kappa$, $\epsilon$, $\gamma$ is consistent with SMFT$^{(1)}$, i.e. considering $K=E$, although exponents $\kappa$ and $\gamma$ are slightly higher than expected by SMFT. Notice that this agreement holds for both loose and dense assemblies, even when the macromechanical evolutionary path is different from SMFT in the latter.
The correspondence with the SMFT predictions is progressively lost for higher confining pressures. In the extreme case of  $\sigma_x=10^7$, the exponent $\gamma=1.48(7)$ is still close to the mean field prediction, and in agreement with recent results in 2D EPM simulations (e.g. $\gamma \approx 1.45$ in \cite{Budrikis2017}). However,  $\kappa$ and $\epsilon$ decrease towards values close to 1.0. 
Note that $\gamma$ plays a minor role in the relation between $\kappa$ and $\varepsilon$ when both exponents are close to $1$, see Eq.~(\ref{eq:gamma}).
The exponents $\kappa$ and $\varepsilon$ at high confining pressures, which depart from SMFT predictions, are closer to the predictions from EPM within 2D lattices, where lower exponents are reported: $\kappa \sim 1.28$ and $\varepsilon \sim 1.20 $~\cite{Budrikis2017,Liu2016}, $\kappa \approx 1.1 - 1.2$~\cite{Jagla2015} or even down to $\kappa = 0.98(1)$~\cite{Shang2019}. 
The ranges found in $\kappa$ are also consistent with previous reports comparing experiments and DEM simulations \cite{Bares2017}, and simulations with variable heterogeneity~\cite{Ma2020}.
It has been argued that this change of behavior can be interpreted as a transition between different universality classes \cite{Budrikis2017}. In that case, our results might indicate a transition between conceptual limits.

To sum up, the statistics of avalanches in non-cohesive assemblies of stiff particles are relatively close to the expectations of the SMFT, specially for the most robust exponent $\varepsilon$, whilst the avalanches for soft particles are close to the predictions of elastoplasticity with short range interactions or geometric propagators. This might signal a dependence to softness of the propagation of interactions, as already predicted \cite{Cates1998} and observed in impact experiments \cite{Clark2015}. A plausible explanation is that the force chains formations are more prominent at the stiff-particle limit, and as such, the effective interactions are long-ranged \cite{clark2015nonlinear} and avalanches can be represented in terms of isotropic homogeneous and independent branching processes. On the contrary, at the soft-particle limit, the deformability of particles accommodates the stress variation in finite structures well modelled by Eshelby-like propagators in the coarse-grained mesoscale.
Therefore, we argue that the non-equilibrium statistical physics of fragile matter do not significantly differ from those of amorphous materials with  attractive potentials as long as correlations, or influence of force-chain structures, are damped at short length scales. Similar parameter-dependent exponents have been found in other simulations and linked to changes in the effective dimension of avalanche location \cite{Amitrano2003,Niemann2012}. From standard renormalization theory it is known that, in general, effective system dimensions and the range of interactions condition the universality class, i.e. the critical exponents. The behavior at intermediate softness could therefore represent a cross-over phenomenon, giving rise to an ``effective'' set of exponents that would depend on the system size.

Finally, following the hyperscaling relation between avalanche-size distribution and the pseudogap \cite{Lin2016}:
\begin{equation}
\kappa = 2 - \frac{\theta}{\theta + 1 }\frac{d}{d_f}.
\end{equation}
Assuming $d_f$ as constant, the exponent $\theta$, determining the distribution of residual stresses, Eq.~(\ref{e:pseudogap}), and interevent times, transitions from $\theta \approx 0.5$ for stiff particles to $\theta \approx 1$ for soft particles if we use the fitted values of $\kappa$ from Table~\ref{tb:exponents}. These two extreme cases (represented in Fig.~\ref{fig:WT}) are in agreement, respectively, with the expected $\theta=0.5$ found in mean field models \cite{Lin2016} and $\theta \approx 0.9-1$ found in DEM simulations considering LJ potentials \cite{Liu2016, Jagla2015,Shang2019}. Overall, we observe a general correspondence between the exponents fitted at low confining pressure with the predictions of SMFT, and we recover the usual ranges of values found in EPM and other simulations of amorphous materials at higher confining pressures.

\subsection{Avalanche statistics on the SES are stationary and scale free}
\label{sec:critSES}

The interpretation of the SES manifold as a stability limit solves the problem of the non-monotonous evolution from a macroscopic point of view, once structural configurational aspects are taken into account.
Beyond that, and as far as we can tell, the SES defines the 'state' of the assembly where avalanche statistics is non-degenerated. 
Upon reaching the SES at different stress values, and once the variations in the avalanche activity rate are accounted for, no difference in the statistics of avalanches is observed (Fig.~\ref{fig:PUKsplit}).
The distributions of avalanche sizes are virtually identical, independently of the $\sigma_y$ values, the evolutionary path, and the initial configuration.
Focusing on the result at the stiff-particle limit, and despite the complexity of the internal processes involved, the macromechanical evolution of loose assemblies, such as L5kSc5 in Fig.~\ref{fig:macromech}, is in agreement with the predictions of the SMFT (Eq.~\eqref{eq:critSMFT}). In particular, the SMFT predictions extend way beyond the proximity-limit of $\sigma_c$, and apply to the entire observation interval. In contrast, initially dense assemblies, such as D5kSc5 in Fig.~\ref{fig:macromech}, display the typical peak in stress and post-peak strain softening which is not captured by Eq.~\eqref{eq:critSMFT} nor the scaling relations in SMFT. 
However, once SES is reached, the distributions in Fig.~\ref{fig:PUKsplit} are already compatible with the scale-free behavior with SMFT universality, besides the aforementioned finite size effects, way before the steady state flow regime.
Therefore, we argue that the SES is not only an attractor manifold to the internal state of the system, but might as well correspond to the self-organized critical (SOC) state \cite{Bak1991,Chen1991,Zapperi1995,Watkins2016} from the perspective of out-of-equilibrium statistical physics. This is also demonstrated in Ref.~\cite{Pouragha2018mu} where a preliminary measure for a distance to criticality is suggested based on the concept of SES. Further studies are nevertheless necessary to determine whether the SES is indeed an exact representation of avalanche criticality or yet another approximation. This question can be addressed in principle through a finite size scaling analysis of avalanche distributions in terms of the parameters representing the distance to SES. Yet, it is a comptuationally cumbersome task if one wants to obtain statistically reliable conclusions~\cite{Budrikis2017}, which remains a challenge for the future.

\section{Conclusions:}
\label{sec:conclude}

We presented a statistical analysis of kinetic avalanches recorded during DEM simulations of granular assemblies with elastic potentials under AQS conditions. It showed that criticality, in the form of power-law statistics, is degenerated in the space of the commonly used macroscopic state variables such as stress and strains. Instead, internal measurements, such as the fabric anisotropy, the rigidity ratio, and average coordination number, in which the manifold herein referred to as stable evolutionary state (SES) is defined, appear to provide a more appropriate representation of criticality, and closeness to criticality, in terms of avalanche statistics and susceptibilities. In particular, we have presented evidence that points towards the existence of a link between (i) the SES, identified as an attractor of the evolution of granular systems in the space of internal measurements \cite{Pouragha2018mu} and (ii) criticality characterized by a divergence of susceptibilities and scale-free avalanches. Therefore, we argue that the SES is itself a self-organized critical (SOC) state, representing a basin of attraction to the state of the system with scale-free independent avalanches before reaching the stationary flow regime.\\

We interpret the observed time predictability as evidence that the SES corresponds to a stability limit under AQS conditions.  
In terms of the residual-stress formulation, we conjecture that the SES, determined by the potential energy landscape, dominates the dynamics during driving as considered in EPM, although the high susceptibility of the system might introduce non-affine stress rearrangements through the force-chain network. Conversely, based on the higher correlation between states at the onset of avalanches than between states arresting avalanches, we believe that the avalanche propagation is strongly affected by the transient kinetic energy in underdamped inertial systems. 
Therefore, the stability of the assembly during the compression process is determined by the combined effect of equilibrium constraints governing the quasi-static stability and the transient addition of kinetic energy during avalanche propagation. Testing this hypothesis further remains a challenge for the future, potentially involving simulations with alternative control protocols, the inclusion of thermal effects and/or the tuning of frictional and damping parameters.
On the other hand, the reported avalanche interevent times are compatible with the power-law pseudogap hypothesis, rendering a Weibull renewal process for large system sizes, while the Poisson hypothesis is still valid for small systems. This is in agreement with the possible existence of a system-size dependent plateau in the distribution of residual stresses recently discussed in the literature~\cite{Ruscher2019,Korchinski2021, Ferrero2019}.\\

Despite the particular aspects of our simulations, the reported avalanche statistics do not differ significantly from the expectations of EPM and can, in fact, bridge the gap between the depinning universality class associated with the slip mean field theory (SMFT) and the extended non-universal exponents in yielding \cite{Lin2014} that could represent a cross-over towards a new universality class of elastoplasticity and, therefore, be invariant with respect to the dimensionality of the system~\cite{Budrikis2017}.
We found that the external stress drop ($\Delta\sigma$) is proportional to the internal energy drop $\Delta U$ and a good proxy for the internal displacement, or $S$ in terms of the slip mean field theory (SMFT)  \cite{Dahmen2011,Dahmen2017}. The peak of kinetic energy ($K$) serves as a good proxy for the avalanche energy $E$. The observed exponents characterizing the critical state are non-universal and display a transition from (i) the expected values for SMFT, for stiff particles or low confining pressures in our simulations, to (ii) the lower exponents observed in 2d simulations of amorphous materials \cite{Budrikis2017,Liu2016,Jagla2015,Shang2019}, for soft particles or high confining pressures in our simulations. This could be linked to the known change in spatial distribution and anisotropy of correlations in the two limits, which depends on the structure and robustness of the internal force-chain network connecting the contact points \cite{Cates1998,Clark2015}. In particular, for stiff particles, small displacements in a single contact can propagate through the force-chain and span the whole system, resulting in a homogeneous and isotropic long-ranged response at the coarse-grained scale, similar to the interactions in the SMFT. In contrast, soft-particles accommodate variations of stress at a short-range scale, reminiscent of the typical Eshelby inclusion modeled in 2d EPM \cite{Albaret2016}.  On the other hand, intertial effects are found to depend on stiffness and are known to affect avalanche statistics as well~\cite{Salerno2012,Salerno2013,Karimi2016,Karimi2017}. Finally, the relation between stiffness and the universality class is further reinforced by the indirect estimation of the exponent $\theta$ defining the distribution of residual stresses, which satisfies the expectations of mean-field models \cite{Lin2016} for stiff particles and 2D EPM for soft particles \cite{Liu2016, Jagla2015,Shang2019}.\\

To summarize, based on our results we propose a static stability limit (SES) as a candidate for a SOC state reached before the traditional flow regime, and a link between the collection of critical exponents defining the critical state and the stiffness of the particles. These two novel conjectures should serve as a starting point to extend our understanding of the discussed non-universality in the deformation of amorphous materials, and the role of the internal structure and its dynamics during the deformation. Indeed, we strongly believe that this might be a fundamental aspect to understand the fit of fragile matter within elastoplastic theory.\\

\acknowledgements

Daniel Korchinski for fruitful discussions. 
J.B. thanks financial support from the Banco de Santander - María de Maetzu fellowship program, and the AXA Research Fund through the project RheMechFail.
This work was financially supported by the Natural Sciences and Engineering Research Council of Canada (NSERC).


\begin{thebibliography}{138}%
\makeatletter
\providecommand \@ifxundefined [1]{%
 \@ifx{#1\undefined}
}%
\providecommand \@ifnum [1]{%
 \ifnum #1\expandafter \@firstoftwo
 \else \expandafter \@secondoftwo
 \fi
}%
\providecommand \@ifx [1]{%
 \ifx #1\expandafter \@firstoftwo
 \else \expandafter \@secondoftwo
 \fi
}%
\providecommand \natexlab [1]{#1}%
\providecommand \enquote  [1]{``#1''}%
\providecommand \bibnamefont  [1]{#1}%
\providecommand \bibfnamefont [1]{#1}%
\providecommand \citenamefont [1]{#1}%
\providecommand \href@noop [0]{\@secondoftwo}%
\providecommand \href [0]{\begingroup \@sanitize@url \@href}%
\providecommand \@href[1]{\@@startlink{#1}\@@href}%
\providecommand \@@href[1]{\endgroup#1\@@endlink}%
\providecommand \@sanitize@url [0]{\catcode `\\12\catcode `\$12\catcode
  `\&12\catcode `\#12\catcode `\^12\catcode `\_12\catcode `\%12\relax}%
\providecommand \@@startlink[1]{}%
\providecommand \@@endlink[0]{}%
\providecommand \url  [0]{\begingroup\@sanitize@url \@url }%
\providecommand \@url [1]{\endgroup\@href {#1}{\urlprefix }}%
\providecommand \urlprefix  [0]{URL }%
\providecommand \Eprint [0]{\href }%
\providecommand \doibase [0]{http://dx.doi.org/}%
\providecommand \selectlanguage [0]{\@gobble}%
\providecommand \bibinfo  [0]{\@secondoftwo}%
\providecommand \bibfield  [0]{\@secondoftwo}%
\providecommand \translation [1]{[#1]}%
\providecommand \BibitemOpen [0]{}%
\providecommand \bibitemStop [0]{}%
\providecommand \bibitemNoStop [0]{.\EOS\space}%
\providecommand \EOS [0]{\spacefactor3000\relax}%
\providecommand \BibitemShut  [1]{\csname bibitem#1\endcsname}%
\let\auto@bib@innerbib\@empty
%
\bibitem [{\citenamefont {Sethna}\ \emph {et~al.}(2001)\citenamefont {Sethna},
  \citenamefont {Dahmen},\ and\ \citenamefont {Myers}}]{Sethna2001}%
  \BibitemOpen
  \bibfield  {author} {\bibinfo {author} {\bibfnamefont {J.~P.}\ \bibnamefont
  {Sethna}}, \bibinfo {author} {\bibfnamefont {K.~A.}\ \bibnamefont {Dahmen}},
  \ and\ \bibinfo {author} {\bibfnamefont {C.~R.}\ \bibnamefont {Myers}},\
  }\href {\doibase 10.1038/35065675} {\bibfield  {journal} {\bibinfo  {journal}
  {Nature}\ }\textbf {\bibinfo {volume} {410}},\ \bibinfo {pages} {242}
  (\bibinfo {year} {2001})}\BibitemShut {NoStop}%
\bibitem [{\citenamefont {Salje}\ and\ \citenamefont
  {Dahmen}(2014)}]{Salje2014}%
  \BibitemOpen
  \bibfield  {author} {\bibinfo {author} {\bibfnamefont {E.~K.}\ \bibnamefont
  {Salje}}\ and\ \bibinfo {author} {\bibfnamefont {K.~A.}\ \bibnamefont
  {Dahmen}},\ }\href {\doibase 10.1146/annurev-conmatphys-031113-133838}
  {\bibfield  {journal} {\bibinfo  {journal} {Annual Review of Condensed Matter
  Physics}\ }\textbf {\bibinfo {volume} {5}},\ \bibinfo {pages} {233} (\bibinfo
  {year} {2014})},\ %
\bibitem [{\citenamefont {Nicolas}\ \emph {et~al.}(2018)\citenamefont
  {Nicolas}, \citenamefont {Ferrero}, \citenamefont {Martens},\ and\
  \citenamefont {Barrat}}]{Nicolas2018}%
  \BibitemOpen
  \bibfield  {author} {\bibinfo {author} {\bibfnamefont {A.}~\bibnamefont
  {Nicolas}}, \bibinfo {author} {\bibfnamefont {E.~E.}\ \bibnamefont
  {Ferrero}}, \bibinfo {author} {\bibfnamefont {K.}~\bibnamefont {Martens}}, \
  and\ \bibinfo {author} {\bibfnamefont {J.-L.}\ \bibnamefont {Barrat}},\
  }\href {\doibase 10.1103/RevModPhys.90.045006} {\bibfield  {journal}
  {\bibinfo  {journal} {Rev. Mod. Phys.}\ }\textbf {\bibinfo {volume} {90}},\
  \bibinfo {pages} {045006} (\bibinfo {year} {2018})}\BibitemShut {NoStop}%
\bibitem [{\citenamefont {Miguel}\ \emph {et~al.}(2001)\citenamefont {Miguel},
  \citenamefont {Vespignani}, \citenamefont {Zapperi}, \citenamefont {Weiss},\
  and\ \citenamefont {Grasso}}]{Miguel2001}%
  \BibitemOpen
  \bibfield  {author} {\bibinfo {author} {\bibfnamefont {M.~C.}\ \bibnamefont
  {Miguel}}, \bibinfo {author} {\bibfnamefont {A.}~\bibnamefont {Vespignani}},
  \bibinfo {author} {\bibfnamefont {S.}~\bibnamefont {Zapperi}}, \bibinfo
  {author} {\bibfnamefont {J.}~\bibnamefont {Weiss}}, \ and\ \bibinfo {author}
  {\bibfnamefont {J.-R.}\ \bibnamefont {Grasso}},\ }\href {\doibase
  10.1038/35070524} {\bibfield  {journal} {\bibinfo  {journal} {Nature}\
  }\textbf {\bibinfo {volume} {410}},\ \bibinfo {pages} {667} (\bibinfo {year}
  {2001})}\BibitemShut {NoStop}%
\bibitem [{\citenamefont {Zaiser}(2006)}]{Zaiser2006}%
  \BibitemOpen
  \bibfield  {author} {\bibinfo {author} {\bibfnamefont {M.}~\bibnamefont
  {Zaiser}},\ }\href {\doibase 10.1080/00018730600583514} {\bibfield  {journal}
  {\bibinfo  {journal} {Advances in Physics}\ }\textbf {\bibinfo {volume}
  {55}},\ \bibinfo {pages} {185} (\bibinfo {year} {2006})},\  \BibitemShut {NoStop}%
\bibitem [{\citenamefont {Friedman}\ \emph {et~al.}(2012)\citenamefont
  {Friedman}, \citenamefont {Jennings}, \citenamefont {Tsekenis}, \citenamefont
  {Kim}, \citenamefont {Tao}, \citenamefont {Uhl}, \citenamefont {Greer},\ and\
  \citenamefont {Dahmen}}]{Friedman2012}%
  \BibitemOpen
  \bibfield  {author} {\bibinfo {author} {\bibfnamefont {N.}~\bibnamefont
  {Friedman}}, \bibinfo {author} {\bibfnamefont {A.~T.}\ \bibnamefont
  {Jennings}}, \bibinfo {author} {\bibfnamefont {G.}~\bibnamefont {Tsekenis}},
  \bibinfo {author} {\bibfnamefont {J.-Y.}\ \bibnamefont {Kim}}, \bibinfo
  {author} {\bibfnamefont {M.}~\bibnamefont {Tao}}, \bibinfo {author}
  {\bibfnamefont {J.~T.}\ \bibnamefont {Uhl}}, \bibinfo {author} {\bibfnamefont
  {J.~R.}\ \bibnamefont {Greer}}, \ and\ \bibinfo {author} {\bibfnamefont
  {K.~A.}\ \bibnamefont {Dahmen}},\ }\href {\doibase
  10.1103/PhysRevLett.109.095507} {\bibfield  {journal} {\bibinfo  {journal}
  {Phys. Rev. Lett.}\ }\textbf {\bibinfo {volume} {109}},\ \bibinfo {pages}
  {095507} (\bibinfo {year} {2012})}\BibitemShut {NoStop}%
\bibitem [{\citenamefont {Wang}\ \emph {et~al.}(2009)\citenamefont {Wang},
  \citenamefont {Chan}, \citenamefont {Xia}, \citenamefont {Yu}, \citenamefont
  {Shen},\ and\ \citenamefont {Wang}}]{Wang2009}%
  \BibitemOpen
  \bibfield  {author} {\bibinfo {author} {\bibfnamefont {G.}~\bibnamefont
  {Wang}}, \bibinfo {author} {\bibfnamefont {K.}~\bibnamefont {Chan}}, \bibinfo
  {author} {\bibfnamefont {L.}~\bibnamefont {Xia}}, \bibinfo {author}
  {\bibfnamefont {P.}~\bibnamefont {Yu}}, \bibinfo {author} {\bibfnamefont
  {J.}~\bibnamefont {Shen}}, \ and\ \bibinfo {author} {\bibfnamefont
  {W.}~\bibnamefont {Wang}},\ }\href {\doibase
  https://doi.org/10.1016/j.actamat.2009.08.040} {\bibfield  {journal}
  {\bibinfo  {journal} {Acta Materialia}\ }\textbf {\bibinfo {volume} {57}},\
  \bibinfo {pages} {6146 } (\bibinfo {year} {2009})}\BibitemShut {NoStop}%
\bibitem [{\citenamefont {Antonaglia}\ \emph {et~al.}(2014)\citenamefont
  {Antonaglia}, \citenamefont {Wright}, \citenamefont {Gu}, \citenamefont
  {Byer}, \citenamefont {Hufnagel}, \citenamefont {LeBlanc}, \citenamefont
  {Uhl},\ and\ \citenamefont {Dahmen}}]{Antonaglia2014}%
  \BibitemOpen
  \bibfield  {author} {\bibinfo {author} {\bibfnamefont {J.}~\bibnamefont
  {Antonaglia}}, \bibinfo {author} {\bibfnamefont {W.~J.}\ \bibnamefont
  {Wright}}, \bibinfo {author} {\bibfnamefont {X.}~\bibnamefont {Gu}}, \bibinfo
  {author} {\bibfnamefont {R.~R.}\ \bibnamefont {Byer}}, \bibinfo {author}
  {\bibfnamefont {T.~C.}\ \bibnamefont {Hufnagel}}, \bibinfo {author}
  {\bibfnamefont {M.}~\bibnamefont {LeBlanc}}, \bibinfo {author} {\bibfnamefont
  {J.~T.}\ \bibnamefont {Uhl}}, \ and\ \bibinfo {author} {\bibfnamefont
  {K.~A.}\ \bibnamefont {Dahmen}},\ }\href {\doibase
  10.1103/PhysRevLett.112.155501} {\bibfield  {journal} {\bibinfo  {journal}
  {Phys. Rev. Lett.}\ }\textbf {\bibinfo {volume} {112}},\ \bibinfo {pages}
  {155501} (\bibinfo {year} {2014})}\BibitemShut {NoStop}%
\bibitem [{\citenamefont {Schall}\ \emph {et~al.}(2007)\citenamefont {Schall},
  \citenamefont {Weitz},\ and\ \citenamefont {Spaepen}}]{Schall2007}%
  \BibitemOpen
  \bibfield  {author} {\bibinfo {author} {\bibfnamefont {P.}~\bibnamefont
  {Schall}}, \bibinfo {author} {\bibfnamefont {D.~A.}\ \bibnamefont {Weitz}}, \
  and\ \bibinfo {author} {\bibfnamefont {F.}~\bibnamefont {Spaepen}},\ }\href
  {\doibase 10.1126/science.1149308} {\bibfield  {journal} {\bibinfo  {journal}
  {Science}\ }\textbf {\bibinfo {volume} {318}},\ \bibinfo {pages} {1895}
  (\bibinfo {year} {2007})},\  \BibitemShut
  {NoStop}%
\bibitem [{\citenamefont {Hu}\ \emph {et~al.}(2018)\citenamefont {Hu},
  \citenamefont {Shu}, \citenamefont {Yang}, \citenamefont {Guo}, \citenamefont
  {Liaw}, \citenamefont {Dahmen},\ and\ \citenamefont {Zuo}}]{Hu2018}%
  \BibitemOpen
  \bibfield  {author} {\bibinfo {author} {\bibfnamefont {Y.}~\bibnamefont
  {Hu}}, \bibinfo {author} {\bibfnamefont {L.}~\bibnamefont {Shu}}, \bibinfo
  {author} {\bibfnamefont {Q.}~\bibnamefont {Yang}}, \bibinfo {author}
  {\bibfnamefont {W.}~\bibnamefont {Guo}}, \bibinfo {author} {\bibfnamefont
  {P.~K.}\ \bibnamefont {Liaw}}, \bibinfo {author} {\bibfnamefont {K.~A.}\
  \bibnamefont {Dahmen}}, \ and\ \bibinfo {author} {\bibfnamefont {J.-M.}\
  \bibnamefont {Zuo}},\ }\href {\doibase 10.1038/s42005-018-0062-z} {\bibfield
  {journal} {\bibinfo  {journal} {Communications Physics}\ }\textbf {\bibinfo
  {volume} {1}},\ \bibinfo {pages} {61} (\bibinfo {year} {2018})}\BibitemShut
  {NoStop}%
\bibitem [{\citenamefont {Rizzardi}\ \emph {et~al.}(2018)\citenamefont
  {Rizzardi}, \citenamefont {Sparks},\ and\ \citenamefont
  {Maa{\ss}}}]{Rizzardi2018}%
  \BibitemOpen
  \bibfield  {author} {\bibinfo {author} {\bibfnamefont {Q.}~\bibnamefont
  {Rizzardi}}, \bibinfo {author} {\bibfnamefont {G.}~\bibnamefont {Sparks}}, \
  and\ \bibinfo {author} {\bibfnamefont {R.}~\bibnamefont {Maa{\ss}}},\ }\href
  {\doibase 10.1007/s11837-018-2856-6} {\bibfield  {journal} {\bibinfo
  {journal} {JOM}\ }\textbf {\bibinfo {volume} {70}},\ \bibinfo {pages} {1088}
  (\bibinfo {year} {2018})}\BibitemShut {NoStop}%
\bibitem [{\citenamefont {Dahmen}\ \emph {et~al.}(2011)\citenamefont {Dahmen},
  \citenamefont {Ben-Zion},\ and\ \citenamefont {Uhl}}]{Dahmen2011}%
  \BibitemOpen
  \bibfield  {author} {\bibinfo {author} {\bibfnamefont {K.~a.}\ \bibnamefont
  {Dahmen}}, \bibinfo {author} {\bibfnamefont {Y.}~\bibnamefont {Ben-Zion}}, \
  and\ \bibinfo {author} {\bibfnamefont {J.~T.}\ \bibnamefont {Uhl}},\ }\href
  {\doibase 10.1038/nphys1957} {\bibfield  {journal} {\bibinfo  {journal}
  {Nature Physics}\ }\textbf {\bibinfo {volume} {7}},\ \bibinfo {pages} {554}
  (\bibinfo {year} {2011})}\BibitemShut {NoStop}%
\bibitem [{\citenamefont {Denisov}\ \emph {et~al.}(2016)\citenamefont
  {Denisov}, \citenamefont {L{\"o}rincz}, \citenamefont {Uhl}, \citenamefont
  {Dahmen},\ and\ \citenamefont {Schall}}]{Denisov2016}%
  \BibitemOpen
  \bibfield  {author} {\bibinfo {author} {\bibfnamefont {D.~V.}\ \bibnamefont
  {Denisov}}, \bibinfo {author} {\bibfnamefont {K.~A.}\ \bibnamefont
  {L{\"o}rincz}}, \bibinfo {author} {\bibfnamefont {J.~T.}\ \bibnamefont
  {Uhl}}, \bibinfo {author} {\bibfnamefont {K.~A.}\ \bibnamefont {Dahmen}}, \
  and\ \bibinfo {author} {\bibfnamefont {P.}~\bibnamefont {Schall}},\ }\href
  {\doibase 10.1038/ncomms10641} {\bibfield  {journal} {\bibinfo  {journal}
  {Nature Communications}\ }\textbf {\bibinfo {volume} {7}},\ \bibinfo {pages}
  {10641} (\bibinfo {year} {2016})}\BibitemShut {NoStop}%
\bibitem [{\citenamefont {Berthier}\ \emph
  {et~al.}(2019{\natexlab{a}})\citenamefont {Berthier}, \citenamefont
  {Kollmer}, \citenamefont {Henkes}, \citenamefont {Liu}, \citenamefont
  {Schwarz},\ and\ \citenamefont {Daniels}}]{Berthier2019a}%
  \BibitemOpen
  \bibfield  {author} {\bibinfo {author} {\bibfnamefont {E.}~\bibnamefont
  {Berthier}}, \bibinfo {author} {\bibfnamefont {J.~E.}\ \bibnamefont
  {Kollmer}}, \bibinfo {author} {\bibfnamefont {S.~E.}\ \bibnamefont {Henkes}},
  \bibinfo {author} {\bibfnamefont {K.}~\bibnamefont {Liu}}, \bibinfo {author}
  {\bibfnamefont {J.~M.}\ \bibnamefont {Schwarz}}, \ and\ \bibinfo {author}
  {\bibfnamefont {K.~E.}\ \bibnamefont {Daniels}},\ }\href {\doibase
  10.1103/PhysRevMaterials.3.075602} {\bibfield  {journal} {\bibinfo  {journal}
  {Phys. Rev. Materials}\ }\textbf {\bibinfo {volume} {3}},\ \bibinfo {pages}
  {075602} (\bibinfo {year} {2019}{\natexlab{a}})}\BibitemShut {NoStop}%
\bibitem [{\citenamefont {Maloney}\ and\ \citenamefont
  {Lema\^{\i}tre}(2004)}]{Maloney2004}%
  \BibitemOpen
  \bibfield  {author} {\bibinfo {author} {\bibfnamefont {C.}~\bibnamefont
  {Maloney}}\ and\ \bibinfo {author} {\bibfnamefont {A.}~\bibnamefont
  {Lema\^{\i}tre}},\ }\href {\doibase 10.1103/PhysRevLett.93.016001} {\bibfield
   {journal} {\bibinfo  {journal} {Phys. Rev. Lett.}\ }\textbf {\bibinfo
  {volume} {93}},\ \bibinfo {pages} {016001} (\bibinfo {year}
  {2004})}\BibitemShut {NoStop}%
\bibitem [{\citenamefont {Salerno}\ \emph {et~al.}(2012)\citenamefont
  {Salerno}, \citenamefont {Maloney},\ and\ \citenamefont
  {Robbins}}]{Salerno2012}%
  \BibitemOpen
  \bibfield  {author} {\bibinfo {author} {\bibfnamefont {K.~M.}\ \bibnamefont
  {Salerno}}, \bibinfo {author} {\bibfnamefont {C.~E.}\ \bibnamefont
  {Maloney}}, \ and\ \bibinfo {author} {\bibfnamefont {M.~O.}\ \bibnamefont
  {Robbins}},\ }\href {\doibase 10.1103/PhysRevLett.109.105703} {\bibfield
  {journal} {\bibinfo  {journal} {Phys. Rev. Lett.}\ }\textbf {\bibinfo
  {volume} {109}},\ \bibinfo {pages} {105703} (\bibinfo {year}
  {2012})}\BibitemShut {NoStop}%
\bibitem [{\citenamefont {Salerno}\ and\ \citenamefont
  {Robbins}(2013)}]{Salerno2013}%
  \BibitemOpen
  \bibfield  {author} {\bibinfo {author} {\bibfnamefont {K.~M.}\ \bibnamefont
  {Salerno}}\ and\ \bibinfo {author} {\bibfnamefont {M.~O.}\ \bibnamefont
  {Robbins}},\ }\href {\doibase 10.1103/PhysRevE.88.062206} {\bibfield
  {journal} {\bibinfo  {journal} {Phys. Rev. E}\ }\textbf {\bibinfo {volume}
  {88}},\ \bibinfo {pages} {062206} (\bibinfo {year} {2013})}\BibitemShut
  {NoStop}%
\bibitem [{\citenamefont {Parisi}\ \emph {et~al.}(2017)\citenamefont {Parisi},
  \citenamefont {Procaccia}, \citenamefont {Rainone},\ and\ \citenamefont
  {Singh}}]{Parisi2017}%
  \BibitemOpen
  \bibfield  {author} {\bibinfo {author} {\bibfnamefont {G.}~\bibnamefont
  {Parisi}}, \bibinfo {author} {\bibfnamefont {I.}~\bibnamefont {Procaccia}},
  \bibinfo {author} {\bibfnamefont {C.}~\bibnamefont {Rainone}}, \ and\
  \bibinfo {author} {\bibfnamefont {M.}~\bibnamefont {Singh}},\ }\href
  {\doibase 10.1073/pnas.1700075114} {\bibfield  {journal} {\bibinfo  {journal}
  {Proceedings of the National Academy of Sciences}\ }\textbf {\bibinfo
  {volume} {114}},\ \bibinfo {pages} {5577} (\bibinfo {year} {2017})},\ %
\bibitem [{\citenamefont {Zhang}\ \emph {et~al.}(2017)\citenamefont {Zhang},
  \citenamefont {Dahmen},\ and\ \citenamefont {Ostoja-Starzewski}}]{Zhang2017}%
  \BibitemOpen
  \bibfield  {author} {\bibinfo {author} {\bibfnamefont {D.}~\bibnamefont
  {Zhang}}, \bibinfo {author} {\bibfnamefont {K.~A.}\ \bibnamefont {Dahmen}}, \
  and\ \bibinfo {author} {\bibfnamefont {M.}~\bibnamefont
  {Ostoja-Starzewski}},\ }\href {\doibase 10.1103/PhysRevE.95.032902}
  {\bibfield  {journal} {\bibinfo  {journal} {Phys. Rev. E}\ }\textbf {\bibinfo
  {volume} {95}},\ \bibinfo {pages} {032902} (\bibinfo {year}
  {2017})}\BibitemShut {NoStop}%
\bibitem [{\citenamefont {Shang}\ \emph {et~al.}(2020)\citenamefont {Shang},
  \citenamefont {Guan},\ and\ \citenamefont {Barrat}}]{Shang2019}%
  \BibitemOpen
  \bibfield  {author} {\bibinfo {author} {\bibfnamefont {B.}~\bibnamefont
  {Shang}}, \bibinfo {author} {\bibfnamefont {P.}~\bibnamefont {Guan}}, \ and\
  \bibinfo {author} {\bibfnamefont {J.-L.}\ \bibnamefont {Barrat}},\ }\href
  {\doibase 10.1073/pnas.1915070117} {\bibfield  {journal} {\bibinfo  {journal}
  {Proceedings of the National Academy of Sciences}\ }\textbf {\bibinfo
  {volume} {117}},\ \bibinfo {pages} {86} (\bibinfo {year} {2020})},\ %
\bibitem [{\citenamefont {Lerner}\ and\ \citenamefont
  {Procaccia}(2009)}]{Lerner2009}%
  \BibitemOpen
  \bibfield  {author} {\bibinfo {author} {\bibfnamefont {E.}~\bibnamefont
  {Lerner}}\ and\ \bibinfo {author} {\bibfnamefont {I.}~\bibnamefont
  {Procaccia}},\ }\href {\doibase 10.1103/PhysRevE.79.066109} {\bibfield
  {journal} {\bibinfo  {journal} {Phys. Rev. E}\ }\textbf {\bibinfo {volume}
  {79}},\ \bibinfo {pages} {066109} (\bibinfo {year} {2009})}\BibitemShut
  {NoStop}%
\bibitem [{\citenamefont {Karmakar}\ \emph {et~al.}(2010)\citenamefont
  {Karmakar}, \citenamefont {Lerner},\ and\ \citenamefont
  {Procaccia}}]{Karmakar2010}%
  \BibitemOpen
  \bibfield  {author} {\bibinfo {author} {\bibfnamefont {S.}~\bibnamefont
  {Karmakar}}, \bibinfo {author} {\bibfnamefont {E.}~\bibnamefont {Lerner}}, \
  and\ \bibinfo {author} {\bibfnamefont {I.}~\bibnamefont {Procaccia}},\ }\href
  {\doibase 10.1103/PhysRevE.82.055103} {\bibfield  {journal} {\bibinfo
  {journal} {Phys. Rev. E}\ }\textbf {\bibinfo {volume} {82}},\ \bibinfo
  {pages} {055103(R)} (\bibinfo {year} {2010})}\BibitemShut {NoStop}%
\bibitem [{\citenamefont {Gimbert}\ \emph {et~al.}(2013)\citenamefont
  {Gimbert}, \citenamefont {Amitrano},\ and\ \citenamefont
  {Weiss}}]{Gimbert2013}%
  \BibitemOpen
  \bibfield  {author} {\bibinfo {author} {\bibfnamefont {F.}~\bibnamefont
  {Gimbert}}, \bibinfo {author} {\bibfnamefont {D.}~\bibnamefont {Amitrano}}, \
  and\ \bibinfo {author} {\bibfnamefont {J.}~\bibnamefont {Weiss}},\ }\href
  {\doibase 10.1209/0295-5075/104/46001} {\bibfield  {journal} {\bibinfo
  {journal} {{EPL} (Europhysics Letters)}\ }\textbf {\bibinfo {volume} {104}},\
  \bibinfo {pages} {46001} (\bibinfo {year} {2013})}\BibitemShut {NoStop}%
\bibitem [{\citenamefont {Pouragha}\ and\ \citenamefont
  {Wan}(2016)}]{Pouragha2016}%
  \BibitemOpen
  \bibfield  {author} {\bibinfo {author} {\bibfnamefont {M.}~\bibnamefont
  {Pouragha}}\ and\ \bibinfo {author} {\bibfnamefont {R.}~\bibnamefont {Wan}},\
  }\href {\doibase 10.1007/s10035-016-0640-2} {\bibfield  {journal} {\bibinfo
  {journal} {Granular Matter}\ }\textbf {\bibinfo {volume} {18}},\ \bibinfo
  {pages} {38} (\bibinfo {year} {2016})}\BibitemShut {NoStop}%
\bibitem [{\citenamefont {Karimi}\ and\ \citenamefont
  {Barrat}(2018)}]{Karimi2018}%
  \BibitemOpen
  \bibfield  {author} {\bibinfo {author} {\bibfnamefont {K.}~\bibnamefont
  {Karimi}}\ and\ \bibinfo {author} {\bibfnamefont {J.-L.}\ \bibnamefont
  {Barrat}},\ }\href {\doibase 10.1038/s41598-018-22310-z} {\bibfield
  {journal} {\bibinfo  {journal} {Scientific Reports}\ }\textbf {\bibinfo
  {volume} {8}},\ \bibinfo {pages} {4021} (\bibinfo {year} {2018})}\BibitemShut
  {NoStop}%
\bibitem [{\citenamefont {Cates}\ \emph {et~al.}(1998)\citenamefont {Cates},
  \citenamefont {Wittmer}, \citenamefont {Bouchaud},\ and\ \citenamefont
  {Claudin}}]{Cates1998}%
  \BibitemOpen
  \bibfield  {author} {\bibinfo {author} {\bibfnamefont {M.~E.}\ \bibnamefont
  {Cates}}, \bibinfo {author} {\bibfnamefont {J.~P.}\ \bibnamefont {Wittmer}},
  \bibinfo {author} {\bibfnamefont {J.-P.}\ \bibnamefont {Bouchaud}}, \ and\
  \bibinfo {author} {\bibfnamefont {P.}~\bibnamefont {Claudin}},\ }\href
  {\doibase 10.1103/PhysRevLett.81.1841} {\bibfield  {journal} {\bibinfo
  {journal} {Phys. Rev. Lett.}\ }\textbf {\bibinfo {volume} {81}},\ \bibinfo
  {pages} {1841} (\bibinfo {year} {1998})}\BibitemShut {NoStop}%
\bibitem [{\citenamefont {Liu}\ and\ \citenamefont {Nagel}(1998)}]{Liu1998}%
  \BibitemOpen
  \bibfield  {author} {\bibinfo {author} {\bibfnamefont {A.~J.}\ \bibnamefont
  {Liu}}\ and\ \bibinfo {author} {\bibfnamefont {S.~R.}\ \bibnamefont
  {Nagel}},\ }\href {\doibase 10.1038/23819} {\bibfield  {journal} {\bibinfo
  {journal} {Nature}\ }\textbf {\bibinfo {volume} {396}},\ \bibinfo {pages}
  {21} (\bibinfo {year} {1998})}\BibitemShut {NoStop}%
\bibitem [{\citenamefont {Bi}\ \emph {et~al.}(2011)\citenamefont {Bi},
  \citenamefont {Zhang}, \citenamefont {Chakraborty},\ and\ \citenamefont
  {Behringer}}]{Bi2011}%
  \BibitemOpen
  \bibfield  {author} {\bibinfo {author} {\bibfnamefont {D.}~\bibnamefont
  {Bi}}, \bibinfo {author} {\bibfnamefont {J.}~\bibnamefont {Zhang}}, \bibinfo
  {author} {\bibfnamefont {B.}~\bibnamefont {Chakraborty}}, \ and\ \bibinfo
  {author} {\bibfnamefont {R.~P.}\ \bibnamefont {Behringer}},\ }\href {\doibase
  10.1038/nature10667} {\bibfield  {journal} {\bibinfo  {journal} {Nature}\
  }\textbf {\bibinfo {volume} {480}},\ \bibinfo {pages} {355} (\bibinfo {year}
  {2011})}\BibitemShut {NoStop}%
\bibitem [{\citenamefont {Bak}\ and\ \citenamefont {Chen}(1991)}]{Bak1991}%
  \BibitemOpen
  \bibfield  {author} {\bibinfo {author} {\bibfnamefont {P.}~\bibnamefont
  {Bak}}\ and\ \bibinfo {author} {\bibfnamefont {K.}~\bibnamefont {Chen}},\
  }\href {\doibase 10.2307/24936753} {\bibfield  {journal} {\bibinfo  {journal}
  {Scientific American}\ }\textbf {\bibinfo {volume} {264}},\ \bibinfo {pages}
  {46} (\bibinfo {year} {1991})}\BibitemShut {NoStop}%
\bibitem [{\citenamefont {Chen}\ \emph {et~al.}(1991)\citenamefont {Chen},
  \citenamefont {Bak},\ and\ \citenamefont {Obukhov}}]{Chen1991}%
  \BibitemOpen
  \bibfield  {author} {\bibinfo {author} {\bibfnamefont {K.}~\bibnamefont
  {Chen}}, \bibinfo {author} {\bibfnamefont {P.}~\bibnamefont {Bak}}, \ and\
  \bibinfo {author} {\bibfnamefont {S.~P.}\ \bibnamefont {Obukhov}},\ }\href
  {\doibase 10.1103/PhysRevA.43.625} {\bibfield  {journal} {\bibinfo  {journal}
  {Phys. Rev. A}\ }\textbf {\bibinfo {volume} {43}},\ \bibinfo {pages} {625}
  (\bibinfo {year} {1991})}\BibitemShut {NoStop}%
\bibitem [{\citenamefont {Zapperi}\ \emph {et~al.}(1995)\citenamefont
  {Zapperi}, \citenamefont {Lauritsen},\ and\ \citenamefont
  {Stanley}}]{Zapperi1995}%
  \BibitemOpen
  \bibfield  {author} {\bibinfo {author} {\bibfnamefont {S.}~\bibnamefont
  {Zapperi}}, \bibinfo {author} {\bibfnamefont {K.~B.}\ \bibnamefont
  {Lauritsen}}, \ and\ \bibinfo {author} {\bibfnamefont {H.~E.}\ \bibnamefont
  {Stanley}},\ }\href {\doibase 10.1103/PhysRevLett.75.4071} {\bibfield
  {journal} {\bibinfo  {journal} {Phys. Rev. Lett.}\ }\textbf {\bibinfo
  {volume} {75}},\ \bibinfo {pages} {4071} (\bibinfo {year}
  {1995})}\BibitemShut {NoStop}%
\bibitem [{\citenamefont {Watkins}\ \emph {et~al.}(2016)\citenamefont
  {Watkins}, \citenamefont {Pruessner}, \citenamefont {Chapman}, \citenamefont
  {Crosby},\ and\ \citenamefont {Jensen}}]{Watkins2016}%
  \BibitemOpen
  \bibfield  {author} {\bibinfo {author} {\bibfnamefont {N.~W.}\ \bibnamefont
  {Watkins}}, \bibinfo {author} {\bibfnamefont {G.}~\bibnamefont {Pruessner}},
  \bibinfo {author} {\bibfnamefont {S.~C.}\ \bibnamefont {Chapman}}, \bibinfo
  {author} {\bibfnamefont {N.~B.}\ \bibnamefont {Crosby}}, \ and\ \bibinfo
  {author} {\bibfnamefont {H.~J.}\ \bibnamefont {Jensen}},\ }
  {\bibfield  {journal} {\bibinfo  {journal} {Space Science Reviews}\ }\textbf
  {\bibinfo {volume} {198}},\ \bibinfo {pages} {3} (\bibinfo {year}
  {2016})}\BibitemShut {NoStop}%
\bibitem [{\citenamefont {H\'ebraud}\ and\ \citenamefont
  {Lequeux}(1998)}]{Hebraud1998}%
  \BibitemOpen
  \bibfield  {author} {\bibinfo {author} {\bibfnamefont {P.}~\bibnamefont
  {H\'ebraud}}\ and\ \bibinfo {author} {\bibfnamefont {F.}~\bibnamefont
  {Lequeux}},\ }\href {\doibase 10.1103/PhysRevLett.81.2934} {\bibfield
  {journal} {\bibinfo  {journal} {Phys. Rev. Lett.}\ }\textbf {\bibinfo
  {volume} {81}},\ \bibinfo {pages} {2934} (\bibinfo {year}
  {1998})}\BibitemShut {NoStop}%
\bibitem [{\citenamefont {Ben-Zion}\ and\ \citenamefont
  {Rice}(1993)}]{Ben-Zion1993}%
  \BibitemOpen
  \bibfield  {author} {\bibinfo {author} {\bibfnamefont {Y.}~\bibnamefont
  {Ben-Zion}}\ and\ \bibinfo {author} {\bibfnamefont {J.~R.}\ \bibnamefont
  {Rice}},\ }\href {\doibase 10.1029/93JB01096} {\bibfield  {journal} {\bibinfo
   {journal} {Journal of Geophysical Research: Solid Earth}\ }\textbf {\bibinfo
  {volume} {98}},\ \bibinfo {pages} {14109} (\bibinfo {year} {1993})},\ %
\bibitem [{\citenamefont {Dahmen}\ \emph {et~al.}(2009)\citenamefont {Dahmen},
  \citenamefont {Ben-Zion},\ and\ \citenamefont {Uhl}}]{Dahmen2009}%
  \BibitemOpen
  \bibfield  {author} {\bibinfo {author} {\bibfnamefont {K.~A.}\ \bibnamefont
  {Dahmen}}, \bibinfo {author} {\bibfnamefont {Y.}~\bibnamefont {Ben-Zion}}, \
  and\ \bibinfo {author} {\bibfnamefont {J.~T.}\ \bibnamefont {Uhl}},\ }\href
  {\doibase 10.1103/PhysRevLett.102.175501} {\bibfield  {journal} {\bibinfo
  {journal} {Phys. Rev. Lett.}\ }\textbf {\bibinfo {volume} {102}},\ \bibinfo
  {pages} {175501} (\bibinfo {year} {2009})}\BibitemShut {NoStop}%
\bibitem [{\citenamefont {Dahmen}(2017)}]{Dahmen2017}%
  \BibitemOpen
  \bibfield  {author} {\bibinfo {author} {\bibfnamefont {K.~A.}\ \bibnamefont
  {Dahmen}},\ } {\bibinfo {title} {Mean field theory of slip
  statistics},}\ in\ \href {\doibase 10.1007/978-3-319-45612-6_2} {\emph
  {\bibinfo {booktitle} {Avalanches in Functional Materials and Geophysics}}},\
  \bibinfo {editor} {edited by\ \bibinfo {editor} {\bibfnamefont {E.~K.}\
  \bibnamefont {Salje}}, \bibinfo {editor} {\bibfnamefont {A.}~\bibnamefont
  {Saxena}}, \ and\ \bibinfo {editor} {\bibfnamefont {A.}~\bibnamefont
  {Planes}}}\ (\bibinfo  {publisher} {Springer International Publishing},\
  \bibinfo {address} {Cham},\ \bibinfo {year} {2017})\ pp.\ \bibinfo {pages}
  {19--30}\BibitemShut {NoStop}%
\bibitem [{\citenamefont {Dahmen}\ \emph {et~al.}(1998)\citenamefont {Dahmen},
  \citenamefont {Erta\ifmmode~\mbox{\c{s}}\else \c{s}\fi{}},\ and\
  \citenamefont {Ben-Zion}}]{Dahmen1998}%
  \BibitemOpen
  \bibfield  {author} {\bibinfo {author} {\bibfnamefont {K.}~\bibnamefont
  {Dahmen}}, \bibinfo {author} {\bibfnamefont {D.}~\bibnamefont
  {Erta\ifmmode~\mbox{\c{s}}\else \c{s}\fi{}}}, \ and\ \bibinfo {author}
  {\bibfnamefont {Y.}~\bibnamefont {Ben-Zion}},\ }\href {\doibase
  10.1103/PhysRevE.58.1494} {\bibfield  {journal} {\bibinfo  {journal} {Phys.
  Rev. E}\ }\textbf {\bibinfo {volume} {58}},\ \bibinfo {pages} {1494}
  (\bibinfo {year} {1998})}\BibitemShut {NoStop}%
\bibitem [{\citenamefont {Mehta}\ \emph {et~al.}(2006)\citenamefont {Mehta},
  \citenamefont {Dahmen},\ and\ \citenamefont {Ben-Zion}}]{Mehta2006}%
  \BibitemOpen
  \bibfield  {author} {\bibinfo {author} {\bibfnamefont {A.~P.}\ \bibnamefont
  {Mehta}}, \bibinfo {author} {\bibfnamefont {K.~A.}\ \bibnamefont {Dahmen}}, \
  and\ \bibinfo {author} {\bibfnamefont {Y.}~\bibnamefont {Ben-Zion}},\ }\href
  {\doibase 10.1103/PhysRevE.73.056104} {\bibfield  {journal} {\bibinfo
  {journal} {Phys. Rev. E}\ }\textbf {\bibinfo {volume} {73}},\ \bibinfo
  {pages} {056104} (\bibinfo {year} {2006})}\BibitemShut {NoStop}%
\bibitem [{\citenamefont {Geller}\ \emph {et~al.}(2015)\citenamefont {Geller},
  \citenamefont {Ecke}, \citenamefont {Dahmen},\ and\ \citenamefont
  {Backhaus}}]{Geller2015}%
  \BibitemOpen
  \bibfield  {author} {\bibinfo {author} {\bibfnamefont {D.~A.}\ \bibnamefont
  {Geller}}, \bibinfo {author} {\bibfnamefont {R.~E.}\ \bibnamefont {Ecke}},
  \bibinfo {author} {\bibfnamefont {K.~A.}\ \bibnamefont {Dahmen}}, \ and\
  \bibinfo {author} {\bibfnamefont {S.}~\bibnamefont {Backhaus}},\ }\href
  {\doibase 10.1103/PhysRevE.92.060201} {\bibfield  {journal} {\bibinfo
  {journal} {Phys. Rev. E}\ }\textbf {\bibinfo {volume} {92}},\ \bibinfo
  {pages} {060201(R)} (\bibinfo {year} {2015})}\BibitemShut {NoStop}%
\bibitem [{\citenamefont {Corwin}\ \emph {et~al.}(2005)\citenamefont {Corwin},
  \citenamefont {Jaeger},\ and\ \citenamefont {Nagel}}]{Corwin2005}%
  \BibitemOpen
  \bibfield  {author} {\bibinfo {author} {\bibfnamefont {E.~I.}\ \bibnamefont
  {Corwin}}, \bibinfo {author} {\bibfnamefont {H.~M.}\ \bibnamefont {Jaeger}},
  \ and\ \bibinfo {author} {\bibfnamefont {S.~R.}\ \bibnamefont {Nagel}},\
  }\href {\doibase 10.1038/nature03698} {\bibfield  {journal} {\bibinfo
  {journal} {Nature}\ }\textbf {\bibinfo {volume} {435}},\ \bibinfo {pages}
  {1075} (\bibinfo {year} {2005})}\BibitemShut {NoStop}%
\bibitem [{\citenamefont {Lin}\ \emph {et~al.}(2014)\citenamefont {Lin},
  \citenamefont {Saade}, \citenamefont {Lerner}, \citenamefont {Rosso},\ and\
  \citenamefont {Wyart}}]{Lin2014}%
  \BibitemOpen
  \bibfield  {author} {\bibinfo {author} {\bibfnamefont {J.}~\bibnamefont
  {Lin}}, \bibinfo {author} {\bibfnamefont {A.}~\bibnamefont {Saade}}, \bibinfo
  {author} {\bibfnamefont {E.}~\bibnamefont {Lerner}}, \bibinfo {author}
  {\bibfnamefont {A.}~\bibnamefont {Rosso}}, \ and\ \bibinfo {author}
  {\bibfnamefont {M.}~\bibnamefont {Wyart}},\ }\href {\doibase
  10.1209/0295-5075/105/26003} {\bibfield  {journal} {\bibinfo  {journal}
  {{EPL} (Europhysics Letters)}\ }\textbf {\bibinfo {volume} {105}},\ \bibinfo
  {pages} {26003} (\bibinfo {year} {2014})}\BibitemShut {NoStop}%
\bibitem [{\citenamefont {Budrikis}\ \emph {et~al.}(2017)\citenamefont
  {Budrikis}, \citenamefont {Castellanos}, \citenamefont {Sandfeld},
  \citenamefont {Zaiser},\ and\ \citenamefont {Zapperi}}]{Budrikis2017}%
  \BibitemOpen
  \bibfield  {author} {\bibinfo {author} {\bibfnamefont {Z.}~\bibnamefont
  {Budrikis}}, \bibinfo {author} {\bibfnamefont {D.~F.}\ \bibnamefont
  {Castellanos}}, \bibinfo {author} {\bibfnamefont {S.}~\bibnamefont
  {Sandfeld}}, \bibinfo {author} {\bibfnamefont {M.}~\bibnamefont {Zaiser}}, \
  and\ \bibinfo {author} {\bibfnamefont {S.}~\bibnamefont {Zapperi}},\ }\href
  {\doibase 10.1038/ncomms15928} {\bibfield  {journal} {\bibinfo  {journal}
  {Nature Communications}\ }\textbf {\bibinfo {volume} {8}},\ \bibinfo {pages}
  {15928} (\bibinfo {year} {2017})}\BibitemShut {NoStop}%
\bibitem [{\citenamefont {Korchinski}\ \emph {et~al.}(2021)\citenamefont
  {Korchinski}, \citenamefont {Ruscher},\ and\ \citenamefont
  {Rottler}}]{Korchinski2021}%
  \BibitemOpen
  \bibfield  {author} {\bibinfo {author} {\bibfnamefont {D.}~\bibnamefont
  {Korchinski}}, \bibinfo {author} {\bibfnamefont {C.}~\bibnamefont {Ruscher}},
  \ and\ \bibinfo {author} {\bibfnamefont {J.}~\bibnamefont {Rottler}},\
  } {\bibfield  {journal} {\bibinfo  {journal} {arXiv preprint
  arXiv:2101.10688}\ } (\bibinfo {year} {2021})}\BibitemShut {NoStop}%
\bibitem [{\citenamefont {Baret}\ \emph {et~al.}(2002)\citenamefont {Baret},
  \citenamefont {Vandembroucq},\ and\ \citenamefont {Roux}}]{Baret2002}%
  \BibitemOpen
  \bibfield  {author} {\bibinfo {author} {\bibfnamefont {J.-C.}\ \bibnamefont
  {Baret}}, \bibinfo {author} {\bibfnamefont {D.}~\bibnamefont {Vandembroucq}},
  \ and\ \bibinfo {author} {\bibfnamefont {S.}~\bibnamefont {Roux}},\ }\href
  {\doibase 10.1103/PhysRevLett.89.195506} {\bibfield  {journal} {\bibinfo
  {journal} {Phys. Rev. Lett.}\ }\textbf {\bibinfo {volume} {89}},\ \bibinfo
  {pages} {195506} (\bibinfo {year} {2002})}\BibitemShut {NoStop}%
\bibitem [{\citenamefont {Picard}\ \emph {et~al.}(2005)\citenamefont {Picard},
  \citenamefont {Ajdari},\ and\ \citenamefont {Lequeux}}]{Picard2005}%
  \BibitemOpen
  \bibfield  {author} {\bibinfo {author} {\bibfnamefont {G.}~\bibnamefont
  {Picard}}, \bibinfo {author} {\bibfnamefont {A.}~\bibnamefont {Ajdari}}, \
  and\ \bibinfo {author} {\bibfnamefont {F.}~\bibnamefont {Lequeux}},
  {\bibfnamefont {L.}~\bibfnamefont {Bocquet}},\
  }\href {\doibase 10.1103/PhysRevE.71.010501} {\bibfield  {journal} {\bibinfo
  {journal} {Phys. Rev. E}\ }\textbf {\bibinfo {volume} {71}},\ \bibinfo
  {pages} {010501(R)} (\bibinfo {year} {2005})}\BibitemShut {NoStop}%
\bibitem [{\citenamefont {Agoritsas}\ \emph {et~al.}(2015)\citenamefont
  {Agoritsas}, \citenamefont {Bertin}, \citenamefont {Martens},\ and\
  \citenamefont {Barrat}}]{Agoritsas2015}%
  \BibitemOpen
  \bibfield  {author} {\bibinfo {author} {\bibfnamefont {E.}~\bibnamefont
  {Agoritsas}}, \bibinfo {author} {\bibfnamefont {E.}~\bibnamefont {Bertin}},
  \bibinfo {author} {\bibfnamefont {K.}~\bibnamefont {Martens}}, \ and\
  \bibinfo {author} {\bibfnamefont {J.-L.}\ \bibnamefont {Barrat}},\ }\href
  {\doibase 10.1140/epje/i2015-15071-x} {\bibfield  {journal} {\bibinfo
  {journal} {The European Physical Journal E}\ }\textbf {\bibinfo {volume}
  {38}},\ \bibinfo {pages} {71} (\bibinfo {year} {2015})}\BibitemShut {NoStop}%
\bibitem [{\citenamefont {Lin}\ and\ \citenamefont {Wyart}(2016)}]{Lin2016}%
  \BibitemOpen
  \bibfield  {author} {\bibinfo {author} {\bibfnamefont {J.}~\bibnamefont
  {Lin}}\ and\ \bibinfo {author} {\bibfnamefont {M.}~\bibnamefont {Wyart}},\
  }\href {\doibase 10.1103/PhysRevX.6.011005} {\bibfield  {journal} {\bibinfo
  {journal} {Phys. Rev. X}\ }\textbf {\bibinfo {volume} {6}},\ \bibinfo {pages}
  {011005} (\bibinfo {year} {2016})}\BibitemShut {NoStop}%
\bibitem [{\citenamefont {Ferrero}\ and\ \citenamefont
  {Jagla}(2019)}]{Ferrero2019}%
  \BibitemOpen
  \bibfield  {author} {\bibinfo {author} {\bibfnamefont {E.~E.}\ \bibnamefont
  {Ferrero}}\ and\ \bibinfo {author} {\bibfnamefont {E.~A.}\ \bibnamefont
  {Jagla}},\ }\href {\doibase 10.1039/C9SM01073D} {\bibfield  {journal}
  {\bibinfo  {journal} {Soft Matter}\ }\textbf {\bibinfo {volume} {15}},\
  \bibinfo {pages} {9041} (\bibinfo {year} {2019})}\BibitemShut {NoStop}%
\bibitem [{\citenamefont {Rodney}\ \emph {et~al.}(2011)\citenamefont {Rodney},
  \citenamefont {Tanguy},\ and\ \citenamefont {Vandembroucq}}]{Rodney2011}%
  \BibitemOpen
  \bibfield  {author} {\bibinfo {author} {\bibfnamefont {D.}~\bibnamefont
  {Rodney}}, \bibinfo {author} {\bibfnamefont {A.}~\bibnamefont {Tanguy}}, \
  and\ \bibinfo {author} {\bibfnamefont {D.}~\bibnamefont {Vandembroucq}},\
  }\href {\doibase 10.1088/0965-0393/19/8/083001} {\bibfield  {journal}
  {\bibinfo  {journal} {Modelling and Simulation in Materials Science and
  Engineering}\ }\textbf {\bibinfo {volume} {19}},\ \bibinfo {pages} {083001}
  (\bibinfo {year} {2011})}\BibitemShut {NoStop}%
\bibitem [{\citenamefont {Miller}\ \emph {et~al.}(1996)\citenamefont {Miller},
  \citenamefont {O'Hern},\ and\ \citenamefont {Behringer}}]{Miller1996}%
  \BibitemOpen
  \bibfield  {author} {\bibinfo {author} {\bibfnamefont {B.}~\bibnamefont
  {Miller}}, \bibinfo {author} {\bibfnamefont {C.}~\bibnamefont {O'Hern}}, \
  and\ \bibinfo {author} {\bibfnamefont {R.~P.}\ \bibnamefont {Behringer}},\
  }\href {\doibase 10.1103/PhysRevLett.77.3110} {\bibfield  {journal} {\bibinfo
   {journal} {Phys. Rev. Lett.}\ }\textbf {\bibinfo {volume} {77}},\ \bibinfo
  {pages} {3110} (\bibinfo {year} {1996})}\BibitemShut {NoStop}%
\bibitem [{\citenamefont {Mueth}\ \emph {et~al.}(1998)\citenamefont {Mueth},
  \citenamefont {Jaeger},\ and\ \citenamefont {Nagel}}]{Mueth1998}%
  \BibitemOpen
  \bibfield  {author} {\bibinfo {author} {\bibfnamefont {D.~M.}\ \bibnamefont
  {Mueth}}, \bibinfo {author} {\bibfnamefont {H.~M.}\ \bibnamefont {Jaeger}}, \
  and\ \bibinfo {author} {\bibfnamefont {S.~R.}\ \bibnamefont {Nagel}},\
  } {\bibfield  {journal} {\bibinfo  {journal} {Physical Review
  E}\ }\textbf {\bibinfo {volume} {57}},\ \bibinfo {pages} {3164} (\bibinfo
  {year} {1998})}\BibitemShut {NoStop}%
\bibitem [{\citenamefont {Liu}\ \emph {et~al.}(1995)\citenamefont {Liu},
  \citenamefont {Nagel}, \citenamefont {Schecter}, \citenamefont {Coppersmith},
  \citenamefont {Majumdar}, \citenamefont {Narayan},\ and\ \citenamefont
  {Witten}}]{Liu1995}%
  \BibitemOpen
  \bibfield  {author} {\bibinfo {author} {\bibfnamefont {C.~h.}\ \bibnamefont
  {Liu}}, \bibinfo {author} {\bibfnamefont {S.~R.}\ \bibnamefont {Nagel}},
  \bibinfo {author} {\bibfnamefont {D.~A.}\ \bibnamefont {Schecter}}, \bibinfo
  {author} {\bibfnamefont {S.~N.}\ \bibnamefont {Coppersmith}}, \bibinfo
  {author} {\bibfnamefont {S.}~\bibnamefont {Majumdar}}, \bibinfo {author}
  {\bibfnamefont {O.}~\bibnamefont {Narayan}}, \ and\ \bibinfo {author}
  {\bibfnamefont {T.~A.}\ \bibnamefont {Witten}},\ }\href {\doibase
  10.1126/science.269.5223.513} {\bibfield  {journal} {\bibinfo  {journal}
  {Science}\ }\textbf {\bibinfo {volume} {269}},\ \bibinfo {pages} {513}
  (\bibinfo {year} {1995})},\ %
\bibitem [{\citenamefont {Howell}\ \emph {et~al.}(1999)\citenamefont {Howell},
  \citenamefont {Behringer},\ and\ \citenamefont {Veje}}]{Howell1999}%
  \BibitemOpen
  \bibfield  {author} {\bibinfo {author} {\bibfnamefont {D.}~\bibnamefont
  {Howell}}, \bibinfo {author} {\bibfnamefont {R.~P.}\ \bibnamefont
  {Behringer}}, \ and\ \bibinfo {author} {\bibfnamefont {C.}~\bibnamefont
  {Veje}},\ }\href {\doibase 10.1103/PhysRevLett.82.5241} {\bibfield  {journal}
  {\bibinfo  {journal} {Phys. Rev. Lett.}\ }\textbf {\bibinfo {volume} {82}},\
  \bibinfo {pages} {5241} (\bibinfo {year} {1999})}\BibitemShut {NoStop}%
\bibitem [{\citenamefont {And{\`o}}\ \emph {et~al.}(2012)\citenamefont
  {And{\`o}}, \citenamefont {Hall}, \citenamefont {Viggiani}, \citenamefont
  {Desrues},\ and\ \citenamefont {B{\'e}suelle}}]{Ando2012experimental}%
  \BibitemOpen
  \bibfield  {author} {\bibinfo {author} {\bibfnamefont {E.}~\bibnamefont
  {And{\`o}}}, \bibinfo {author} {\bibfnamefont {S.}~\bibnamefont {Hall}},
  \bibinfo {author} {\bibfnamefont {G.}~\bibnamefont {Viggiani}}, \bibinfo
  {author} {\bibfnamefont {J.}~\bibnamefont {Desrues}}, \ and\ \bibinfo
  {author} {\bibfnamefont {P.}~\bibnamefont {B{\'e}suelle}},\ }
  {\bibfield  {journal} {\bibinfo  {journal} {G{\'e}otechnique Letters}\
  }\textbf {\bibinfo {volume} {2}},\ \bibinfo {pages} {107} (\bibinfo {year}
  {2012})}\BibitemShut {NoStop}%
\bibitem [{\citenamefont {Gendelman}\ \emph {et~al.}(2016)\citenamefont
  {Gendelman}, \citenamefont {Pollack}, \citenamefont {Procaccia},
  \citenamefont {Sengupta},\ and\ \citenamefont {Zylberg}}]{Gendelman2016}%
  \BibitemOpen
  \bibfield  {author} {\bibinfo {author} {\bibfnamefont {O.}~\bibnamefont
  {Gendelman}}, \bibinfo {author} {\bibfnamefont {Y.~G.}\ \bibnamefont
  {Pollack}}, \bibinfo {author} {\bibfnamefont {I.}~\bibnamefont {Procaccia}},
  \bibinfo {author} {\bibfnamefont {S.}~\bibnamefont {Sengupta}}, \ and\
  \bibinfo {author} {\bibfnamefont {J.}~\bibnamefont {Zylberg}},\ }\href
  {\doibase 10.1103/PhysRevLett.116.078001} {\bibfield  {journal} {\bibinfo
  {journal} {Physical Review Letters}\ }\textbf {\bibinfo {volume} {116}},\
  \bibinfo {pages} {078001} (\bibinfo {year} {2016})},\ %
\bibitem [{\citenamefont {Berthier}\ \emph
  {et~al.}(2019{\natexlab{b}})\citenamefont {Berthier}, \citenamefont
  {Porter},\ and\ \citenamefont {Daniels}}]{Berthier2019b}%
  \BibitemOpen
  \bibfield  {author} {\bibinfo {author} {\bibfnamefont {E.}~\bibnamefont
  {Berthier}}, \bibinfo {author} {\bibfnamefont {M.~A.}\ \bibnamefont
  {Porter}}, \ and\ \bibinfo {author} {\bibfnamefont {K.~E.}\ \bibnamefont
  {Daniels}},\ }\href {\doibase 10.1073/pnas.1900272116} {\bibfield  {journal}
  {\bibinfo  {journal} {Proceedings of the National Academy of Sciences}\
  }\textbf {\bibinfo {volume} {116}},\ \bibinfo {pages} {16742} (\bibinfo
  {year} {2019}{\natexlab{b}})},\ %
\bibitem [{\citenamefont {Tordesillas}(2007)}]{Tordesillas2007}%
  \BibitemOpen
  \bibfield  {author} {\bibinfo {author} {\bibfnamefont {A.}~\bibnamefont
  {Tordesillas}},\ }\href {\doibase 10.1080/14786430701594848} {\bibfield
  {journal} {\bibinfo  {journal} {Philosophical Magazine}\ }\textbf {\bibinfo
  {volume} {87}},\ \bibinfo {pages} {4987} (\bibinfo {year} {2007})},\ %
\bibitem [{\citenamefont {Guo}\ and\ \citenamefont
  {Zhao}(2013)}]{Guo2013signature}%
  \BibitemOpen
  \bibfield  {author} {\bibinfo {author} {\bibfnamefont {N.}~\bibnamefont
  {Guo}}\ and\ \bibinfo {author} {\bibfnamefont {J.}~\bibnamefont {Zhao}},\
  } {\bibfield  {journal} {\bibinfo  {journal} {Computers and
  Geotechnics}\ }\textbf {\bibinfo {volume} {47}},\ \bibinfo {pages} {1}
  (\bibinfo {year} {2013})}\BibitemShut {NoStop}%
\bibitem [{\citenamefont {Kuhn}(1999)}]{Kuhn1999structured}%
  \BibitemOpen
  \bibfield  {author} {\bibinfo {author} {\bibfnamefont {M.~R.}\ \bibnamefont
  {Kuhn}},\ } {\bibfield  {journal} {\bibinfo  {journal}
  {Mechanics of materials}\ }\textbf {\bibinfo {volume} {31}},\ \bibinfo
  {pages} {407} (\bibinfo {year} {1999})}\BibitemShut {NoStop}%
\bibitem [{\citenamefont {Pouragha}\ and\ \citenamefont
  {Wan}(2017{\natexlab{a}})}]{Pouragha2017non}%
  \BibitemOpen
  \bibfield  {author} {\bibinfo {author} {\bibfnamefont {M.}~\bibnamefont
  {Pouragha}}\ and\ \bibinfo {author} {\bibfnamefont {R.}~\bibnamefont {Wan}},\
  } {\bibfield  {journal} {\bibinfo  {journal} {International
  Journal of Solids and Structures}\ }\textbf {\bibinfo {volume} {110}},\
  \bibinfo {pages} {94} (\bibinfo {year} {2017}{\natexlab{a}})}\BibitemShut
  {NoStop}%
\bibitem [{\citenamefont {Edwards}\ and\ \citenamefont
  {Oakeshott}(1989)}]{Edwards1989}%
  \BibitemOpen
  \bibfield  {author} {\bibinfo {author} {\bibfnamefont {S.}~\bibnamefont
  {Edwards}}\ and\ \bibinfo {author} {\bibfnamefont {R.}~\bibnamefont
  {Oakeshott}},\ }\href {\doibase https://doi.org/10.1016/0378-4371(89)90034-4}
  {\bibfield  {journal} {\bibinfo  {journal} {Physica A: Statistical Mechanics
  and its Applications}\ }\textbf {\bibinfo {volume} {157}},\ \bibinfo {pages}
  {1080 } (\bibinfo {year} {1989})}\BibitemShut {NoStop}%
\bibitem [{\citenamefont {Blumenfeld}\ and\ \citenamefont
  {Edwards}(2009)}]{Blumenfeld2009}%
  \BibitemOpen
  \bibfield  {author} {\bibinfo {author} {\bibfnamefont {R.}~\bibnamefont
  {Blumenfeld}}\ and\ \bibinfo {author} {\bibfnamefont {S.~F.}\ \bibnamefont
  {Edwards}},\ }\href {\doibase 10.1021/jp809768y} {\bibfield  {journal}
  {\bibinfo  {journal} {The Journal of Physical Chemistry B}\ }\textbf
  {\bibinfo {volume} {113}},\ \bibinfo {pages} {3981} (\bibinfo {year}
  {2009})},\ %
\bibitem [{\citenamefont {Blumenfeld}\ \emph {et~al.}(2012)\citenamefont
  {Blumenfeld}, \citenamefont {Jordan},\ and\ \citenamefont
  {Edwards}}]{Blumenfeld2012}%
  \BibitemOpen
  \bibfield  {author} {\bibinfo {author} {\bibfnamefont {R.}~\bibnamefont
  {Blumenfeld}}, \bibinfo {author} {\bibfnamefont {J.~F.}\ \bibnamefont
  {Jordan}}, \ and\ \bibinfo {author} {\bibfnamefont {S.~F.}\ \bibnamefont
  {Edwards}},\ }\href {\doibase 10.1103/PhysRevLett.109.238001} {\bibfield
  {journal} {\bibinfo  {journal} {Phys. Rev. Lett.}\ }\textbf {\bibinfo
  {volume} {109}},\ \bibinfo {pages} {238001} (\bibinfo {year}
  {2012})}\BibitemShut {NoStop}%
\bibitem [{\citenamefont {Bi}\ \emph {et~al.}(2015)\citenamefont {Bi},
  \citenamefont {Henkes}, \citenamefont {Daniels},\ and\ \citenamefont
  {Chakraborty}}]{Bi2015}%
  \BibitemOpen
  \bibfield  {author} {\bibinfo {author} {\bibfnamefont {D.}~\bibnamefont
  {Bi}}, \bibinfo {author} {\bibfnamefont {S.}~\bibnamefont {Henkes}}, \bibinfo
  {author} {\bibfnamefont {K.~E.}\ \bibnamefont {Daniels}}, \ and\ \bibinfo
  {author} {\bibfnamefont {B.}~\bibnamefont {Chakraborty}},\ }\href {\doibase
  10.1146/annurev-conmatphys-031214-014336} {\bibfield  {journal} {\bibinfo
  {journal} {Annual Review of Condensed Matter Physics}\ }\textbf {\bibinfo
  {volume} {6}},\ \bibinfo {pages} {63} (\bibinfo {year} {2015})},\ %
\bibitem [{\citenamefont {Baule}\ \emph {et~al.}(2018)\citenamefont {Baule},
  \citenamefont {Morone}, \citenamefont {Herrmann},\ and\ \citenamefont
  {Makse}}]{Baule2018}%
  \BibitemOpen
  \bibfield  {author} {\bibinfo {author} {\bibfnamefont {A.}~\bibnamefont
  {Baule}}, \bibinfo {author} {\bibfnamefont {F.}~\bibnamefont {Morone}},
  \bibinfo {author} {\bibfnamefont {H.~J.}\ \bibnamefont {Herrmann}}, \ and\
  \bibinfo {author} {\bibfnamefont {H.~A.}\ \bibnamefont {Makse}},\ }\href
  {\doibase 10.1103/RevModPhys.90.015006} {\bibfield  {journal} {\bibinfo
  {journal} {Rev. Mod. Phys.}\ }\textbf {\bibinfo {volume} {90}},\ \bibinfo
  {pages} {015006} (\bibinfo {year} {2018})}\BibitemShut {NoStop}%
\bibitem [{\citenamefont {Bililign}\ \emph {et~al.}(2019)\citenamefont
  {Bililign}, \citenamefont {Kollmer},\ and\ \citenamefont
  {Daniels}}]{Bililign2019}%
  \BibitemOpen
  \bibfield  {author} {\bibinfo {author} {\bibfnamefont {E.~S.}\ \bibnamefont
  {Bililign}}, \bibinfo {author} {\bibfnamefont {J.~E.}\ \bibnamefont
  {Kollmer}}, \ and\ \bibinfo {author} {\bibfnamefont {K.~E.}\ \bibnamefont
  {Daniels}},\ }\href {\doibase 10.1103/PhysRevLett.122.038001} {\bibfield
  {journal} {\bibinfo  {journal} {Phys. Rev. Lett.}\ }\textbf {\bibinfo
  {volume} {122}},\ \bibinfo {pages} {038001} (\bibinfo {year}
  {2019})}\BibitemShut {NoStop}%
\bibitem [{\citenamefont {Ball}(2019)}]{Ball2019}%
  \BibitemOpen
  \bibfield  {author} {\bibinfo {author} {\bibfnamefont {P.}~\bibnamefont
  {Ball}},\ }\href {\doibase 10.1038/s41563-019-0306-7} {\bibfield  {journal}
  {\bibinfo  {journal} {Nature Materials}\ }\textbf {\bibinfo {volume} {18}},\
  \bibinfo {pages} {196} (\bibinfo {year} {2019})}\BibitemShut {NoStop}%
\bibitem [{\citenamefont {Masson}\ and\ \citenamefont
  {Martinez}(2001)}]{masson2001}%
  \BibitemOpen
  \bibfield  {author} {\bibinfo {author} {\bibfnamefont {S.}~\bibnamefont
  {Masson}}\ and\ \bibinfo {author} {\bibfnamefont {J.}~\bibnamefont
  {Martinez}},\ } {\bibfield  {journal} {\bibinfo  {journal}
  {Journal of engineering mechanics}\ }\textbf {\bibinfo {volume} {127}},\
  \bibinfo {pages} {1007} (\bibinfo {year} {2001})}\BibitemShut {NoStop}%
\bibitem [{\citenamefont {Desrues}\ and\ \citenamefont
  {Viggiani}(2004)}]{Desrues2004strain}%
  \BibitemOpen
  \bibfield  {author} {\bibinfo {author} {\bibfnamefont {J.}~\bibnamefont
  {Desrues}}\ and\ \bibinfo {author} {\bibfnamefont {G.}~\bibnamefont
  {Viggiani}},\ } {\bibfield  {journal} {\bibinfo  {journal}
  {International Journal for Numerical and Analytical Methods in Geomechanics}\
  }\textbf {\bibinfo {volume} {28}},\ \bibinfo {pages} {279} (\bibinfo {year}
  {2004})}\BibitemShut {NoStop}%
\bibitem [{\citenamefont {Desrues}\ and\ \citenamefont
  {Hammad}(1989)}]{Desrues1989shear}%
  \BibitemOpen
  \bibfield  {author} {\bibinfo {author} {\bibfnamefont {J.}~\bibnamefont
  {Desrues}}\ and\ \bibinfo {author} {\bibfnamefont {W.}~\bibnamefont
  {Hammad}},\ }in\  {\emph {\bibinfo {booktitle} {Proc. of the
  Int. Workshop on Numerical Methods for Localization and Bifurcation of
  Granular Bodies}}}\ (\bibinfo {year} {1989})\ pp.\ \bibinfo {pages}
  {57--67}\BibitemShut {NoStop}%
\bibitem [{\citenamefont {Pouragha}\ and\ \citenamefont
  {Wan}(2017{\natexlab{b}})}]{Pouragha2017strain}%
  \BibitemOpen
  \bibfield  {author} {\bibinfo {author} {\bibfnamefont {M.}~\bibnamefont
  {Pouragha}}\ and\ \bibinfo {author} {\bibfnamefont {R.}~\bibnamefont {Wan}},\
  } {\bibfield  {journal} {\bibinfo  {journal} {Journal of
  Engineering Mechanics}\ }\textbf {\bibinfo {volume} {143}},\ \bibinfo {pages}
  {C4016002} (\bibinfo {year} {2017}{\natexlab{b}})}\BibitemShut {NoStop}%
\bibitem [{\citenamefont {Inc}(2008)}]{itasca2008version}%
  \BibitemOpen
  \bibfield  {author} {\bibinfo {author} {\bibfnamefont {I.~C.~G.}\
  \bibnamefont {Inc}},\ } {\bibfield  {journal} {\bibinfo
  {journal} {Minneapolis: Itasca Consulting Group Inc.}\ } (\bibinfo {year}
  {2008})}\BibitemShut {NoStop}%
\bibitem [{\citenamefont {Da~Cruz}\ \emph {et~al.}(2005)\citenamefont
  {Da~Cruz}, \citenamefont {Emam}, \citenamefont {Prochnow}, \citenamefont
  {Roux},\ and\ \citenamefont {Chevoir}}]{Da2005rheophysics}%
  \BibitemOpen
  \bibfield  {author} {\bibinfo {author} {\bibfnamefont {F.}~\bibnamefont
  {da~Cruz}}, \bibinfo {author} {\bibfnamefont {S.}~\bibnamefont {Emam}},
  \bibinfo {author} {\bibfnamefont {M.}~\bibnamefont {Prochnow}}, \bibinfo
  {author} {\bibfnamefont {J.-N.}\ \bibnamefont {Roux}}, \ and\ \bibinfo
  {author} {\bibfnamefont {F.}~\bibnamefont {Chevoir}},\ }
  {\bibfield  {journal} {\bibinfo  {journal} {Physical Review E}\ }\textbf
  {\bibinfo {volume} {72}},\ \bibinfo {pages} {021309} (\bibinfo {year}
  {2005})}\BibitemShut {NoStop}%
\bibitem [{\citenamefont {Radjai}\ and\ \citenamefont
  {Roux}(2002)}]{Radjai2002}%
  \BibitemOpen
  \bibfield  {author} {\bibinfo {author} {\bibfnamefont {F.}~\bibnamefont
  {Radjai}}\ and\ \bibinfo {author} {\bibfnamefont {S.}~\bibnamefont {Roux}},\
  }\href {\doibase 10.1103/PhysRevLett.89.064302} {\bibfield  {journal}
  {\bibinfo  {journal} {Phys. Rev. Lett.}\ }\textbf {\bibinfo {volume} {89}},\
  \bibinfo {pages} {064302} (\bibinfo {year} {2002})}\BibitemShut {NoStop}%
\bibitem [{\citenamefont {Bonfanti}\ \emph {et~al.}(2019)\citenamefont
  {Bonfanti}, \citenamefont {Guerra}, \citenamefont {Mondal}, \citenamefont
  {Procaccia},\ and\ \citenamefont {Zapperi}}]{Bonfanti2019}%
  \BibitemOpen
  \bibfield  {author} {\bibinfo {author} {\bibfnamefont {S.}~\bibnamefont
  {Bonfanti}}, \bibinfo {author} {\bibfnamefont {R.}~\bibnamefont {Guerra}},
  \bibinfo {author} {\bibfnamefont {C.}~\bibnamefont {Mondal}}, \bibinfo
  {author} {\bibfnamefont {I.}~\bibnamefont {Procaccia}}, \ and\ \bibinfo
  {author} {\bibfnamefont {S.}~\bibnamefont {Zapperi}},\ }\href {\doibase
  10.1103/PhysRevE.100.060602} {\bibfield  {journal} {\bibinfo  {journal}
  {Phys. Rev. E}\ }\textbf {\bibinfo {volume} {100}},\ \bibinfo {pages}
  {060602(R)} (\bibinfo {year} {2019})}\BibitemShut {NoStop}%
\bibitem [{\citenamefont {Nicot}\ \emph {et~al.}(2017)\citenamefont {Nicot},
  \citenamefont {Xiong}, \citenamefont {Wautier}, \citenamefont {Lerbet},\ and\
  \citenamefont {Darve}}]{Nicot2017force}%
  \BibitemOpen
  \bibfield  {author} {\bibinfo {author} {\bibfnamefont {F.}~\bibnamefont
  {Nicot}}, \bibinfo {author} {\bibfnamefont {H.}~\bibnamefont {Xiong}},
  \bibinfo {author} {\bibfnamefont {A.}~\bibnamefont {Wautier}}, \bibinfo
  {author} {\bibfnamefont {J.}~\bibnamefont {Lerbet}}, \ and\ \bibinfo {author}
  {\bibfnamefont {F.}~\bibnamefont {Darve}},\ } {\bibfield
  {journal} {\bibinfo  {journal} {Granular Matter}\ }\textbf {\bibinfo {volume}
  {19}},\ \bibinfo {pages} {18} (\bibinfo {year} {2017})}\BibitemShut {NoStop}%
\bibitem [{\citenamefont {Pouragha}\ and\ \citenamefont
  {Wan}(2018{\natexlab{a}})}]{Pouragha2018mu}%
  \BibitemOpen
  \bibfield  {author} {\bibinfo {author} {\bibfnamefont {M.}~\bibnamefont
  {Pouragha}}\ and\ \bibinfo {author} {\bibfnamefont {R.}~\bibnamefont {Wan}},\
  } {\bibfield  {journal} {\bibinfo  {journal} {Mechanics of
  Materials}\ }\textbf {\bibinfo {volume} {126}},\ \bibinfo {pages} {57}
  (\bibinfo {year} {2018}{\natexlab{a}})}\BibitemShut {NoStop}%
\bibitem [{\citenamefont {Lema\^{\i}tre}\ and\ \citenamefont
  {Caroli}(2009)}]{Lamaitre2009}%
  \BibitemOpen
  \bibfield  {author} {\bibinfo {author} {\bibfnamefont {A.}~\bibnamefont
  {Lema\^{\i}tre}}\ and\ \bibinfo {author} {\bibfnamefont {C.}~\bibnamefont
  {Caroli}},\ }\href {\doibase 10.1103/PhysRevLett.103.065501} {\bibfield
  {journal} {\bibinfo  {journal} {Phys. Rev. Lett.}\ }\textbf {\bibinfo
  {volume} {103}},\ \bibinfo {pages} {065501} (\bibinfo {year}
  {2009})}\BibitemShut {NoStop}%
\bibitem [{\citenamefont {LU}\ \emph {et~al.}(2007)\citenamefont {LU},
  \citenamefont {Brodsky},\ and\ \citenamefont {Kavehpour}}]{Lu2007}%
  \BibitemOpen
  \bibfield  {author} {\bibinfo {author} {\bibfnamefont {K.}~\bibnamefont
  {LU}}, \bibinfo {author} {\bibfnamefont {E.~E.}\ \bibnamefont {Brodsky}}, \
  and\ \bibinfo {author} {\bibfnamefont {H.~P.}\ \bibnamefont {Kavehpour}},\
  }\href {\doibase 10.1017/S0022112007007331} {\bibfield  {journal} {\bibinfo
  {journal} {Journal of Fluid Mechanics}\ }\textbf {\bibinfo {volume} {587}},\
  \bibinfo {pages} {347–372} (\bibinfo {year} {2007})}\BibitemShut {NoStop}%
\bibitem [{\citenamefont {Nicot}\ \emph {et~al.}(2012)\citenamefont {Nicot},
  \citenamefont {Hadda}, \citenamefont {Bourrier}, \citenamefont {Sibille},
  \citenamefont {Wan},\ and\ \citenamefont {Darve}}]{Nicot2012inertia}%
  \BibitemOpen
  \bibfield  {author} {\bibinfo {author} {\bibfnamefont {F.}~\bibnamefont
  {Nicot}}, \bibinfo {author} {\bibfnamefont {N.}~\bibnamefont {Hadda}},
  \bibinfo {author} {\bibfnamefont {F.}~\bibnamefont {Bourrier}}, \bibinfo
  {author} {\bibfnamefont {L.}~\bibnamefont {Sibille}}, \bibinfo {author}
  {\bibfnamefont {R.}~\bibnamefont {Wan}}, \ and\ \bibinfo {author}
  {\bibfnamefont {F.}~\bibnamefont {Darve}},\ } {\bibfield
  {journal} {\bibinfo  {journal} {International Journal of solids and
  Structures}\ }\textbf {\bibinfo {volume} {49}},\ \bibinfo {pages} {1252}
  (\bibinfo {year} {2012})}\BibitemShut {NoStop}%
\bibitem [{\citenamefont {Maimon}\ and\ \citenamefont
  {Schwarz}(2004)}]{Maimon2004}%
  \BibitemOpen
  \bibfield  {author} {\bibinfo {author} {\bibfnamefont {R.}~\bibnamefont
  {Maimon}}\ and\ \bibinfo {author} {\bibfnamefont {J.~M.}\ \bibnamefont
  {Schwarz}},\ }\href {\doibase 10.1103/PhysRevLett.92.255502} {\bibfield
  {journal} {\bibinfo  {journal} {Phys. Rev. Lett.}\ }\textbf {\bibinfo
  {volume} {92}},\ \bibinfo {pages} {255502} (\bibinfo {year}
  {2004})}\BibitemShut {NoStop}%
\bibitem [{\citenamefont {Chang}\ \emph {et~al.}(1995)\citenamefont {Chang},
  \citenamefont {Chao},\ and\ \citenamefont {Chang}}]{Chang1995estimates}%
  \BibitemOpen
  \bibfield  {author} {\bibinfo {author} {\bibfnamefont {C.~S.}\ \bibnamefont
  {Chang}}, \bibinfo {author} {\bibfnamefont {S.~J.}\ \bibnamefont {Chao}}, \
  and\ \bibinfo {author} {\bibfnamefont {Y.}~\bibnamefont {Chang}},\
  } {\bibfield  {journal} {\bibinfo  {journal} {International
  journal of solids and structures}\ }\textbf {\bibinfo {volume} {32}},\
  \bibinfo {pages} {1989} (\bibinfo {year} {1995})}\BibitemShut {NoStop}%
\bibitem [{\citenamefont {Pouragha}\ and\ \citenamefont
  {Wan}(2018{\natexlab{b}})}]{Pouragha2018elastic}%
  \BibitemOpen
  \bibfield  {author} {\bibinfo {author} {\bibfnamefont {M.}~\bibnamefont
  {Pouragha}}\ and\ \bibinfo {author} {\bibfnamefont {R.}~\bibnamefont {Wan}},\
  } {\bibfield  {journal} {\bibinfo  {journal} {International
  journal of solids and structures}\ }\textbf {\bibinfo {volume} {138}},\
  \bibinfo {pages} {97} (\bibinfo {year} {2018}{\natexlab{b}})}\BibitemShut
  {NoStop}%
\bibitem [{\citenamefont {Privman}(1990)}]{privman1990}%
  \BibitemOpen
  \bibfield  {author} {\bibinfo {author} {\bibfnamefont {V.}~\bibnamefont
  {Privman}},\ } {\emph {\bibinfo {title} {Finite size scaling and
  numerical simulation of statistical systems}}}\ (\bibinfo  {publisher} {World
  Scientific Singapore},\ \bibinfo {year} {1990})\BibitemShut {NoStop}%
\bibitem [{\citenamefont {Liu}\ \emph {et~al.}(2016)\citenamefont {Liu},
  \citenamefont {Ferrero}, \citenamefont {Puosi}, \citenamefont {Barrat},\ and\
  \citenamefont {Martens}}]{Liu2016}%
  \BibitemOpen
  \bibfield  {author} {\bibinfo {author} {\bibfnamefont {C.}~\bibnamefont
  {Liu}}, \bibinfo {author} {\bibfnamefont {E.~E.}\ \bibnamefont {Ferrero}},
  \bibinfo {author} {\bibfnamefont {F.}~\bibnamefont {Puosi}}, \bibinfo
  {author} {\bibfnamefont {J.-L.}\ \bibnamefont {Barrat}}, \ and\ \bibinfo
  {author} {\bibfnamefont {K.}~\bibnamefont {Martens}},\ }\href {\doibase
  10.1103/PhysRevLett.116.065501} {\bibfield  {journal} {\bibinfo  {journal}
  {Phys. Rev. Lett.}\ }\textbf {\bibinfo {volume} {116}},\ \bibinfo {pages}
  {065501} (\bibinfo {year} {2016})}\BibitemShut {NoStop}%
\bibitem [{\citenamefont {Zapperi}\ \emph {et~al.}(1997)\citenamefont
  {Zapperi}, \citenamefont {Vespignani},\ and\ \citenamefont
  {Stanley}}]{Zapperi1997}%
  \BibitemOpen
  \bibfield  {author} {\bibinfo {author} {\bibfnamefont {S.}~\bibnamefont
  {Zapperi}}, \bibinfo {author} {\bibfnamefont {A.}~\bibnamefont {Vespignani}},
  \ and\ \bibinfo {author} {\bibfnamefont {H.~E.}\ \bibnamefont {Stanley}},\
  }\href {\doibase 10.1038/41737} {\bibfield  {journal} {\bibinfo  {journal}
  {Nature}\ }\textbf {\bibinfo {volume} {388}},\ \bibinfo {pages} {658}
  (\bibinfo {year} {1997})}\BibitemShut {NoStop}%
\bibitem [{\citenamefont {Bar\'o}\ and\ \citenamefont
  {Vives}(2012)}]{Baro2012}%
  \BibitemOpen
  \bibfield  {author} {\bibinfo {author} {\bibfnamefont {J.}~\bibnamefont
  {Bar\'o}}\ and\ \bibinfo {author} {\bibfnamefont {E.}~\bibnamefont {Vives}},\
  }\href {\doibase 10.1103/PhysRevE.85.066121} {\bibfield  {journal} {\bibinfo
  {journal} {Phys. Rev. E}\ }\textbf {\bibinfo {volume} {85}},\ \bibinfo
  {pages} {066121} (\bibinfo {year} {2012})}\BibitemShut {NoStop}%
\bibitem [{\citenamefont {Christensen}\ \emph {et~al.}(2008)\citenamefont
  {Christensen}, \citenamefont {Farid}, \citenamefont {Pruessner},\ and\
  \citenamefont {Stapleton}}]{Christensen2008}%
  \BibitemOpen
  \bibfield  {author} {\bibinfo {author} {\bibfnamefont {K.}~\bibnamefont
  {Christensen}}, \bibinfo {author} {\bibfnamefont {N.}~\bibnamefont {Farid}},
  \bibinfo {author} {\bibfnamefont {G.}~\bibnamefont {Pruessner}}, \ and\
  \bibinfo {author} {\bibfnamefont {M.}~\bibnamefont {Stapleton}},\ }\href
  {\doibase 10.1140/epjb/e2008-00173-2} {\bibfield  {journal} {\bibinfo
  {journal} {The European Physical Journal B}\ }\textbf {\bibinfo {volume}
  {62}},\ \bibinfo {pages} {331} (\bibinfo {year} {2008})}\BibitemShut
  {NoStop}%
\bibitem [{\citenamefont {Karimi}\ and\ \citenamefont
  {Barrat}(2016)}]{Karimi2016}%
  \BibitemOpen
  \bibfield  {author} {\bibinfo {author} {\bibfnamefont {K.}~\bibnamefont
  {Karimi}}\ and\ \bibinfo {author} {\bibfnamefont {J.-L.}\ \bibnamefont
  {Barrat}},\ }\href {\doibase 10.1103/PhysRevE.93.022904} {\bibfield
  {journal} {\bibinfo  {journal} {Phys. Rev. E}\ }\textbf {\bibinfo {volume}
  {93}},\ \bibinfo {pages} {022904} (\bibinfo {year} {2016})}\BibitemShut
  {NoStop}%
\bibitem [{\citenamefont {Amitrano}(2003)}]{Amitrano2003}%
  \BibitemOpen
  \bibfield  {author} {\bibinfo {author} {\bibfnamefont {D.}~\bibnamefont
  {Amitrano}},\ }\href {\doibase 10.1029/2001JB000680} {\bibfield  {journal}
  {\bibinfo  {journal} {Journal of Geophysical Research: Solid Earth}\ }\textbf
  {\bibinfo {volume} {108}} (\bibinfo {year} {2003}),\ 10.1029/2001JB000680},\ %
\bibitem [{\citenamefont {Lord-May}\ \emph {et~al.}(2020)\citenamefont
  {Lord-May}, \citenamefont {Bar\'o}, \citenamefont {Eaton},\ and\
  \citenamefont {Davidsen}}]{Lord-May2020}%
  \BibitemOpen
  \bibfield  {author} {\bibinfo {author} {\bibfnamefont {C.}~\bibnamefont
  {Lord-May}}, \bibinfo {author} {\bibfnamefont {J.}~\bibnamefont {Bar\'o}},
  \bibinfo {author} {\bibfnamefont {D.~W.}\ \bibnamefont {Eaton}}, \ and\
  \bibinfo {author} {\bibfnamefont {J.}~\bibnamefont {Davidsen}},\ }\href
  {\doibase 10.1103/PhysRevResearch.2.043324} {\bibfield  {journal} {\bibinfo
  {journal} {Phys. Rev. Research}\ }\textbf {\bibinfo {volume} {2}},\ \bibinfo
  {pages} {043324} (\bibinfo {year} {2020})}\BibitemShut {NoStop}%
\bibitem [{\citenamefont {{Didier Sornette}}(1994)}]{Sornette1994}%
  \BibitemOpen
  \bibfield  {author} {\bibinfo {author} {\bibnamefont {{Didier Sornette}}},\
  }\href {\doibase 10.1051/jp1:1994133} {\bibfield  {journal} {\bibinfo
  {journal} {J. Phys. I France}\ }\textbf {\bibinfo {volume} {4}},\ \bibinfo
  {pages} {209} (\bibinfo {year} {1994})}\BibitemShut {NoStop}%
\bibitem [{\citenamefont {Vazquez}\ \emph {et~al.}(2011)\citenamefont
  {Vazquez}, \citenamefont {Bonachela}, \citenamefont {L\'opez},\ and\
  \citenamefont {Mu\~noz}}]{Vazquez2011}%
  \BibitemOpen
  \bibfield  {author} {\bibinfo {author} {\bibfnamefont {F.}~\bibnamefont
  {Vazquez}}, \bibinfo {author} {\bibfnamefont {J.~A.}\ \bibnamefont
  {Bonachela}}, \bibinfo {author} {\bibfnamefont {C.}~\bibnamefont {L\'opez}},
  \ and\ \bibinfo {author} {\bibfnamefont {M.~A.}\ \bibnamefont {Mu\~noz}},\
  }\href {\doibase 10.1103/PhysRevLett.106.235702} {\bibfield  {journal}
  {\bibinfo  {journal} {Phys. Rev. Lett.}\ }\textbf {\bibinfo {volume} {106}},\
  \bibinfo {pages} {235702} (\bibinfo {year} {2011})}\BibitemShut {NoStop}%
\bibitem [{\citenamefont {Utsu}\ \emph {et~al.}(1995)\citenamefont {Utsu},
  \citenamefont {Ogata} \emph {et~al.}}]{Utsu1995}%
  \BibitemOpen
  \bibfield  {author} {\bibinfo {author} {\bibfnamefont {T.}~\bibnamefont
  {Utsu}}, \bibinfo {author} {\bibfnamefont {Y.}~\bibnamefont {Ogata}},  \emph
  {et~al.},\ }\href {\doibase https://doi.org/10.4294/jpe1952.43.1} {\bibfield
  {journal} {\bibinfo  {journal} {Journal of Physics of the Earth}\ }\textbf
  {\bibinfo {volume} {43}},\ \bibinfo {pages} {1} (\bibinfo {year}
  {1995})}\BibitemShut {NoStop}%
\bibitem [{\citenamefont {Weiss}\ and\ \citenamefont {{Carmen
  Miguel}}(2004)}]{Weiss2004}%
  \BibitemOpen
  \bibfield  {author} {\bibinfo {author} {\bibfnamefont {J.}~\bibnamefont
  {Weiss}}\ and\ \bibinfo {author} {\bibfnamefont {M.}~\bibnamefont {{Carmen
  Miguel}}},\ }\href {\doibase https://doi.org/10.1016/j.msea.2004.01.101}
  {\bibfield  {journal} {\bibinfo  {journal} {Materials Science and
  Engineering: A}\ }\textbf {\bibinfo {volume} {387-389}},\ \bibinfo {pages}
  {292 } (\bibinfo {year} {2004})},\ \bibinfo {note} {13th International
  Conference on the Strength of Materials}\BibitemShut {NoStop}%
\bibitem [{\citenamefont {Crane}\ and\ \citenamefont
  {Sornette}(2008)}]{Crane2008}%
  \BibitemOpen
  \bibfield  {author} {\bibinfo {author} {\bibfnamefont {R.}~\bibnamefont
  {Crane}}\ and\ \bibinfo {author} {\bibfnamefont {D.}~\bibnamefont
  {Sornette}},\ }\href {\doibase 10.1073/pnas.0803685105} {\bibfield  {journal}
  {\bibinfo  {journal} {Proceedings of the National Academy of Sciences}\
  }\textbf {\bibinfo {volume} {105}},\ \bibinfo {pages} {15649} (\bibinfo
  {year} {2008})},\ %
\bibitem [{\citenamefont {Sornette}\ and\ \citenamefont
  {Utkin}(2009)}]{Sornette2009}%
  \BibitemOpen
  \bibfield  {author} {\bibinfo {author} {\bibfnamefont {D.}~\bibnamefont
  {Sornette}}\ and\ \bibinfo {author} {\bibfnamefont {S.}~\bibnamefont
  {Utkin}},\ }\href {\doibase 10.1103/PhysRevE.79.061110} {\bibfield  {journal}
  {\bibinfo  {journal} {Phys. Rev. E}\ }\textbf {\bibinfo {volume} {79}},\
  \bibinfo {pages} {061110} (\bibinfo {year} {2009})}\BibitemShut {NoStop}%
\bibitem [{\citenamefont {Bar\'o}\ \emph {et~al.}(2013)\citenamefont {Bar\'o},
  \citenamefont {Corral}, \citenamefont {Illa}, \citenamefont {Planes},
  \citenamefont {Salje}, \citenamefont {Schranz}, \citenamefont {Soto-Parra},\
  and\ \citenamefont {Vives}}]{Baro2013}%
  \BibitemOpen
  \bibfield  {author} {\bibinfo {author} {\bibfnamefont {J.}~\bibnamefont
  {Bar\'o}}, \bibinfo {author} {\bibfnamefont {A.}~\bibnamefont {Corral}},
  \bibinfo {author} {\bibfnamefont {X.}~\bibnamefont {Illa}}, \bibinfo {author}
  {\bibfnamefont {A.}~\bibnamefont {Planes}}, \bibinfo {author} {\bibfnamefont
  {E.~K.~H.}\ \bibnamefont {Salje}}, \bibinfo {author} {\bibfnamefont
  {W.}~\bibnamefont {Schranz}}, \bibinfo {author} {\bibfnamefont {D.~E.}\
  \bibnamefont {Soto-Parra}}, \ and\ \bibinfo {author} {\bibfnamefont
  {E.}~\bibnamefont {Vives}},\ }\href {\doibase 10.1103/PhysRevLett.110.088702}
  {\bibfield  {journal} {\bibinfo  {journal} {Phys. Rev. Lett.}\ }\textbf
  {\bibinfo {volume} {110}},\ \bibinfo {pages} {088702} (\bibinfo {year}
  {2013})}\BibitemShut {NoStop}%
\bibitem [{\citenamefont {Hainzl}\ \emph {et~al.}(2014)\citenamefont {Hainzl},
  \citenamefont {Moradpour},\ and\ \citenamefont {Davidsen}}]{Hainzl2014}%
  \BibitemOpen
  \bibfield  {author} {\bibinfo {author} {\bibfnamefont {S.}~\bibnamefont
  {Hainzl}}, \bibinfo {author} {\bibfnamefont {J.}~\bibnamefont {Moradpour}}, \
  and\ \bibinfo {author} {\bibfnamefont {J.}~\bibnamefont {Davidsen}},\ }\href
  {\doibase https://doi.org/10.1002/2014GL061975} {\bibfield  {journal}
  {\bibinfo  {journal} {Geophysical Research Letters}\ }\textbf {\bibinfo
  {volume} {41}},\ \bibinfo {pages} {8818} (\bibinfo {year} {2014})},\ %
\bibitem [{\citenamefont {Bar{\'{o}}}\ \emph {et~al.}(2014)\citenamefont
  {Bar{\'{o}}}, \citenamefont {Mart{\'{\i}}n-Olalla}, \citenamefont {Romero},
  \citenamefont {Gallardo}, \citenamefont {Salje}, \citenamefont {Vives},\ and\
  \citenamefont {Planes}}]{Baro2014}%
  \BibitemOpen
  \bibfield  {author} {\bibinfo {author} {\bibfnamefont {J.}~\bibnamefont
  {Bar{\'{o}}}}, \bibinfo {author} {\bibfnamefont {J.-M.}\ \bibnamefont
  {Mart{\'{\i}}n-Olalla}}, \bibinfo {author} {\bibfnamefont {F.~J.}\
  \bibnamefont {Romero}}, \bibinfo {author} {\bibfnamefont {M.~C.}\
  \bibnamefont {Gallardo}}, \bibinfo {author} {\bibfnamefont {E.~K.~H.}\
  \bibnamefont {Salje}}, \bibinfo {author} {\bibfnamefont {E.}~\bibnamefont
  {Vives}}, \ and\ \bibinfo {author} {\bibfnamefont {A.}~\bibnamefont
  {Planes}},\ }\href {\doibase 10.1088/0953-8984/26/12/125401} {\bibfield
  {journal} {\bibinfo  {journal} {Journal of Physics: Condensed Matter}\
  }\textbf {\bibinfo {volume} {26}},\ \bibinfo {pages} {125401} (\bibinfo
  {year} {2014})}\BibitemShut {NoStop}%
\bibitem [{\citenamefont {Davidsen}\ \emph {et~al.}(2017)\citenamefont
  {Davidsen}, \citenamefont {Kwiatek}, \citenamefont {Charalampidou},
  \citenamefont {Goebel}, \citenamefont {Stanchits}, \citenamefont {R\"uck},\
  and\ \citenamefont {Dresen}}]{Davidsen2017}%
  \BibitemOpen
  \bibfield  {author} {\bibinfo {author} {\bibfnamefont {J.}~\bibnamefont
  {Davidsen}}, \bibinfo {author} {\bibfnamefont {G.}~\bibnamefont {Kwiatek}},
  \bibinfo {author} {\bibfnamefont {E.-M.}\ \bibnamefont {Charalampidou}},
  \bibinfo {author} {\bibfnamefont {T.}~\bibnamefont {Goebel}}, \bibinfo
  {author} {\bibfnamefont {S.}~\bibnamefont {Stanchits}}, \bibinfo {author}
  {\bibfnamefont {M.}~\bibnamefont {R\"uck}}, \ and\ \bibinfo {author}
  {\bibfnamefont {G.}~\bibnamefont {Dresen}},\ }\href {\doibase
  10.1103/PhysRevLett.119.068501} {\bibfield  {journal} {\bibinfo  {journal}
  {Phys. Rev. Lett.}\ }\textbf {\bibinfo {volume} {119}},\ \bibinfo {pages}
  {068501} (\bibinfo {year} {2017})}\BibitemShut {NoStop}%
\bibitem [{\citenamefont {Kumar}\ \emph {et~al.}(2020)\citenamefont {Kumar},
  \citenamefont {Korkolis}, \citenamefont {Benzi}, \citenamefont {Denisov},
  \citenamefont {Niemeijer}, \citenamefont {Schall}, \citenamefont {Toschi},\
  and\ \citenamefont {Trampert}}]{Kumar2020}%
  \BibitemOpen
  \bibfield  {author} {\bibinfo {author} {\bibfnamefont {P.}~\bibnamefont
  {Kumar}}, \bibinfo {author} {\bibfnamefont {E.}~\bibnamefont {Korkolis}},
  \bibinfo {author} {\bibfnamefont {R.}~\bibnamefont {Benzi}}, \bibinfo
  {author} {\bibfnamefont {D.}~\bibnamefont {Denisov}}, \bibinfo {author}
  {\bibfnamefont {A.}~\bibnamefont {Niemeijer}}, \bibinfo {author}
  {\bibfnamefont {P.}~\bibnamefont {Schall}}, \bibinfo {author} {\bibfnamefont
  {F.}~\bibnamefont {Toschi}}, \ and\ \bibinfo {author} {\bibfnamefont
  {J.}~\bibnamefont {Trampert}},\ }\href {\doibase 10.1038/s41598-019-56764-6}
  {\bibfield  {journal} {\bibinfo  {journal} {Scientific Reports}\ }\textbf
  {\bibinfo {volume} {10}},\ \bibinfo {pages} {626} (\bibinfo {year}
  {2020})}\BibitemShut {NoStop}%
\bibitem [{\citenamefont {Dieterich}(1972)}]{Dieterich1972}%
  \BibitemOpen
  \bibfield  {author} {\bibinfo {author} {\bibfnamefont {J.~H.}\ \bibnamefont
  {Dieterich}},\ }\href {\doibase 10.1029/JB077i020p03771} {\bibfield
  {journal} {\bibinfo  {journal} {Journal of Geophysical Research (1896-1977)}\
  }\textbf {\bibinfo {volume} {77}},\ \bibinfo {pages} {3771} (\bibinfo {year}
  {1972})},\ %
\bibitem [{\citenamefont {Hainzl}\ \emph {et~al.}(1999)\citenamefont {Hainzl},
  \citenamefont {Zöller},\ and\ \citenamefont {Kurths}}]{Hainzl1999}%
  \BibitemOpen
  \bibfield  {author} {\bibinfo {author} {\bibfnamefont {S.}~\bibnamefont
  {Hainzl}}, \bibinfo {author} {\bibfnamefont {G.}~\bibnamefont {Zöller}}, \
  and\ \bibinfo {author} {\bibfnamefont {J.}~\bibnamefont {Kurths}},\ }\href
  {\doibase https://doi.org/10.1029/1998JB900122} {\bibfield  {journal}
  {\bibinfo  {journal} {Journal of Geophysical Research: Solid Earth}\ }\textbf
  {\bibinfo {volume} {104}},\ \bibinfo {pages} {7243} (\bibinfo {year}
  {1999})},\ %
\bibitem [{\citenamefont {Zöller}\ \emph {et~al.}(2006)\citenamefont
  {Zöller}, \citenamefont {Hainzl}, \citenamefont {Ben-Zion},\ and\
  \citenamefont {Holschneider}}]{Zoller2006}%
  \BibitemOpen
  \bibfield  {author} {\bibinfo {author} {\bibfnamefont {G.}~\bibnamefont
  {Zöller}}, \bibinfo {author} {\bibfnamefont {S.}~\bibnamefont {Hainzl}},
  \bibinfo {author} {\bibfnamefont {Y.}~\bibnamefont {Ben-Zion}}, \ and\
  \bibinfo {author} {\bibfnamefont {M.}~\bibnamefont {Holschneider}},\ }\href
  {\doibase https://doi.org/10.1016/j.tecto.2006.03.007} {\bibfield  {journal}
  {\bibinfo  {journal} {Tectonophysics}\ }\textbf {\bibinfo {volume} {423}},\
  \bibinfo {pages} {137} (\bibinfo {year} {2006})},\ \bibinfo {note}
  {spatiotemporal Models of Seismicity and Earthquake Occurrence}\BibitemShut
  {NoStop}%
\bibitem [{\citenamefont {Ben-Zion}\ and\ \citenamefont
  {Lyakhovsky}(2006)}]{Ben-Zion2006}%
  \BibitemOpen
  \bibfield  {author} {\bibinfo {author} {\bibfnamefont {Y.}~\bibnamefont
  {Ben-Zion}}\ and\ \bibinfo {author} {\bibfnamefont {V.}~\bibnamefont
  {Lyakhovsky}},\ }\href {\doibase 10.1111/j.1365-246X.2006.02878.x} {\bibfield
   {journal} {\bibinfo  {journal} {Geophysical Journal International}\ }\textbf
  {\bibinfo {volume} {165}},\ \bibinfo {pages} {197} (\bibinfo {year}
  {2006})},\ %
\bibitem [{\citenamefont {Jagla}(2014)}]{Jagla2014}%
  \BibitemOpen
  \bibfield  {author} {\bibinfo {author} {\bibfnamefont {E.~A.}\ \bibnamefont
  {Jagla}},\ }\href {\doibase 10.1103/PhysRevE.90.042129} {\bibfield  {journal}
  {\bibinfo  {journal} {Phys. Rev. E}\ }\textbf {\bibinfo {volume} {90}},\
  \bibinfo {pages} {042129} (\bibinfo {year} {2014})}\BibitemShut {NoStop}%
\bibitem [{\citenamefont {Zhang}\ and\ \citenamefont
  {Shcherbakov}(2016)}]{Zhang2016}%
  \BibitemOpen
  \bibfield  {author} {\bibinfo {author} {\bibfnamefont {X.}~\bibnamefont
  {Zhang}}\ and\ \bibinfo {author} {\bibfnamefont {R.}~\bibnamefont
  {Shcherbakov}},\ }\href {https://doi.org/10.1038/srep36668} {\bibfield
  {journal} {\bibinfo  {journal} {Scientific Reports}\ }\textbf {\bibinfo
  {volume} {6}},\ \bibinfo {pages} {36668 EP } (\bibinfo {year} {2016})},\
  \bibinfo {note} {article}\BibitemShut {NoStop}%
\bibitem [{\citenamefont {Bar\'o}\ and\ \citenamefont
  {Davidsen}(2018)}]{Baro2018a}%
  \BibitemOpen
  \bibfield  {author} {\bibinfo {author} {\bibfnamefont {J.}~\bibnamefont
  {Bar\'o}}\ and\ \bibinfo {author} {\bibfnamefont {J.}~\bibnamefont
  {Davidsen}},\ }\href {\doibase 10.1103/PhysRevE.97.033002} {\bibfield
  {journal} {\bibinfo  {journal} {Phys. Rev. E}\ }\textbf {\bibinfo {volume}
  {97}},\ \bibinfo {pages} {033002} (\bibinfo {year} {2018})}\BibitemShut
  {NoStop}%
\bibitem [{\citenamefont {Ruscher}\ and\ \citenamefont
  {Rottler}(2020)}]{Ruscher2019}%
  \BibitemOpen
  \bibfield  {author} {\bibinfo {author} {\bibfnamefont {C.}~\bibnamefont
  {Ruscher}}\ and\ \bibinfo {author} {\bibfnamefont {J.}~\bibnamefont
  {Rottler}},\ }\href {\doibase 10.1039/D0SM01155J} {\bibfield  {journal}
  {\bibinfo  {journal} {Soft Matter}\ }\textbf {\bibinfo {volume} {16}},\
  \bibinfo {pages} {8940} (\bibinfo {year} {2020})}\BibitemShut {NoStop}%
\bibitem [{\citenamefont {Touati}\ \emph {et~al.}(2009)\citenamefont {Touati},
  \citenamefont {Naylor},\ and\ \citenamefont {Main}}]{Touati2009}%
  \BibitemOpen
  \bibfield  {author} {\bibinfo {author} {\bibfnamefont {S.}~\bibnamefont
  {Touati}}, \bibinfo {author} {\bibfnamefont {M.}~\bibnamefont {Naylor}}, \
  and\ \bibinfo {author} {\bibfnamefont {I.~G.}\ \bibnamefont {Main}},\ }\href
  {\doibase 10.1103/PhysRevLett.102.168501} {\bibfield  {journal} {\bibinfo
  {journal} {Phys. Rev. Lett.}\ }\textbf {\bibinfo {volume} {102}},\ \bibinfo
  {pages} {168501} (\bibinfo {year} {2009})}\BibitemShut {NoStop}%
\bibitem [{\citenamefont {Bar{\'{o}}}\ and\ \citenamefont
  {Davidsen}(2017)}]{Baro2017}%
  \BibitemOpen
  \bibfield  {author} {\bibinfo {author} {\bibfnamefont {J.}~\bibnamefont
  {Bar{\'{o}}}}\ and\ \bibinfo {author} {\bibfnamefont {J.}~\bibnamefont
  {Davidsen}},\ }\href {\doibase 10.1140/epjst/e2017-70072-4} {\bibfield
  {journal} {\bibinfo  {journal} {European Physical Journal: Special Topics}\
  }\textbf {\bibinfo {volume} {226}},\ \bibinfo {pages} {3211} (\bibinfo {year}
  {2017})}\BibitemShut {NoStop}%
\bibitem [{\citenamefont {Kuhn}\ and\ \citenamefont
  {Daouadji}(2019)}]{Kuhn2019stress}%
  \BibitemOpen
  \bibfield  {author} {\bibinfo {author} {\bibfnamefont {M.~R.}\ \bibnamefont
  {Kuhn}}\ and\ \bibinfo {author} {\bibfnamefont {A.}~\bibnamefont
  {Daouadji}},\ } {\bibfield  {journal} {\bibinfo  {journal}
  {Granular Matter}\ }\textbf {\bibinfo {volume} {21}},\ \bibinfo {pages} {10}
  (\bibinfo {year} {2019})}\BibitemShut {NoStop}%
\bibitem [{\citenamefont {Bi}\ \emph {et~al.}(1989)\citenamefont {Bi},
  \citenamefont {B{\"o}rner},\ and\ \citenamefont {Chu}}]{Bi1989}%
  \BibitemOpen
  \bibfield  {author} {\bibinfo {author} {\bibfnamefont {H.}~\bibnamefont
  {Bi}}, \bibinfo {author} {\bibfnamefont {G.}~\bibnamefont {B{\"o}rner}}, \
  and\ \bibinfo {author} {\bibfnamefont {Y.}~\bibnamefont {Chu}},\ }
  {\bibfield  {journal} {\bibinfo  {journal} {Astronomy and Astrophysics}\
  }\textbf {\bibinfo {volume} {218}},\ \bibinfo {pages} {19} (\bibinfo {year}
  {1989})}\BibitemShut {NoStop}%
\bibitem [{\citenamefont {Reid}(1910)}]{Reid1910}%
  \BibitemOpen
  \bibfield  {author} {\bibinfo {author} {\bibfnamefont {H.~F.}\ \bibnamefont
  {Reid}},\ } {\bibfield  {journal} {\bibinfo  {journal} {The
  California Earthquake of April 18, 1906, Report of the State Earthquake
  Investigation Commission}\ } (\bibinfo {year} {1910})}\BibitemShut {NoStop}%
\bibitem [{\citenamefont {Shimazaki}\ and\ \citenamefont
  {Nakata}(1980)}]{Shimazaki1980}%
  \BibitemOpen
  \bibfield  {author} {\bibinfo {author} {\bibfnamefont {K.}~\bibnamefont
  {Shimazaki}}\ and\ \bibinfo {author} {\bibfnamefont {T.}~\bibnamefont
  {Nakata}},\ }\href {\doibase 10.1029/GL007i004p00279} {\bibfield  {journal}
  {\bibinfo  {journal} {Geophysical Research Letters}\ }\textbf {\bibinfo
  {volume} {7}},\ \bibinfo {pages} {279} (\bibinfo {year} {1980})},\ %
\bibitem [{\citenamefont {Combe}\ and\ \citenamefont
  {Roux}(2000)}]{combe2000strain}%
  \BibitemOpen
  \bibfield  {author} {\bibinfo {author} {\bibfnamefont {G.}~\bibnamefont
  {Combe}}\ and\ \bibinfo {author} {\bibfnamefont {J.-N.}\ \bibnamefont
  {Roux}},\ } {\bibfield  {journal} {\bibinfo  {journal} {Physical
  Review Letters}\ }\textbf {\bibinfo {volume} {85}},\ \bibinfo {pages} {3628}
  (\bibinfo {year} {2000})}\BibitemShut {NoStop}%
\bibitem [{\citenamefont {Picard}\ \emph {et~al.}(2004)\citenamefont {Picard},
  \citenamefont {Ajdari}, \citenamefont {Lequeux},\ and\ \citenamefont
  {Bocquet}}]{Picard2004}%
  \BibitemOpen
  \bibfield  {author} {\bibinfo {author} {\bibfnamefont {G.}~\bibnamefont
  {Picard}}, \bibinfo {author} {\bibfnamefont {A.}~\bibnamefont {Ajdari}},
  \bibinfo {author} {\bibfnamefont {F.}~\bibnamefont {Lequeux}}, \ and\
  \bibinfo {author} {\bibfnamefont {L.}~\bibnamefont {Bocquet}},\ }\href
  {\doibase 10.1140/epje/i2004-10054-8} {\bibfield  {journal} {\bibinfo
  {journal} {The European Physical Journal E}\ }\textbf {\bibinfo {volume}
  {15}},\ \bibinfo {pages} {371} (\bibinfo {year} {2004})}\BibitemShut
  {NoStop}%
\bibitem [{\citenamefont {Albaret}\ \emph {et~al.}(2016)\citenamefont
  {Albaret}, \citenamefont {Tanguy}, \citenamefont {Boioli},\ and\
  \citenamefont {Rodney}}]{Albaret2016}%
  \BibitemOpen
  \bibfield  {author} {\bibinfo {author} {\bibfnamefont {T.}~\bibnamefont
  {Albaret}}, \bibinfo {author} {\bibfnamefont {A.}~\bibnamefont {Tanguy}},
  \bibinfo {author} {\bibfnamefont {F.}~\bibnamefont {Boioli}}, \ and\ \bibinfo
  {author} {\bibfnamefont {D.}~\bibnamefont {Rodney}},\ }\href {\doibase
  10.1103/PhysRevE.93.053002} {\bibfield  {journal} {\bibinfo  {journal} {Phys.
  Rev. E}\ }\textbf {\bibinfo {volume} {93}},\ \bibinfo {pages} {053002}
  (\bibinfo {year} {2016})}\BibitemShut {NoStop}%
\bibitem [{\citenamefont {Lockner}(1993)}]{Lockner1993}%
  \BibitemOpen
  \bibfield  {author} {\bibinfo {author} {\bibfnamefont {D.}~\bibnamefont
  {Lockner}},\ }\href {\doibase https://doi.org/10.1016/0148-9062(93)90041-B}
  {\bibfield  {journal} {\bibinfo  {journal} {International Journal of Rock
  Mechanics and Mining Sciences \& Geomechanics Abstracts}\ }\textbf {\bibinfo
  {volume} {30}},\ \bibinfo {pages} {883} (\bibinfo {year} {1993})}\BibitemShut
  {NoStop}%
\bibitem [{\citenamefont {Lennartz-Sassinek}\ \emph {et~al.}(2014)\citenamefont
  {Lennartz-Sassinek}, \citenamefont {Main}, \citenamefont {Zaiser},\ and\
  \citenamefont {Graham}}]{Lennartz2014}%
  \BibitemOpen
  \bibfield  {author} {\bibinfo {author} {\bibfnamefont {S.}~\bibnamefont
  {Lennartz-Sassinek}}, \bibinfo {author} {\bibfnamefont {I.~G.}\ \bibnamefont
  {Main}}, \bibinfo {author} {\bibfnamefont {M.}~\bibnamefont {Zaiser}}, \ and\
  \bibinfo {author} {\bibfnamefont {C.~C.}\ \bibnamefont {Graham}},\ }\href
  {\doibase 10.1103/PhysRevE.90.052401} {\bibfield  {journal} {\bibinfo
  {journal} {Phys. Rev. E}\ }\textbf {\bibinfo {volume} {90}},\ \bibinfo
  {pages} {052401} (\bibinfo {year} {2014})}\BibitemShut {NoStop}%
\bibitem [{\citenamefont {Bar\'o}\ \emph {et~al.}(2018)\citenamefont {Bar\'o},
  \citenamefont {Dahmen}, \citenamefont {Davidsen}, \citenamefont {Planes},
  \citenamefont {Castillo}, \citenamefont {Nataf}, \citenamefont {Salje},\ and\
  \citenamefont {Vives}}]{Baro2018b}%
  \BibitemOpen
  \bibfield  {author} {\bibinfo {author} {\bibfnamefont {J.}~\bibnamefont
  {Bar\'o}}, \bibinfo {author} {\bibfnamefont {K.~A.}\ \bibnamefont {Dahmen}},
  \bibinfo {author} {\bibfnamefont {J.}~\bibnamefont {Davidsen}}, \bibinfo
  {author} {\bibfnamefont {A.}~\bibnamefont {Planes}}, \bibinfo {author}
  {\bibfnamefont {P.~O.}\ \bibnamefont {Castillo}}, \bibinfo {author}
  {\bibfnamefont {G.~F.}\ \bibnamefont {Nataf}}, \bibinfo {author}
  {\bibfnamefont {E.~K.~H.}\ \bibnamefont {Salje}}, \ and\ \bibinfo {author}
  {\bibfnamefont {E.}~\bibnamefont {Vives}},\ }\href {\doibase
  10.1103/PhysRevLett.120.245501} {\bibfield  {journal} {\bibinfo  {journal}
  {Phys. Rev. Lett.}\ }\textbf {\bibinfo {volume} {120}},\ \bibinfo {pages}
  {245501} (\bibinfo {year} {2018})}\BibitemShut {NoStop}%
\bibitem [{\citenamefont {Kun}\ \emph {et~al.}(2014)\citenamefont {Kun},
  \citenamefont {Varga}, \citenamefont {Lennartz-Sassinek},\ and\ \citenamefont
  {Main}}]{Kun2014}%
  \BibitemOpen
  \bibfield  {author} {\bibinfo {author} {\bibfnamefont {F.}~\bibnamefont
  {Kun}}, \bibinfo {author} {\bibfnamefont {I.}~\bibnamefont {Varga}}, \bibinfo
  {author} {\bibfnamefont {S.}~\bibnamefont {Lennartz-Sassinek}}, \ and\
  \bibinfo {author} {\bibfnamefont {I.~G.}\ \bibnamefont {Main}},\ }\href
  {\doibase 10.1103/PhysRevLett.112.065501} {\bibfield  {journal} {\bibinfo
  {journal} {Phys. Rev. Lett.}\ }\textbf {\bibinfo {volume} {112}},\ \bibinfo
  {pages} {065501} (\bibinfo {year} {2014})}\BibitemShut {NoStop}%
\bibitem [{\citenamefont {Kwiatek}\ \emph {et~al.}(2014)\citenamefont
  {Kwiatek}, \citenamefont {Goebel},\ and\ \citenamefont
  {Dresen}}]{Kwiatek2014}%
  \BibitemOpen
  \bibfield  {author} {\bibinfo {author} {\bibfnamefont {G.}~\bibnamefont
  {Kwiatek}}, \bibinfo {author} {\bibfnamefont {T.~H.~W.}\ \bibnamefont
  {Goebel}}, \ and\ \bibinfo {author} {\bibfnamefont {G.}~\bibnamefont
  {Dresen}},\ }\href {\doibase https://doi.org/10.1002/2014GL060159} {\bibfield
   {journal} {\bibinfo  {journal} {Geophysical Research Letters}\ }\textbf
  {\bibinfo {volume} {41}},\ \bibinfo {pages} {5838} (\bibinfo {year}
  {2014})},\ %
\bibitem [{\citenamefont {Tordesillas}\ \emph {et~al.}(2009)\citenamefont
  {Tordesillas}, \citenamefont {Zhang},\ and\ \citenamefont
  {Behringer}}]{tordesillas2009buckling}%
  \BibitemOpen
  \bibfield  {author} {\bibinfo {author} {\bibfnamefont {A.}~\bibnamefont
  {Tordesillas}}, \bibinfo {author} {\bibfnamefont {J.}~\bibnamefont {Zhang}},
  \ and\ \bibinfo {author} {\bibfnamefont {R.}~\bibnamefont {Behringer}},\
  } {\bibfield  {journal} {\bibinfo  {journal} {Geomechanics and
  Geoengineering: An International Journal}\ }\textbf {\bibinfo {volume} {4}},\
  \bibinfo {pages} {3} (\bibinfo {year} {2009})}\BibitemShut {NoStop}%
\bibitem [{\citenamefont {Vu}\ and\ \citenamefont {Weiss}(2020)}]{Vu2020}%
  \BibitemOpen
  \bibfield  {author} {\bibinfo {author} {\bibfnamefont {C.-C.}\ \bibnamefont
  {Vu}}\ and\ \bibinfo {author} {\bibfnamefont {J.}~\bibnamefont {Weiss}},\
  }\href {\doibase 10.1103/PhysRevLett.125.105502} {\bibfield  {journal}
  {\bibinfo  {journal} {Phys. Rev. Lett.}\ }\textbf {\bibinfo {volume} {125}},\
  \bibinfo {pages} {105502} (\bibinfo {year} {2020})}\BibitemShut {NoStop}%
\bibitem [{\citenamefont {Jani\ifmmode \acute{c}\else
  \'{c}\fi{}evi\ifmmode~\acute{c}\else \'{c}\fi{}}\ \emph
  {et~al.}(2016)\citenamefont {Jani\ifmmode \acute{c}\else
  \'{c}\fi{}evi\ifmmode~\acute{c}\else \'{c}\fi{}}, \citenamefont {Laurson},
  \citenamefont {M\aa{}l\o{}y}, \citenamefont {Santucci},\ and\ \citenamefont
  {Alava}}]{Janicevic2016}%
  \BibitemOpen
  \bibfield  {author} {\bibinfo {author} {\bibfnamefont {S.}~\bibnamefont
  {Jani\ifmmode \acute{c}\else \'{c}\fi{}evi\ifmmode~\acute{c}\else
  \'{c}\fi{}}}, \bibinfo {author} {\bibfnamefont {L.}~\bibnamefont {Laurson}},
  \bibinfo {author} {\bibfnamefont {K.~J.}\ \bibnamefont {M\aa{}l\o{}y}},
  \bibinfo {author} {\bibfnamefont {S.}~\bibnamefont {Santucci}}, \ and\
  \bibinfo {author} {\bibfnamefont {M.~J.}\ \bibnamefont {Alava}},\ }\href
  {\doibase 10.1103/PhysRevLett.117.230601} {\bibfield  {journal} {\bibinfo
  {journal} {Phys. Rev. Lett.}\ }\textbf {\bibinfo {volume} {117}},\ \bibinfo
  {pages} {230601} (\bibinfo {year} {2016})}\BibitemShut {NoStop}%
\bibitem [{\citenamefont {Salditch}\ \emph {et~al.}(2020)\citenamefont
  {Salditch}, \citenamefont {Stein}, \citenamefont {Neely}, \citenamefont
  {Spencer}, \citenamefont {Brooks}, \citenamefont {Agnon},\ and\ \citenamefont
  {Liu}}]{Salditch2020}%
  \BibitemOpen
  \bibfield  {author} {\bibinfo {author} {\bibfnamefont {L.}~\bibnamefont
  {Salditch}}, \bibinfo {author} {\bibfnamefont {S.}~\bibnamefont {Stein}},
  \bibinfo {author} {\bibfnamefont {J.}~\bibnamefont {Neely}}, \bibinfo
  {author} {\bibfnamefont {B.~D.}\ \bibnamefont {Spencer}}, \bibinfo {author}
  {\bibfnamefont {E.~M.}\ \bibnamefont {Brooks}}, \bibinfo {author}
  {\bibfnamefont {A.}~\bibnamefont {Agnon}}, \ and\ \bibinfo {author}
  {\bibfnamefont {M.}~\bibnamefont {Liu}},\ }\href {\doibase
  https://doi.org/10.1016/j.tecto.2019.228289} {\bibfield  {journal} {\bibinfo
  {journal} {Tectonophysics}\ }\textbf {\bibinfo {volume} {774}},\ \bibinfo
  {pages} {228289} (\bibinfo {year} {2020})}\BibitemShut {NoStop}%
\bibitem [{\citenamefont {Karimi}\ \emph {et~al.}(2017)\citenamefont {Karimi},
  \citenamefont {Ferrero},\ and\ \citenamefont {Barrat}}]{Karimi2017}%
  \BibitemOpen
  \bibfield  {author} {\bibinfo {author} {\bibfnamefont {K.}~\bibnamefont
  {Karimi}}, \bibinfo {author} {\bibfnamefont {E.~E.}\ \bibnamefont {Ferrero}},
  \ and\ \bibinfo {author} {\bibfnamefont {J.-L.}\ \bibnamefont {Barrat}},\
  }\href {\doibase 10.1103/PhysRevE.95.013003} {\bibfield  {journal} {\bibinfo
  {journal} {Phys. Rev. E}\ }\textbf {\bibinfo {volume} {95}},\ \bibinfo
  {pages} {013003} (\bibinfo {year} {2017})}\BibitemShut {NoStop}%
\bibitem [{\citenamefont {Anthony}\ and\ \citenamefont
  {Marone}(2005)}]{Anthony2005}%
  \BibitemOpen
  \bibfield  {author} {\bibinfo {author} {\bibfnamefont {J.~L.}\ \bibnamefont
  {Anthony}}\ and\ \bibinfo {author} {\bibfnamefont {C.}~\bibnamefont
  {Marone}},\ }\href {\doibase 10.1029/2004JB003399} {\bibfield  {journal}
  {\bibinfo  {journal} {Journal of Geophysical Research: Solid Earth}\ }\textbf
  {\bibinfo {volume} {110}} (\bibinfo {year} {2005}),\ 10.1029/2004JB003399},\ %
\bibitem [{\citenamefont {Frenkel}\ \emph {et~al.}(2008)\citenamefont
  {Frenkel}, \citenamefont {Blumenfeld}, \citenamefont {Grof},\ and\
  \citenamefont {King}}]{Frenkel2008}%
  \BibitemOpen
  \bibfield  {author} {\bibinfo {author} {\bibfnamefont {G.}~\bibnamefont
  {Frenkel}}, \bibinfo {author} {\bibfnamefont {R.}~\bibnamefont {Blumenfeld}},
  \bibinfo {author} {\bibfnamefont {Z.}~\bibnamefont {Grof}}, \ and\ \bibinfo
  {author} {\bibfnamefont {P.~R.}\ \bibnamefont {King}},\ }\href {\doibase
  10.1103/PhysRevE.77.041304} {\bibfield  {journal} {\bibinfo  {journal} {Phys.
  Rev. E}\ }\textbf {\bibinfo {volume} {77}},\ \bibinfo {pages} {041304}
  (\bibinfo {year} {2008})}\BibitemShut {NoStop}%
\bibitem [{\citenamefont {Henkes}\ \emph {et~al.}(2007)\citenamefont {Henkes},
  \citenamefont {O'Hern},\ and\ \citenamefont {Chakraborty}}]{Henkes2007}%
  \BibitemOpen
  \bibfield  {author} {\bibinfo {author} {\bibfnamefont {S.}~\bibnamefont
  {Henkes}}, \bibinfo {author} {\bibfnamefont {C.~S.}\ \bibnamefont {O'Hern}},
  \ and\ \bibinfo {author} {\bibfnamefont {B.}~\bibnamefont {Chakraborty}},\
  }\href {\doibase 10.1103/PhysRevLett.99.038002} {\bibfield  {journal}
  {\bibinfo  {journal} {Phys. Rev. Lett.}\ }\textbf {\bibinfo {volume} {99}},\
  \bibinfo {pages} {038002} (\bibinfo {year} {2007})}\BibitemShut {NoStop}%
\bibitem [{\citenamefont {Jagla}(2015)}]{Jagla2015}%
  \BibitemOpen
  \bibfield  {author} {\bibinfo {author} {\bibfnamefont {E.~A.}\ \bibnamefont
  {Jagla}},\ }\href {\doibase 10.1103/PhysRevE.92.042135} {\bibfield  {journal}
  {\bibinfo  {journal} {Phys. Rev. E}\ }\textbf {\bibinfo {volume} {92}},\
  \bibinfo {pages} {042135} (\bibinfo {year} {2015})}\BibitemShut {NoStop}%
\bibitem [{\citenamefont {Bar\'es}\ \emph {et~al.}(2017)\citenamefont
  {Bar\'es}, \citenamefont {Wang}, \citenamefont {Wang}, \citenamefont
  {Bertrand}, \citenamefont {O'Hern},\ and\ \citenamefont
  {Behringer}}]{Bares2017}%
  \BibitemOpen
  \bibfield  {author} {\bibinfo {author} {\bibfnamefont {J.}~\bibnamefont
  {Bar\'es}}, \bibinfo {author} {\bibfnamefont {D.}~\bibnamefont {Wang}},
  \bibinfo {author} {\bibfnamefont {D.}~\bibnamefont {Wang}}, \bibinfo {author}
  {\bibfnamefont {T.}~\bibnamefont {Bertrand}}, \bibinfo {author}
  {\bibfnamefont {C.~S.}\ \bibnamefont {O'Hern}}, \ and\ \bibinfo {author}
  {\bibfnamefont {R.~P.}\ \bibnamefont {Behringer}},\ }\href {\doibase
  10.1103/PhysRevE.96.052902} {\bibfield  {journal} {\bibinfo  {journal} {Phys.
  Rev. E}\ }\textbf {\bibinfo {volume} {96}},\ \bibinfo {pages} {052902}
  (\bibinfo {year} {2017})}\BibitemShut {NoStop}%
\bibitem [{\citenamefont {Ma}\ \emph {et~al.}(2020)\citenamefont {Ma},
  \citenamefont {Zou}, \citenamefont {Gao}, \citenamefont {Zhao},\ and\
  \citenamefont {Zhou}}]{Ma2020}%
  \BibitemOpen
  \bibfield  {author} {\bibinfo {author} {\bibfnamefont {G.}~\bibnamefont
  {Ma}}, \bibinfo {author} {\bibfnamefont {Y.}~\bibnamefont {Zou}}, \bibinfo
  {author} {\bibfnamefont {K.}~\bibnamefont {Gao}}, \bibinfo {author}
  {\bibfnamefont {J.}~\bibnamefont {Zhao}}, \ and\ \bibinfo {author}
  {\bibfnamefont {W.}~\bibnamefont {Zhou}},\ }\href {\doibase
  https://doi.org/10.1029/2020GL090458} {\bibfield  {journal} {\bibinfo
  {journal} {Geophysical Research Letters}\ }\textbf {\bibinfo {volume} {47}},\
  \bibinfo {pages} {e2020GL090458} (\bibinfo {year} {2020})},\ \bibinfo {note}
  {e2020GL090458 2020GL090458},\ %
\bibitem [{\citenamefont {Clark}\ \emph
  {et~al.}(2015{\natexlab{a}})\citenamefont {Clark}, \citenamefont {Petersen},
  \citenamefont {Kondic},\ and\ \citenamefont {Behringer}}]{Clark2015}%
  \BibitemOpen
  \bibfield  {author} {\bibinfo {author} {\bibfnamefont {A.~H.}\ \bibnamefont
  {Clark}}, \bibinfo {author} {\bibfnamefont {A.~J.}\ \bibnamefont {Petersen}},
  \bibinfo {author} {\bibfnamefont {L.}~\bibnamefont {Kondic}}, \ and\ \bibinfo
  {author} {\bibfnamefont {R.~P.}\ \bibnamefont {Behringer}},\ }\href {\doibase
  10.1103/PhysRevLett.114.144502} {\bibfield  {journal} {\bibinfo  {journal}
  {Phys. Rev. Lett.}\ }\textbf {\bibinfo {volume} {114}},\ \bibinfo {pages}
  {144502} (\bibinfo {year} {2015}{\natexlab{a}})}\BibitemShut {NoStop}%
\bibitem [{\citenamefont {Clark}\ \emph
  {et~al.}(2015{\natexlab{b}})\citenamefont {Clark}, \citenamefont {Petersen},
  \citenamefont {Kondic},\ and\ \citenamefont
  {Behringer}}]{clark2015nonlinear}%
  \BibitemOpen
  \bibfield  {author} {\bibinfo {author} {\bibfnamefont {A.~H.}\ \bibnamefont
  {Clark}}, \bibinfo {author} {\bibfnamefont {A.~J.}\ \bibnamefont {Petersen}},
  \bibinfo {author} {\bibfnamefont {L.}~\bibnamefont {Kondic}}, \ and\ \bibinfo
  {author} {\bibfnamefont {R.~P.}\ \bibnamefont {Behringer}},\ }
  {\bibfield  {journal} {\bibinfo  {journal} {Physical review letters}\
  }\textbf {\bibinfo {volume} {114}},\ \bibinfo {pages} {144502} (\bibinfo
  {year} {2015}{\natexlab{b}})}\BibitemShut {NoStop}%
\bibitem [{\citenamefont {Niemann}\ \emph {et~al.}(2012)\citenamefont
  {Niemann}, \citenamefont {Bar\'o}, \citenamefont {Heczko}, \citenamefont
  {Schultz}, \citenamefont {F\"ahler}, \citenamefont {Vives}, \citenamefont
  {Ma\~nosa},\ and\ \citenamefont {Planes}}]{Niemann2012}%
  \BibitemOpen
  \bibfield  {author} {\bibinfo {author} {\bibfnamefont {R.}~\bibnamefont
  {Niemann}}, \bibinfo {author} {\bibfnamefont {J.}~\bibnamefont {Bar\'o}},
  \bibinfo {author} {\bibfnamefont {O.}~\bibnamefont {Heczko}}, \bibinfo
  {author} {\bibfnamefont {L.}~\bibnamefont {Schultz}}, \bibinfo {author}
  {\bibfnamefont {S.}~\bibnamefont {F\"ahler}}, \bibinfo {author}
  {\bibfnamefont {E.}~\bibnamefont {Vives}}, \bibinfo {author} {\bibfnamefont
  {L.}~\bibnamefont {Ma\~nosa}}, \ and\ \bibinfo {author} {\bibfnamefont
  {A.}~\bibnamefont {Planes}},\ }\href {\doibase 10.1103/PhysRevB.86.214101}
  {\bibfield  {journal} {\bibinfo  {journal} {Phys. Rev. B}\ }\textbf {\bibinfo
  {volume} {86}},\ \bibinfo {pages} {214101} (\bibinfo {year}
  {2012})}\BibitemShut {NoStop}%
\end{thebibliography}
\end{document}


\title
{
Supplementary material for:
Quasistatic kinetic avalanches and self-organized criticality in deviatorically loaded granular media}

\begin{abstract}
Additional figures complementing the results presented in ``Quasistatic kinetic avalanches and self-organized criticality in deviatorically loaded granular media''. These include: (Fig.~S1) the distribution of avalanche durations ($T$); (Fig.~S2) the effective scaling relation between $\sigma_x$ and the relation $\sigma_x^2(1-\phi))^{-2}/Z$ expected to reproduce the scaling of $\Delta U (\Delta \sigma_y / \sigma_x)$; (Fig.~S3) the visual representation of the temporal correlations for simulations not presented in the main manuscript; (Fig.~S4, (Fig.~S5) the distributions of $\Delta U$ and $K$ in SI units with the used intervals for the exponent estimation; (Fig.~S6) a numerical verification for the prevalence of a pseudogap for small system sizes; (Fig.~S7) the predicted proportionality between $K$ and an auxiliary variable $\Delta (U^2)$.
\end{abstract}

\author{Jordi Bar\'o}
\email{jbaro@crm.cat}
\affiliation{Department of Physics and Astronomy,
University of Calgary.
2500 University Drive NW
Calgary, Alberta  T2N 1N4, Canada}

\affiliation{Centre for Mathematical Research,
Campus de Bellaterra, Edifici C	 
08193 Bellaterra, Barcelona, Spain}

\author{Mehdi Pouragha}
\affiliation{Civil Engineering Department, 
University of Calgary.
2500 University Drive NW
Calgary, Alberta  T2N 1N4, Canada}
\affiliation{Department of Civil and Environmental Engineering, 
Carleton University.
1125 Colonel By Drive, Ottawa, ON K1S 5B6 Canada}

\author{Richard Wan}
\affiliation{Civil Engineering Department, 
University of Calgary.
2500 University Drive NW
Calgary, Alberta  T2N 1N4, Canada}

\author{J\"orn Davidsen}%
\affiliation{Department of Physics and Astronomy,
University of Calgary.
2500 University Drive NW
Calgary, Alberta  T2N 1N4, Canada} 
\affiliation{Hotchkiss Brain Institute,
University of Calgary.
3330 Hospital Drive NW,
Calgary, Alberta  T2N 4N1, Canada}
  
\maketitle

\onecolumngrid

\begin{figure}
\includegraphics[width=0.6\columnwidth]{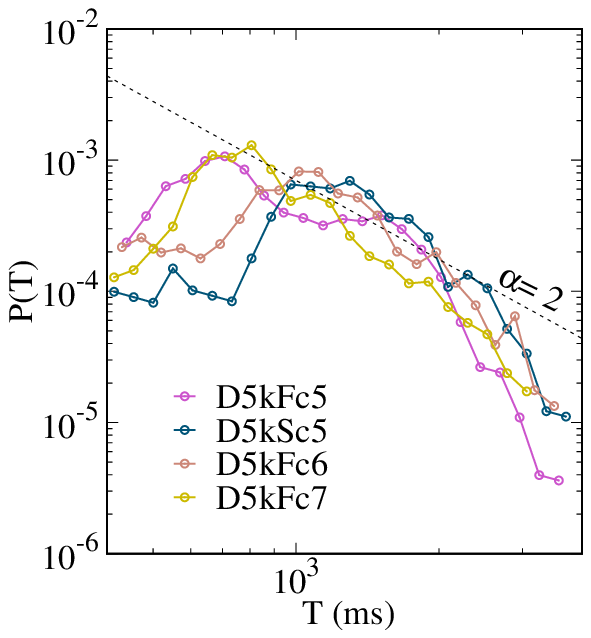} 
\caption{Distribution of avalanche durations (T), taken as the time between the avalanche onset and the first rebound ($d \sigma_y=0$) characterizing the elastic energy drop. The exponent is apparently close to 2, but the power-law range is too short to provide a reliable estimate.}
\end{figure}

\begin{figure}[h]
\centering
\includegraphics[width=0.4\columnwidth]{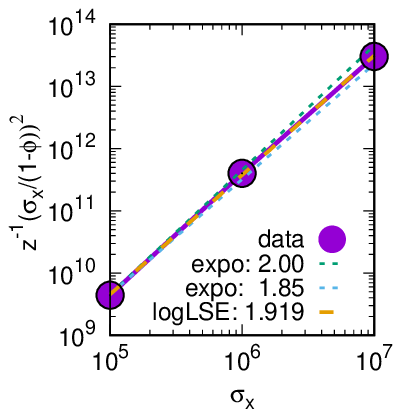} 
\caption{Effective scaling of the term $\sigma_x/(1-\phi)^2/z$ given the corrections in the dependencies of $\phi(\sigma_x)$ and  $z(\sigma_x)$. The best fit is given by exponent $1.91(4)$ (yellow dashed line). Results with exponent $2$ (uncorrected) and $1.85$ (scaling observed in Fig.~\ref{fig:stressEnergy}) are shown for comparisson.}
\end{figure}

\begin{figure}[]
\centering
    \includegraphics[width=0.75\columnwidth]{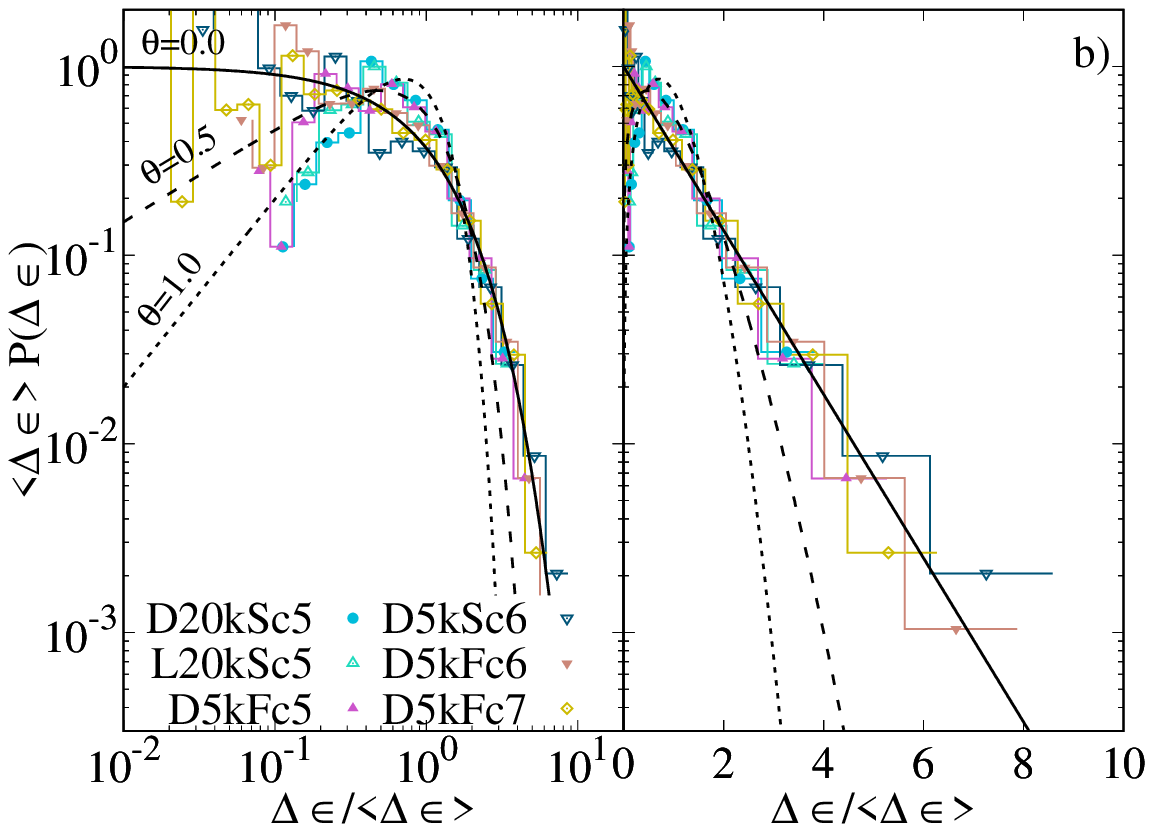}\\
    \includegraphics[width=0.75\columnwidth]{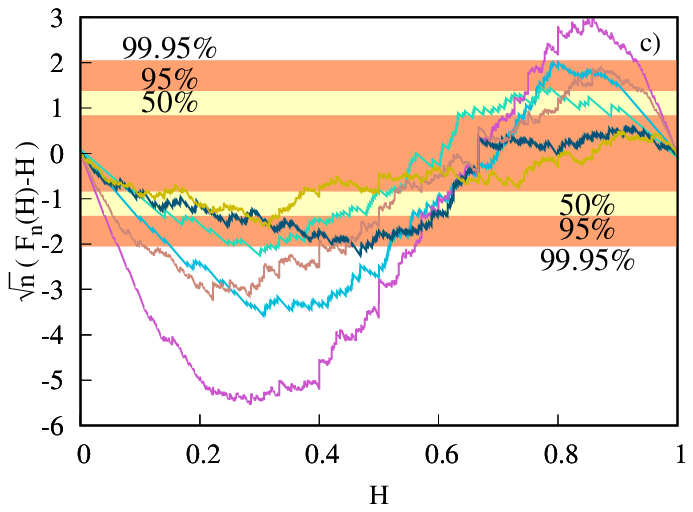}
    \caption{(a,b) Scaled distribution of interevent times and (c) Kolmogorov–Smirnov test over the $H$-values obtained by the Bi-test for Poisson rejection for all the samples not represented in the Fig.~9 of the main text.}
\end{figure}

\begin{figure}[h]
\includegraphics[width=\columnwidth]{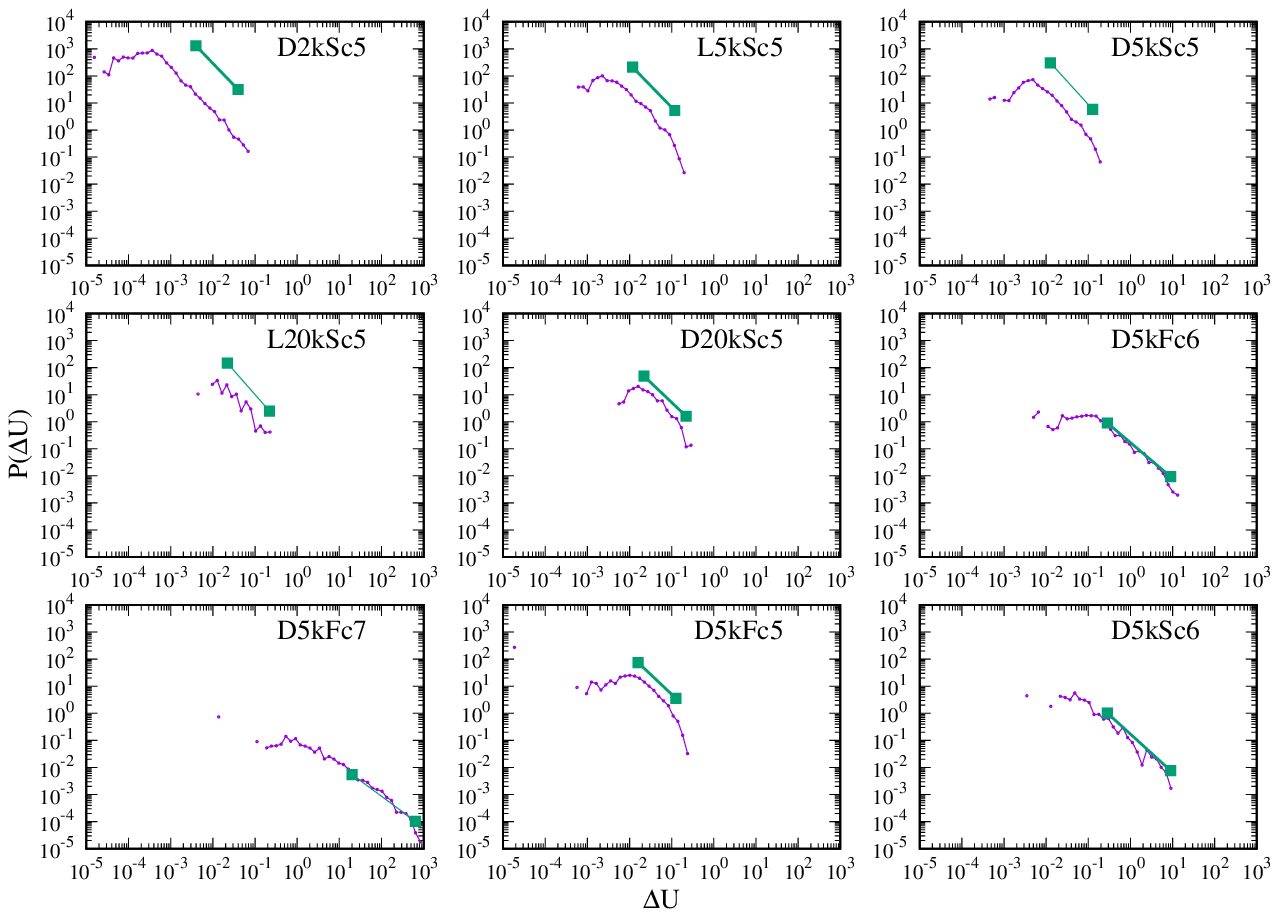} 
\caption{Non-scaled distributions of $\Delta U$ represented in purple histograms, with the selected intervals for the exponent estimation, represented in green lines. The intervals are selected according to the best estimates of the scaling exponents  ${\gamma_\sigma^{\Delta U}}= 1.85$ and ${\gamma_N^{\Delta U}}= 0.50$, as explained in the main text.}
\end{figure}

\begin{figure}[h]
\includegraphics[width=\columnwidth]{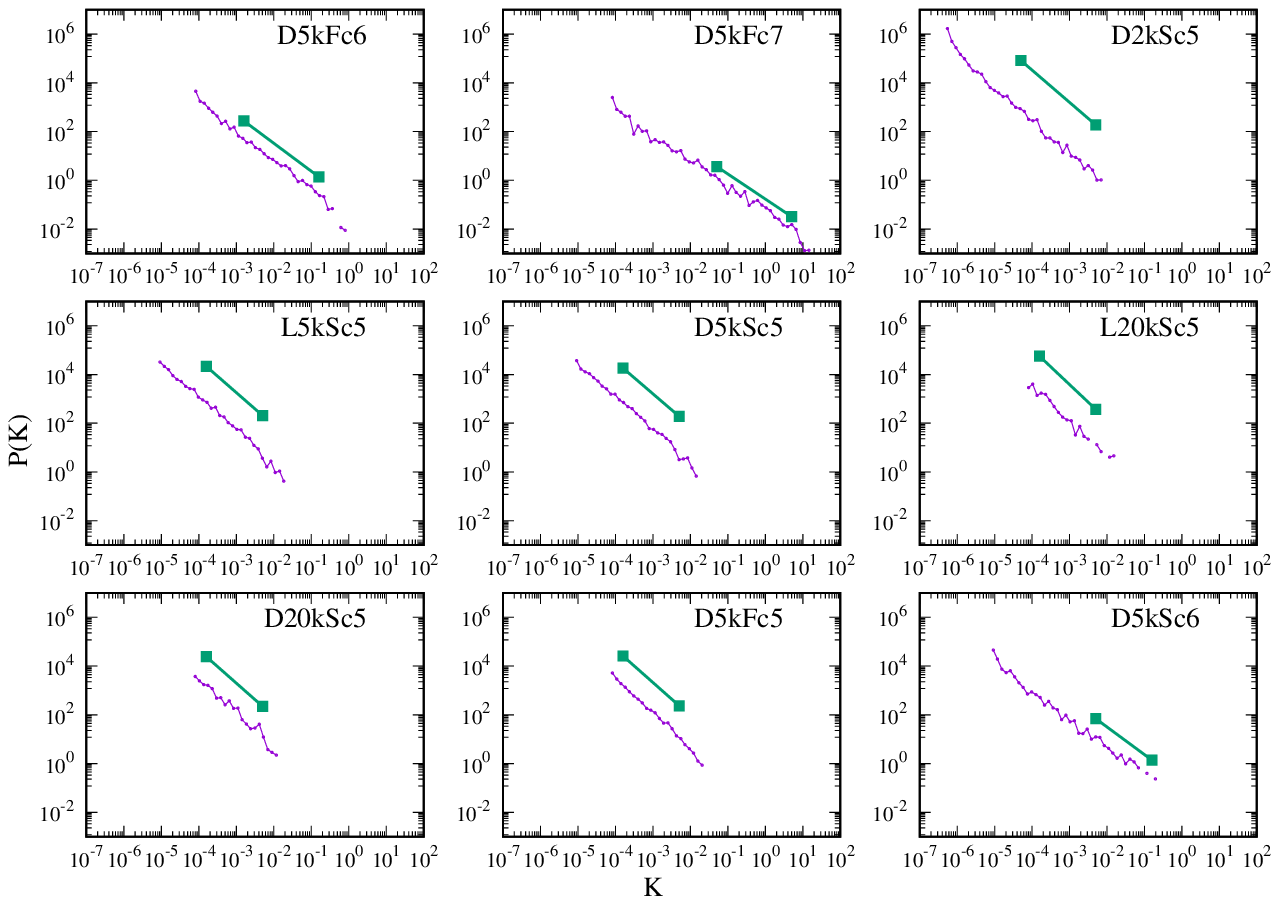} 
\caption{Non-scaled distributions of $K$ represented in purple histograms, with the selected intervals for the exponent estimation, represented in green lines. The intervals are selected according to the best estimates of the scaling exponents  ${\gamma_\sigma^{K}}= 1.50$ and ${\gamma_N^{K}}= 0.00$, as explained in the main text.}
\end{figure}

\begin{figure}[]
\includegraphics[width=0.6\columnwidth]{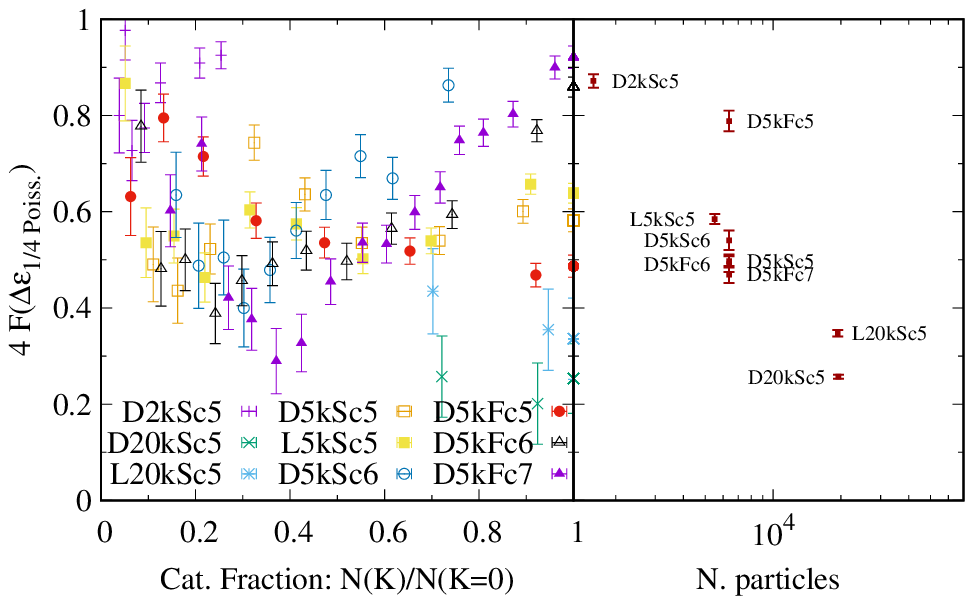} 
\caption{Normalized fraction of events within the expected range  of the first quartile of an exponential distribution in the numerical distributions of scaled waiting times ($\Delta\epsilon / \langle \Delta\epsilon \rangle $). We divide the number of events with $\Delta\epsilon / \langle \Delta\epsilon \rangle \in [0, \log(3/4)]$ by the expected fraction ($1/4$) in an exponential distribution, i.e. the interevent times in a Poisson process. (a) Results considering different energy ($K$) thresholds for all simulations (color-coded). The x-axis represents the catalog completeness, which decreases with the energy threshold ($K$). (b) Averages for all the energy thresholds represented in (a) for each catalog. The x-axis shows the number of particles in each simulation. In general, we observe an increase of the discrepancy with the exponential distribution (lower fraction) with the system size.}
\end{figure}

\begin{figure}[h]
\includegraphics[width=0.6\columnwidth]{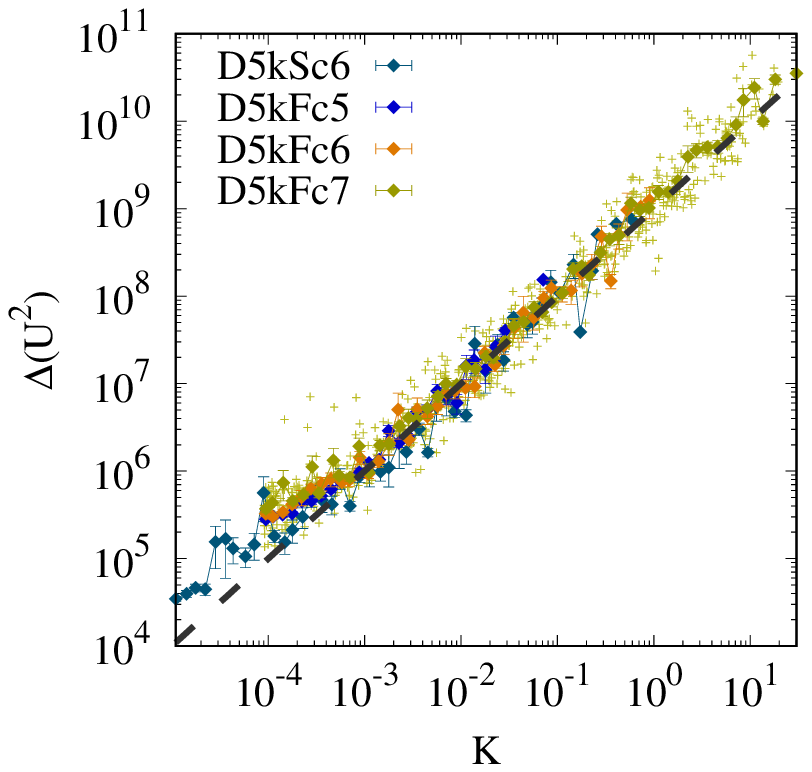} 
\caption{Relation between $K$ and $\Delta (U^2):= \int dt (dU/dt)^2$. Conditional averages are represented in lines with errorbars representing the std. deviation. Scatter plot for D5kf7 shown as an example (yellow dots). The grey dashed line represents a proportionality relation $\Delta(U^2) \approx K$}
\end{figure}